\documentclass[fleqn,usenatbib]{mnras}

\usepackage{newtxtext,newtxmath}
\usepackage[T1]{fontenc}
\usepackage{ae,aecompl}

\usepackage{subcaption}
\usepackage{float}
\usepackage{graphicx}	% Including figure files
\usepackage{amsmath}	% Advanced maths commands
\usepackage{amssymb}	% Extra maths symbols

\title[Spectra and Thermal Inversions due to TiO]{Assessing Spectra and Thermal Inversions due to TiO in Hot Jupiter Atmospheres}

\author[A. A. A. Piette et al.]{Anjali A. A. Piette,$^{1}$\thanks{E-mail: ap763@cam.ac.uk}
Nikku Madhusudhan,$^{1}$\thanks{E-mail: nmadhu@ast.cam.ac.uk}
Laura K. McKemmish,$^{2}$
\newauthor
Siddharth Gandhi,$^{1,3}$
Thomas Masseron,$^{4,5}$
Luis Welbanks$^{1}$
\\
$^{1}$Institute of Astronomy, University of Cambridge, Madingley Road, Cambridge, CB3 0HA, UK\\
$^{2}$School of Chemistry, University of New South Wales, Kensington, Sydney, 2052 Australia\\
$^{3}$Department of Physics, University of Warwick, Gibbet
Hill Road, Coventry CV4 7AL\\
$^{4}$Instituto de Astrof\'isica de Canarias, E-38205 La Laguna, Tenerife, Spain\\
$^{5}$Departamento de Astrof\'isica, Universidad de La Laguna, E-38206 La Laguna, Tenerife, Spain
}

\date{Accepted XXX. Received YYY; in original form ZZZ}

\pubyear{2019}

\begin{document}
\label{firstpage}
\pagerange{\pageref{firstpage}--\pageref{lastpage}}
\maketitle

\begin{abstract}
Recent detections of thermal inversions in the dayside atmospheres of some hot Jupiters are motivating new avenues to understand the interplay between their temperature structures and other atmospheric conditions. In particular, TiO has long been proposed to cause thermal inversions in hot Jupiters, depending on other factors such as stellar irradiation, C/O, and vertical mixing. TiO also has spectral features in the optical and near-infrared that have been detected. However, interpretations of TiO signatures rely on the accuracy of TiO opacity used in the models. The recently reported \textsc{Toto} TiO line list provides a new opportunity to investigate these dependencies, which is the goal of the present work. First, we investigate how the \textsc{Toto} line list affects observable transmission and emission spectra of hot Jupiters at low and high resolution. The improvement in the \textsc{Toto} line list compared to a previous line list results in observable differences in the model spectra, particularly in the optical at high resolution. Secondly, we explore the interplay between temperature structure, irradiation and composition with TiO as the primary source of optical opacity, using 1D self-consistent atmospheric models. Among other trends, we find that the propensity for thermal inversions due to TiO peaks at $\mathrm{C/O}\sim$0.9, consistent with recent studies. Using these models, we further assess metrics to quantify thermal inversions due to TiO, compared to frequently-used Spitzer photometry, over a range in C/O, irradiation, metallicity, gravity and stellar type.
\end{abstract}


%%%%%%%%%%%%%%%%%%%%%%%%%%%%%%%%%%%%%%%%%%%%%%%%%%

%%%%%%%%%%%%%%%%% BODY OF PAPER %%%%%%%%%%%%%%%%%%

\section{Introduction}
\label{sec:intro}
In the quest to understand exoplanets in ever-increasing detail, spectroscopic observations of increasing quality are being used to study exoplanetary atmospheres. In particular, secondary eclipse spectra of transiting exoplanets provide a unique opportunity to constrain both the chemical properties and thermal structures of their dayside atmospheres \citep[e.g.][]{Seager2005,Burrows2008}. Both the chemical and thermal properties of an atmosphere are deeply intertwined and many works to date have aimed to shed light on the various processes which shape exoplanet emission spectra  \citep[e.g.][]{Hubeny2003,Fortney2006,Burrows2008,Zahnle2009,Madhusudhan2012,Molliere2015,Parmentier2018}. In particular, the phenomenon of thermal inversions has been the focus of much work for over a decade, with approaches from both theory and observation  \citep[e.g.][]{Hubeny2003,Burrows2008,Fortney2008,Spiegel2009,Knutson2010,Madhusudhan2012,Menou2012,Molliere2015,Beatty2017,Parmentier2018,Lothringer2019,Gandhi2019}.

Although thermal inversions are known to exist in the atmospheres of solar system planets \citep[e.g.][]{Moses2005,Robinson2014}, the molecules which cause them (e.g. ozone in Earth's atmosphere and hydrocarbon hazes on Jupiter) cannot exist in gas phase at the high temperatures present in hot Jupiter atmospheres. However, \citet{Hubeny2003} suggested that alternative, high-temperature, absorbers such as TiO or VO could produce thermal inversions in strongly-irradiated exoplanets, opening up the field of exoplanet atmospheres to a new class of thermal structures \citep[e.g.][]{Fortney2006,Burrows2007}.

Further developments in the theory of thermal inversions have been driven by both trends observed in exoplanet emission spectra and theoretical considerations. For example, \citet{Fortney2008} suggested a two-way classification of hot Jupiters based on stellar irradiation, arguing that inversion-causing species such as TiO and VO would only be present in the gas phase for the hotter, more strongly-irradiated class. This was in agreement with inferences of thermal inversions at the time, with HD 209458 b at the boundary between the two classes (\citealp{Burrows2008,Knutson2008} but cf. \citealp{Diamond-Lowe2014}). Not all stellar irradiation aids the formation of thermal inversions, however. \citet{Knutson2010} found a negative correlation between host star activity and the presence of thermal inversions, leading them to suggest that the increased UV flux from active stars could result in the photodissociation of inversion-causing compounds and hinder the presence of thermal inversions. However, inferences of thermal inversions in some planets have since been updated, e.g. HD~209458b, see \citet{Diamond-Lowe2014}.

Beyond photodissociation, there are further challenges in creating thermal inversions with TiO/VO. One is the cold trap effect, whereby these compounds condense out either on the nightside or in certain (cooler) regions in the atmosphere, which can deplete the abundance of gaseous TiO/VO \citep{Spiegel2009,Beatty2017}. Another is the fact that TiO and VO are heavy compounds relative to a H$_2$-dominated atmosphere and will gravitationally settle if they are not kept aloft by other means \citep{Spiegel2009}. However, both of these effects could be mitigated by vertical mixing in the most irradiated atmospheres \citep[e.g.][]{Spiegel2009}. Indeed, \citet{Parmentier2013} find that TiO can be kept aloft in the atmosphere by vertical mixing as long as it forms sufficiently small particles when condensed on the night side.

To date, three exoplanets are known to host thermal inversions; WASP-18b \citep{Sheppard2017}, WASP-33b \citep{Haynes2015} and WASP-121b \citep{Evans2017}. These inferences are typically made using a combination of thermal emission spectra obtained with the HST WFC3 spectrograph (1.1-1.7 $\mu$m) and broadband photometry in the Spitzer IRAC bands at 3.6 and 4.5 $\mu$m. Whereas the HST WFC3 band is a good probe of H$_2$O opacity and the spectral continuum, the IRAC 4.5 $\mu$m band has strong opacity due to CO, which is expected to be abundant in hot Jupiter atmospheres. For solar compositions and sufficiently high temperatures, the IRAC 3.6 $\mu$m band has relatively low opacity and is therefore used as a measure of the spectral continuum against which the IRAC 4.5 $\mu$m band can be compared. Under these conditions, a higher brightness temperature in the IRAC 4.5 $\mu$m band relative to the 3.6 $\mu$m band indicates the presence of a CO emission feature and an inverted temperature profile. This metric has commonly been used to assess the presence of thermal inversions in hot Jupiter atmospheres \citep{Burrows2007,Knutson2010,Madhusudhan2010}.

On the other hand, many highly-irradiated ultra-hot Jupiters (with effective temperature $\gtrsim$ 2000 K) have emission spectra consistent with blackbody curves in the near-infrared \citep[e.g.][]{Crossfield2012,Delrez2018}. One explanation for this observation is an isothermal temperature profile \citep[e.g.][]{Crossfield2012}, though an absence of strong near-infrared absorbers such as H$_2$O would also result in a featureless spectrum in the NIR \citep[e.g.][]{Madhusudhan2011}. Such a depletion in H$_2$O could be caused by either a super-solar C/O ratio \citep{Madhusudhan2012,Moses2013} or due to its thermal dissociation in ultra-hot Jupiters \citep{Parmentier2018}. Several recent studies also suggest that the continuum opacity in the near-infrared due to H$^-$ ions, which can exist in ultra-hot Jupiters, could reduce the amplitude of the H$_2$O features in the WFC3 band \citep[e.g.][]{Arcangeli2018,Lothringer2018,Parmentier2018}.

Among the factors which shape the thermal profile of an atmosphere, chemistry is a critical component. In particular, the C/O ratio has an important role in determining atmospheric chemistry, with high C/O ratios limiting the abundance of O-bearing species including H$_2$O, TiO and VO in hot Jupiter atmospheres \citep{Madhusudhan2011b,Madhusudhan2012}. Here we are focusing on typical hot Jupiters orbiting Sun-like stars. Since a thermal inversion requires a high optical opacity relative to the infrared opacity \citep[e.g.][]{Hubeny2003,Hansen2008,Guillot2010}, a decrease in H$_2$O abundance due to a high C/O ratio can to some extent make it easier for a thermal inversion to occur \citep{Molliere2015,Gandhi2019}. Furthermore, it has been found that the optical opacity causing thermal inversions may come from a variety of compounds besides TiO/VO, including sulfur compounds \citep{Zahnle2009}, Na/K \citep{Molliere2015}, H$^-$ ions \citep{Arcangeli2018,Parmentier2018,Lothringer2018}, and various oxides, hydrides and atomic metals \citep{Lothringer2018,Gandhi2019}. State-of-the-art observations and chemical characterisation are allowing the first detections of such compounds in exoplanet atmospheres. Recent examples include detections of TiO in high resolution \citep{Nugroho2017} and low resolution (\citealt{Sedaghati2017}, but cf. \citealt{Espinoza2019}) as well as an indication of AlO \citep{vonEssen2019}, both species thought to be capable of creating thermal inversions \citep[e.g.][]{Hubeny2003,Fortney2008,Gandhi2019}.

These detailed chemical detections rely on the accuracy and completeness of the molecular cross sections used to interpret spectroscopic observations \citep[e.g.][]{Schwenke1998,Patrascu2015}. While in the past, cross sections were not designed specifically for use with exoplanet spectra, the recent need for such cross sections has led to the development of several state-of-the-art line lists for various molecules at temperatures relevant to exoplanetary atmospheres. \citep[e.g.][]{Rothman2010,Rothman2013,Tennyson2016}. Accurate and complete molecular cross sections also play a key role in determining the spectral appearance and thermal profile of an atmosphere. In particular, for high-resolution spectra, line position accuracy is important as chemical detections are typically made using cross-correlation methods, which are very sensitive to line position \citep{Brogi2012,Birkby2018,Nugroho2017}. Conversely, completeness of a line list affects the strength of spectral features in low-resolution spectra.

In this work, our goal is to explore important factors for assessing thermal inversions in the spectra of hot Jupiters. We begin by investigating the importance of up-to-date molecular cross sections of TiO for inferring the effects of this molecule on the thermal structures and spectra of hot Jupiters. To this end, we compare the latest TiO line list, \textsc{Toto} \citep{McKemmish2019}, against a previous line list \citep{Schwenke1998}, assessing the differences they make to both transmission and emission spectra for low-resolution as well as high-resolution spectroscopic observations. We also embark on a reassessment of the criteria used for quantifying thermal inversions in hot Jupiters. Traditionally, thermal inversions have been assessed based on the relative flux differential between the Spitzer IRAC 1 and IRAC 2 bands at 3.6 $\mu$m and 4.5 $\mu$m, respectively \citep[e.g.][]{Knutson2008,Sheppard2017,Haynes2015,Kreidberg2018}, which relies on the assumption that there is strong CO opacity in the 4.5 $\mu$m band relative to the 3.6 $\mu$m band (i.e. the 3.6 $\mu$m band is a continuum) for hot Jupiters. However, factors such as C/O ratio could affect this assumption. As a result, the IRAC 1/IRAC 2 flux differential may not necessarily be a robust or optimal metric across all atmospheric chemistries. Furthermore, chemical properties such as the C/O ratio have also been shown to influence thermal inversions \citep[e.g.][]{Madhusudhan2011b,Molliere2015,Gandhi2019}, so the way in which they affect the performance of the IRAC 1/IRAC 2 metric may be non-trivial. In this study, we explore the performance of the IRAC 1/IRAC 2 metric as well as an alternative metric as a function of C/O ratio, irradiation, metallicity and gravity.

In what follows, we begin with comparisons of different TiO line lists and their effects on model spectra under different conditions in section \ref{sec:crosssec}. Here we show how the choice of line list could impact the interpretation of emission and transmission spectra at low and high resolution. Using semi-analytic considerations of radiative equilibrium, we then explore in section \ref{sec:eqtio} how the optical versus infrared opacity varies as a function of C/O ratio and temperature, thereby assessing the dayside equilibrium temperatures at which thermal inversions could occur when TiO is a primary optical absorber. In section \ref{sec:h2o_metric}, we assess metrics for quantifying the presence and strength of thermal inversions in the context of solar-composition atmospheres. In section \ref{sec:colmaps}, we then use self-consistent atmospheric models to explore the dependence of thermal inversion strength on irradiation, C/O ratio, metallicity and gravity. We also use these models to investigate the performance of the metrics discussed in section \ref{sec:h2o_metric} as a function of these atmospheric parameters. We summarize our findings and discuss our conclusions in section \ref{sec:summary}.

\section{Impact of T\lowercase{i}O line lists on exoplanet opacity}

%\textsc{Toto} data that was obtained from a \textsc{MARVEL} analysis \cite{McKemmish2017} that incorporated all known and assigned TiO rovibronically resolved spectroscopic data to obtain a single set of \textsc{MARVEL} energies.

%\red{Laura to change focus - new metric quantifying reliability of line list positions}
%TiO can also be detected with high-resolution Doppler spectroscopy \citep{Nugroho2017}. This method involves cross-correlating a high-resolution Doppler spectrum with template spectra. Since the observed spectrum is at a very high spectral resolution, the accuracy of line positions in the cross sections are very important.

%Different line lists may therefore lead to different results from high-resolution Doppler spectra, as their accuracies and completeness may differ. We refer the reader to \citet{McKemmish2019} for a comparison of the

\label{sec:crosssec}
%%%%%%%%%%%%%%%%%%%%% -- Figures -- %%%%%%%%%%%%%%%%%%%%%
\begin{figure*}
\centering
	\begin{subfigure}[b]{0.45\textwidth}
    	\includegraphics[width=\textwidth]{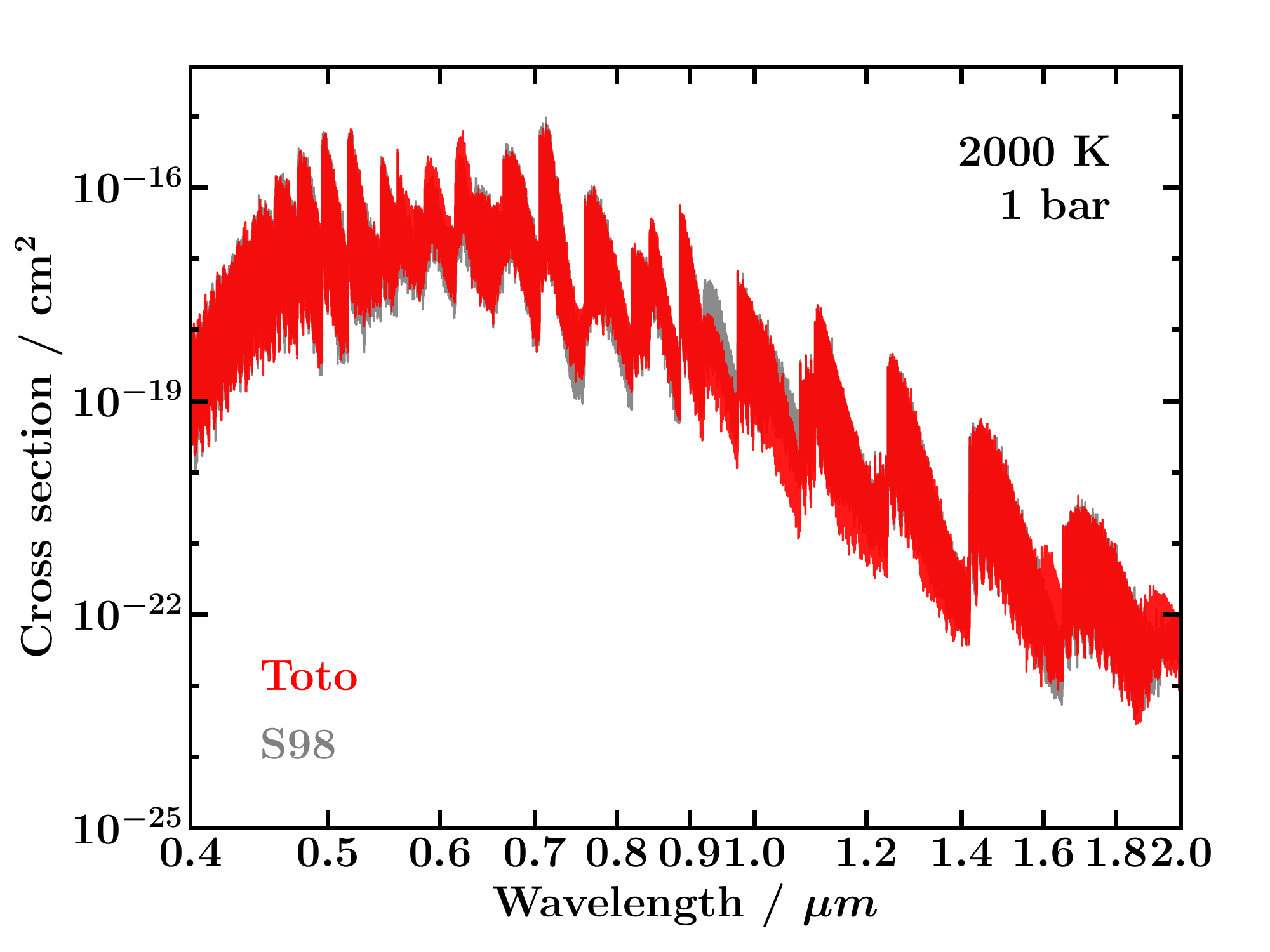}
    \end{subfigure}
    \quad
	\begin{subfigure}[b]{0.45\textwidth}
    	\includegraphics[width=\textwidth]{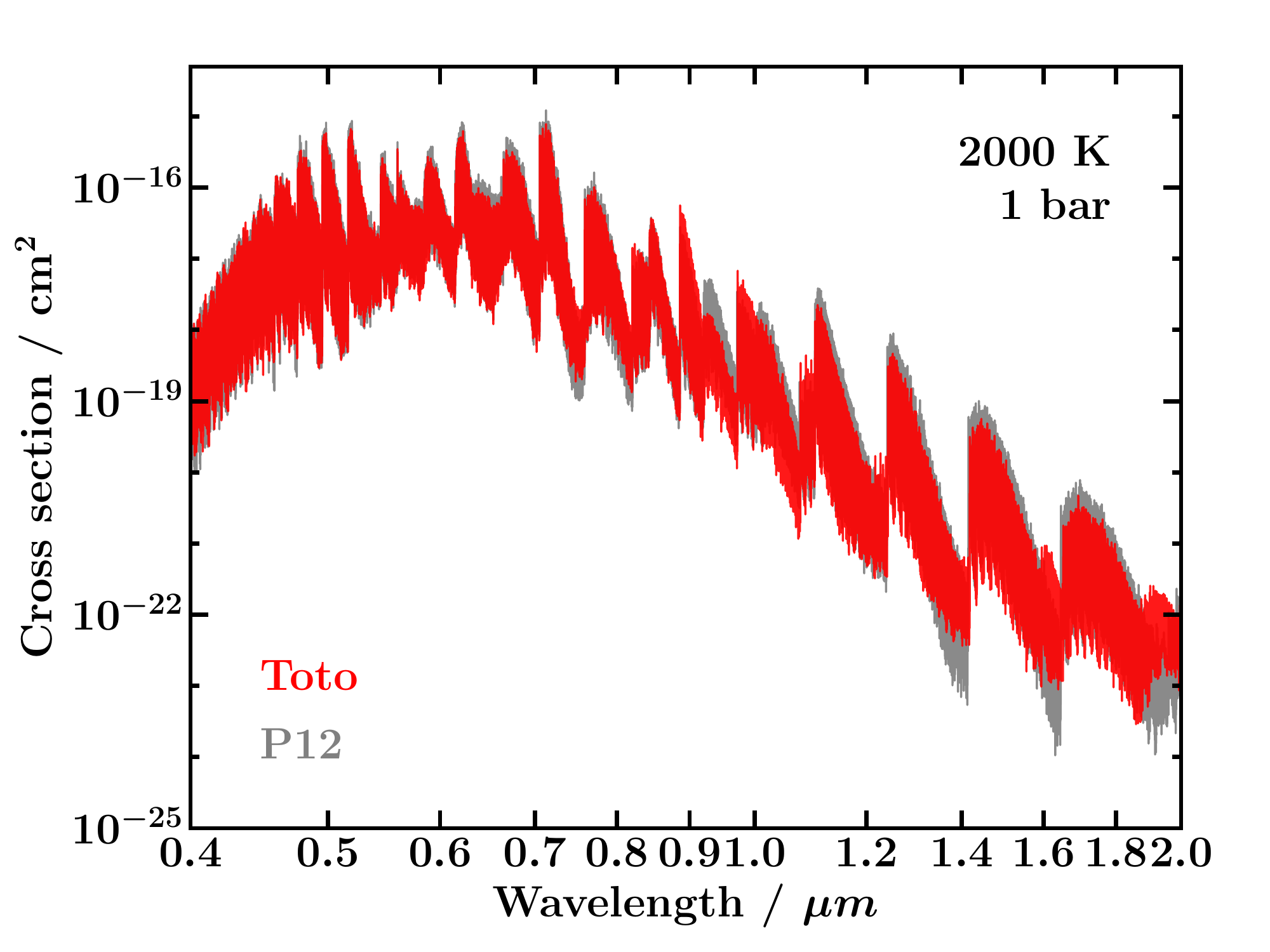}
    \end{subfigure}
	\begin{subfigure}[b]{0.45\textwidth}
    	\includegraphics[width=\textwidth]{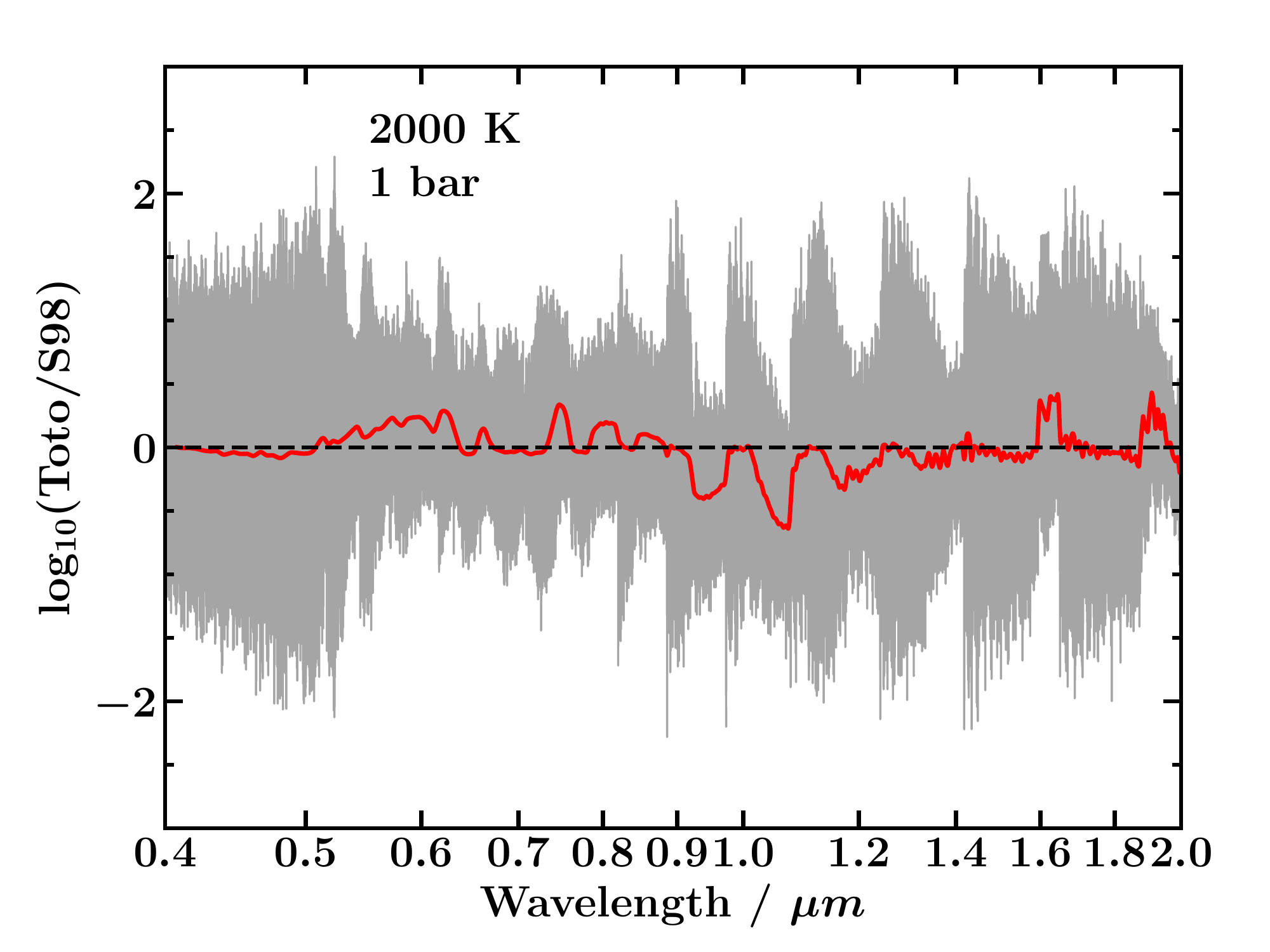}
    \end{subfigure}
	\begin{subfigure}[b]{0.45\textwidth}
    	\includegraphics[width=\textwidth]{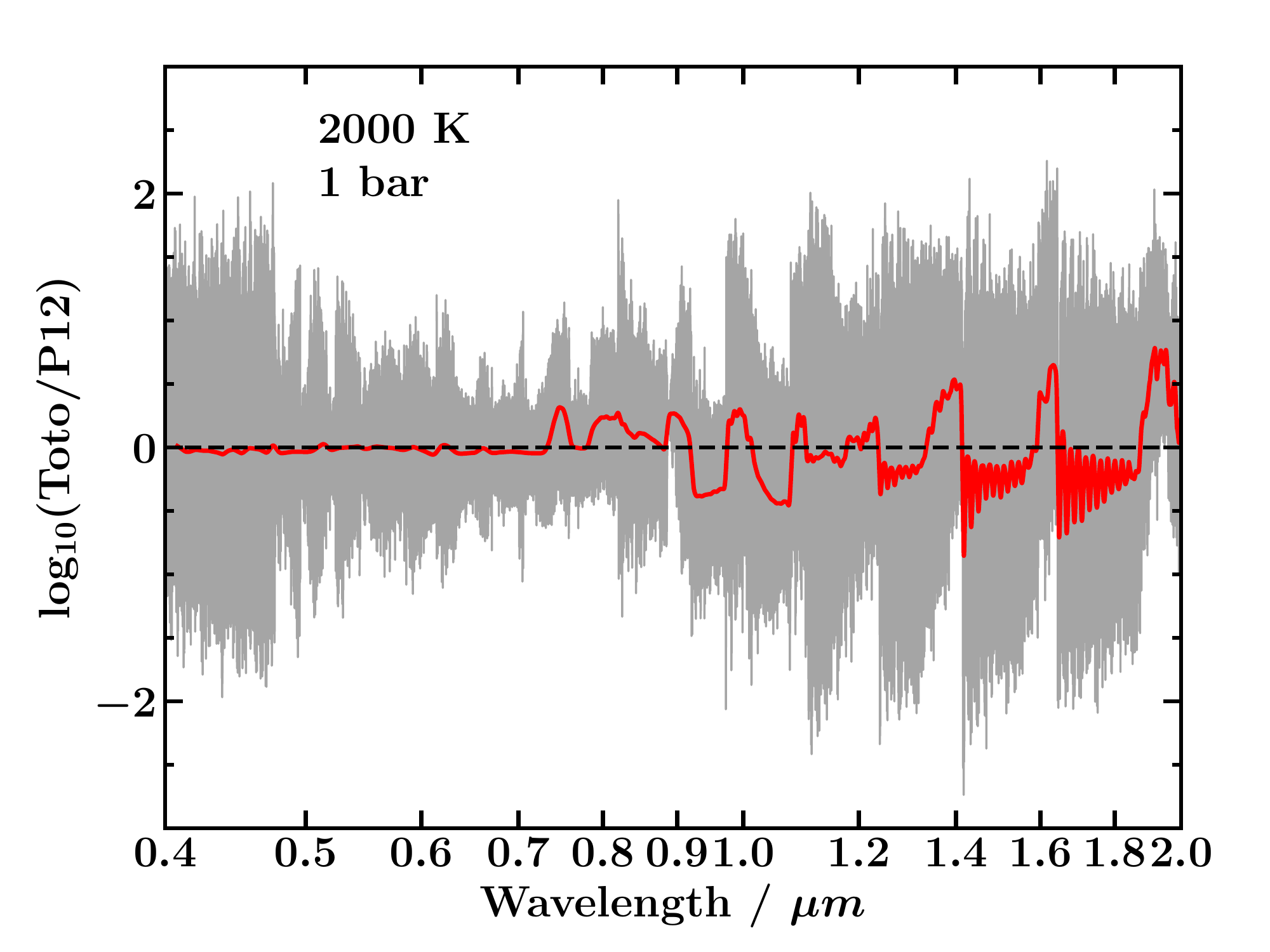}
    \end{subfigure}
\caption{\label{fig:tio_oldnew} Top panels: the S98 (grey, left panel)/ P12 (grey, right panel) and \textsc{Toto} (red) cross sections over-plotted in the range 0.4-2.0$\mu$m. Bottom panels: the log difference between the S98 (left)/ P12 (right) and \textsc{Toto} cross sections in the same wavelength range, at native resolution (grey) and smoothed with a Gaussian of width 2.3 nm (red, similar to the PSF of HST/WFC3). At certain wavelengths, the cross sections differ by almost a factor of $\sim$10 when smoothed. The amplitudes of the residuals decrease for lower resolutions, indicating that differences in line position contribute significantly to the differences between the line lists. In all panels, the cross sections are taken at 2000 K and 1 bar.}
\end{figure*}

\begin{figure}
\centering
    	   \includegraphics[width=0.5\textwidth]{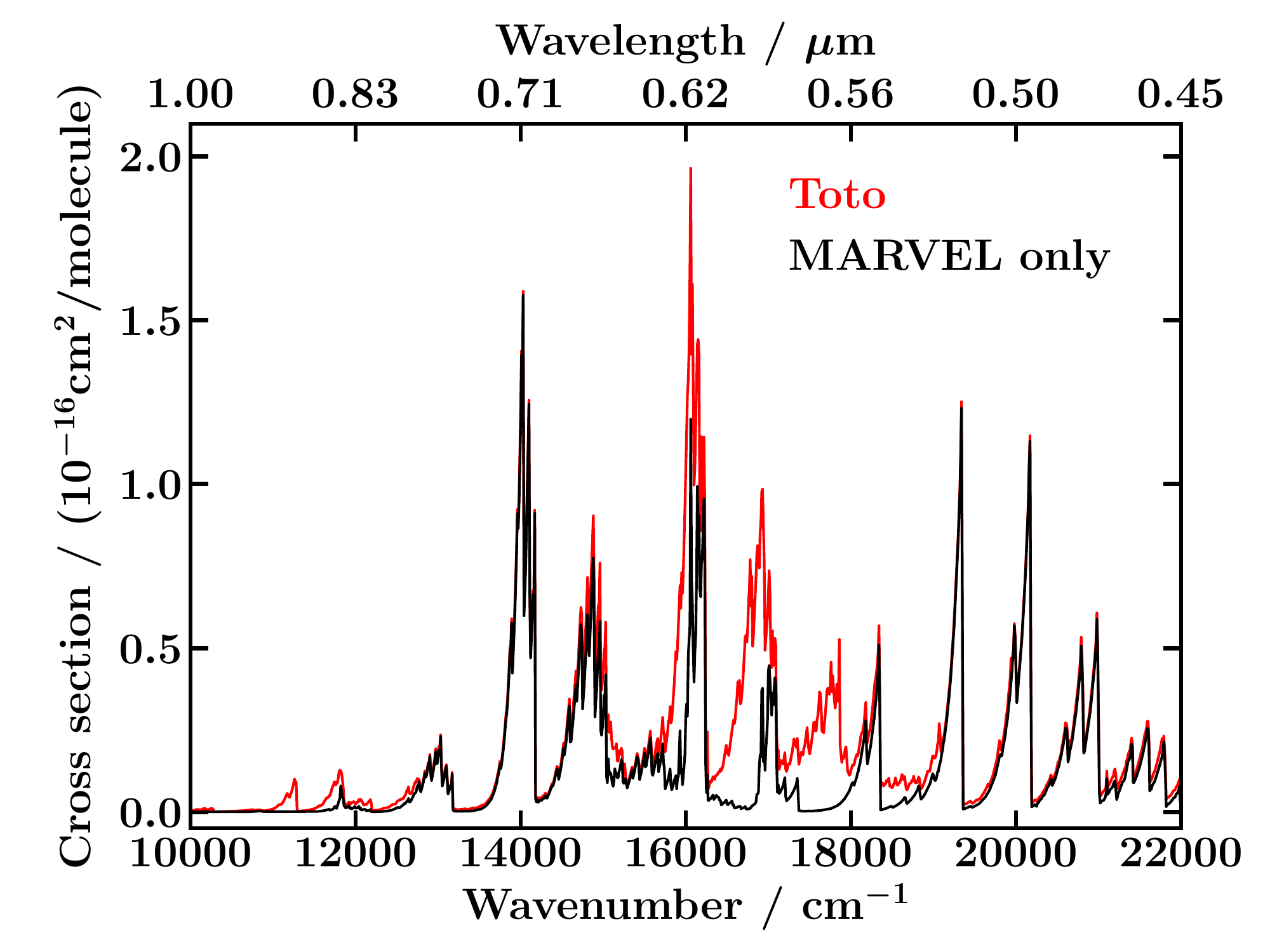}
\caption{\label{fig:hires} The full \textsc{Toto} (red) and \textsc{Marvel}-only (black) cross sections, calculated at 2000 K using Gaussian broadening with a hwhm of 2 cm$^{-1}$ calculated using \textsc{ExoCross}  \citep{ExoCross}. Here, we focus on comparisons between the strongest lines as these are the regions where experimentally derived transitions are most accessible. Furthermore, it is the strongest lines which have the greatest impact on the interpretation of high-resolution Doppler spectroscopy with cross-correlation.}
\end{figure}

%%%%%%%%%%%%%%%%%%%%%%%%%%%%%%%%%%%%%%%%%%%%%%%%%%%%%%%%%
%%%%%%%%%%%%%%%%%%%%% -- Figures -- %%%%%%%%%%%%%%%%%%%%%

\begin{figure*}
    \centering
    \includegraphics[width=0.8\textwidth]{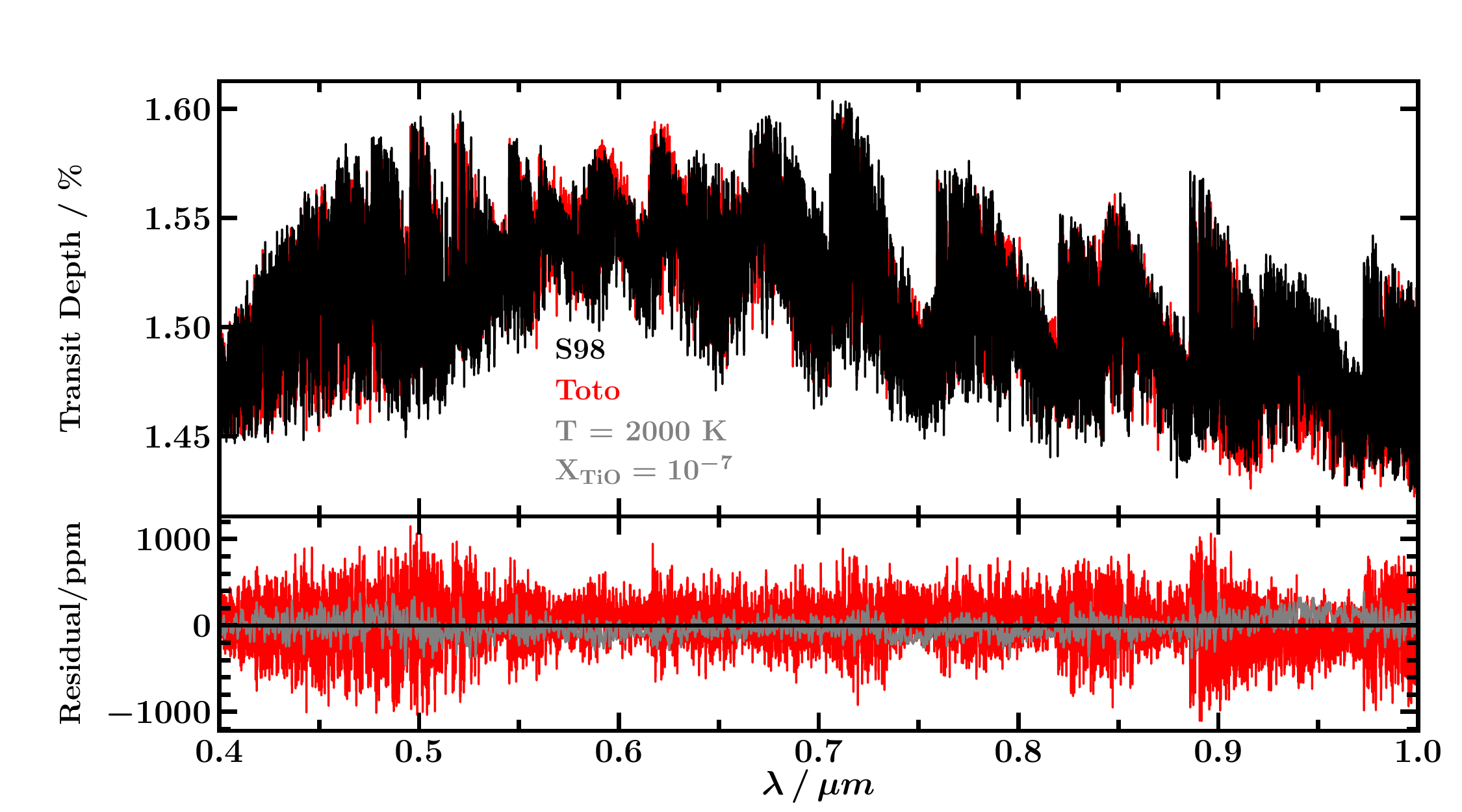}
    \caption{Upper panel: Transmission spectra calculated using the \textsc{Toto} (red) and S98 (black) cross sections at a resolution of R$\sim$10000. The $P$-$T$ profile is assumed to be an isotherm at 2000 K, and the planetary properties are those listed in section \ref{sec:crosssec}. Lower panel: red line shows the residual between the red and black spectra in the top panel (\textsc{Toto} spectrum - S98 spectrum). Grey line is calculated by smoothing the red and black spectra with a Gaussian of width 0.16 nm (to represent the PSF of a ground-based instrument) and finding the residual between them.}
    \label{fig:transmission}%emission_plot_simple
\end{figure*}

\begin{figure}
    \centering
    \includegraphics[width=0.5\textwidth]{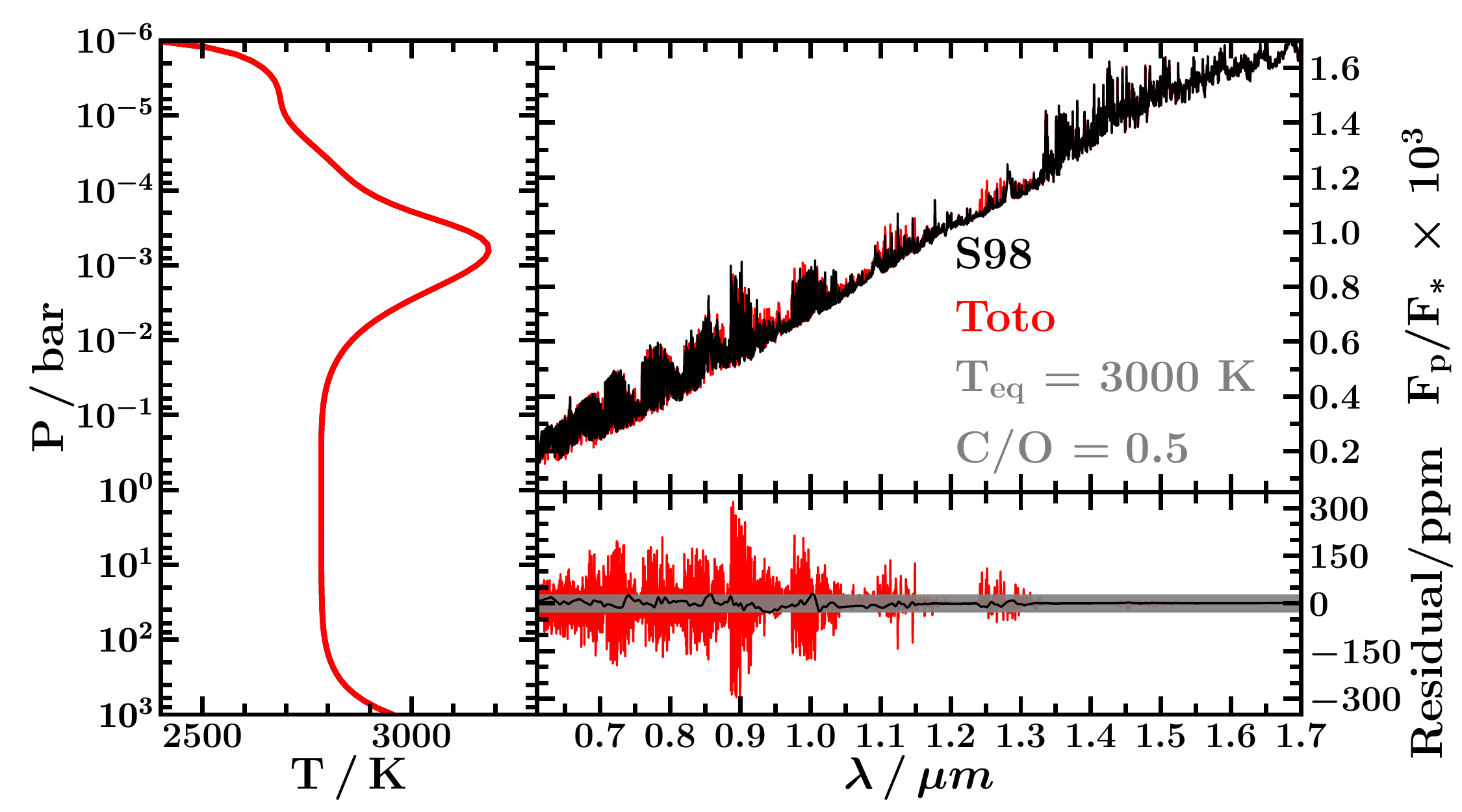}
    \caption{\label{fig:emission_simple} Upper right panel: spectra for two model atmospheres with identical $P$-$T$ profiles, chemical profiles, irradiation and planetary properties, but using the S98 (black) and \textsc{Toto} (red) TiO cross sections. The planetary properties are those listed in section \ref{sec:crosssec} and the semi-major axis is adjusted such that the dayside equilibrium temperature is $3000\,\mathrm{K}$. Lower right panel: the residuals between the two spectra (\textsc{Toto} spectrum - S98 spectrum) at a resolution of $R\sim10000$ (red) and smoothed with a Gaussian of width 2.3 nm (black) to represent the PSF of HST/WFC3.The atmospheric model used is self-consistent for the \textsc{Toto} cross sections, and the $P$-$T$ profile is shown in the left panel.}
\end{figure}

%%%%%%%%%%%%%%%%%%%%%%%%%%%%%%%%%%%%%%%%%%%%%%%%%%%%%%%%%

%%%%%%%%%%%%%%%%%%%%% -- Figures -- %%%%%%%%%%%%%%%%%%%%%
\begin{figure}
\centering
	\begin{subfigure}[b]{0.47\textwidth}
    	\includegraphics[width=\textwidth]{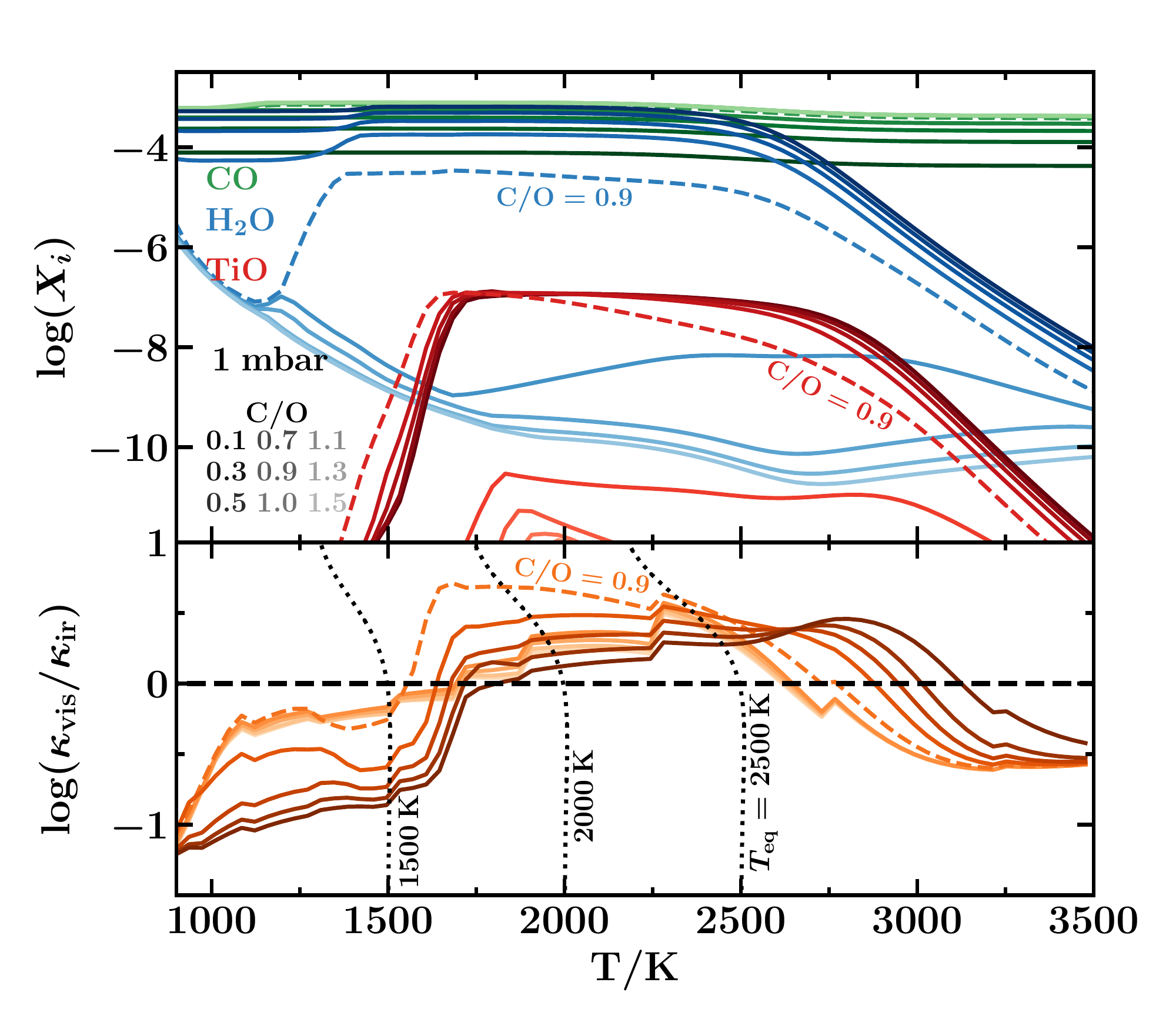}
    \end{subfigure}
	\begin{subfigure}[b]{0.47\textwidth}
    	\includegraphics[width=\textwidth]{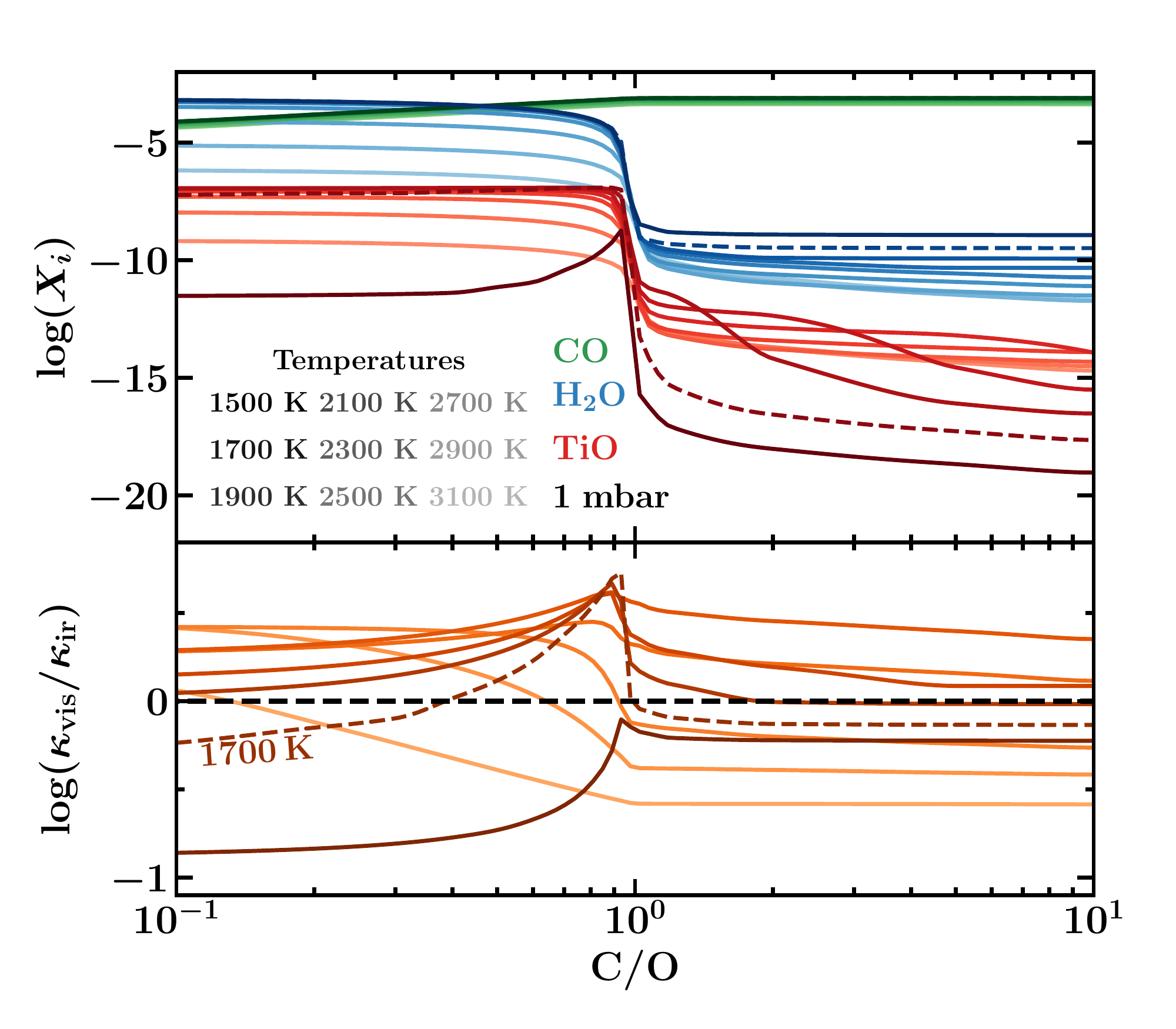}
    \end{subfigure}
\caption{\label{fig:XTCO} Equilibrium abundances of TiO, H$_2$O and CO as a function of temperature (top panel) and C/O ratio (bottom panel), for a pressure of 1~mbar. In each plot, the lower section shows the log of the ratio of optical to infrared opacities, log($\kappa_{\mathrm{vis}}/\kappa_{\mathrm{ir}}$). Bold black dashed lines highlight $\kappa_{\mathrm{vis}}/\kappa_{\mathrm{ir}}$=1. Values of $\kappa_{\mathrm{vis}}/\kappa_{\mathrm{ir}}\gtrsim1$ correspond to a thermal inversion in the atmosphere. In the upper panel, dotted black lines show $\kappa_{\mathrm{vis}}/\kappa_{\mathrm{ir}}$ as a function of photospheric temperature (temperature at an optical depth of 2/3), calculated using the $P$-$T$ profile of \citet{Guillot2010} and assuming equilibrium temperatures of $T_{\mathrm{eq}}$= 1500~K, 2000~K and 2500~K, respectively. In the top (bottom) panels, different line colours indicate different values of the C/O ratio (temperature), with darker shades indicating smaller values as shown by the legend. In the top panel, dashed coloured lines correspond to C/O=0.9, while in the bottom panel dashed coloured lines correspond to a temperature of 1700~K. For these values, $\kappa_{\mathrm{vis}}/\kappa_{\mathrm{ir}}$ has the highest peak in each plot, respectively.}
\end{figure}
%%%%%%%%%%%%%%%%%%%%%%%%%%%%%%%%%%%%%%%%%%%%%%%%%%%%%%%%%

Chemical cross sections play a crucial role in interpreting atmospheric observations, and their accuracy is important to correctly infer the presence and abundance of each molecule. New observational needs, new experimental data and new theoretical techniques have recently motivated a new line list for TiO absorption. Prior to 2019, there were two major line lists available for TiO; the \citep{Schwenke1998} line list developed in 1998 and the Plez line list originally constructed in 1998 \citep{Plez1998} with the latest update in 2012 \citep{Ryabchikova2015}. These line lists were originally designed for applications in modelling M dwarfs, while now the community needs data that can model absorption by TiO in exoplanets including for ultra-high-resolution ground-based Doppler spectroscopy \citep[e.g.][]{Hoeijmakers2015} which demands line position accuracies better than $\sim$0.1 cm$^{-1}$ (R$\gtrsim 10^5$) for strong lines.

Fortunately, new experimental data is available for many TiO bands. This new data is combined with old data using a new algorithm called \textsc{Marvel} \citep{jt412,07CsCzFu.marvel,12FuCsi.method} to produce accurate empirically-derived energies for more than 8000 rovibronic states. Next, the new \textsc{Duo} \citep{Yurchenko2016} software package is used to fit an accurate spectroscopic model of TiO with 13 electronic states based on ab initio data and these \textsc{Marvel} energy levels \citep{McKemmish2017}, then to solve this spectroscopic model to produce a large dataset with over 55 million transitions and 300,000 energy levels. Finally, the energy levels are "\textsc{Marvel}ised", i.e. model energies are replaced with the appropriate \textsc{Marvel} experimentally-derived values to dramatically improve the accuracy of the predicted line positions. The resulting new TiO line list, called \textsc{Toto} \citep{McKemmish2019}, covers the spectral range up to 30,000 cm$^{-1}$ or 0.33 $\mu$m and is suitable for use in both modelling and detecting TiO in exoplanets.

Our results here extend the original results in \citet{McKemmish2019} in two ways: (1) quantifying the extent of "\textsc{MARVEL}isation" in the \textsc{Toto} line list (Section \ref{sec:marvelisation}), and (2) exploring the impact of different line lists on predicted hot Jupiter spectra (Sections \ref{sec:transmission}, \ref{sec:emission}).

\subsection{Quantifying line list suitability for high-resolution spectroscopy}
\label{sec:marvelisation}
The importance of line list choice depends on the resolution of the observations.  Figure \ref{fig:tio_oldnew} shows the \citet{Schwenke1998} and Plez (2012) \citep[sourced from][]{Ryabchikova2015} cross sections (hereafter S98 and P12) alongside the \textsc{Toto} cross sections in the optical and near-infrared at 2000 K, as well as the differences between them. In this spectral range, the two cross sections differ by up to almost 3 orders of magnitude at native resolution and one order of magnitude when smoothed by a Gaussian PSF similar to that of HST/WFC3 (see grey and red lines in lower panel of figure \ref{fig:tio_oldnew}, respectively). This can have a significant impact on the interpretation of both optical and near-infrared spectra at low and high resolutions, but is clearly much more important for high-resolution observations. In sections \ref{sec:transmission} and \ref{sec:emission}, we focus on the differences between the \textsc{Toto} and S98 cross sections as these differ more in the optical (where TiO is most spectrally active) than the \textsc{Toto}-P12 comparison, therefore providing an upper limit on the effects of these differences on hot Jupiter spectra.

Use of high-resolution Doppler spectroscopy to identify molecules in exoplanets relies on the availability of very accurate line positions. For complex molecules like TiO, the necessary accuracy can only usually be obtained if model energy levels are explicitly replaced by experimentally-derived energy levels and/or transition frequencies. This replacement has not been done for the early 1998 Plez and Schwenke line lists, but is done for the Plez (2012) line list \citep[sourced from][]{Ryabchikova2015}  and the new \textsc{Toto} line list; the former two line lists are thus not preferable for use in high-resolution cross-correlation techniques. Clear identification of the line list used in exoplanet high-resolution cross-correlation studies is thus essential  for accurate chemical detections \citep{Brogi2012,Hoeijmakers2015,Birkby2018,Nugroho2017}.

Previous studies \citep{McKemmish2019} have compared the results of the \textsc{Toto} line list to the Plez (2012) line list \citep[sourced from][]{Ryabchikova2015} in the context of high-resolution Doppler spectroscopy, including cross-correlation of both line lists against high-resolution M star spectra in different spectral regions. \textsc{Toto} had superior performance, which can be attributed to the fact that the line list is "\textsc{Marvel}ised".

The extent to which a given line list explicitly includes experimental or experimentally-derived (e.g. \textsc{MARVEL}) energy levels and/or transitions has not previously been quantified, yet this is an essential factor in determining the suitability of a given line list for use in high-resolution cross-correlation studies.
Note that strong lines have the greatest contribution in the cross-correlation so their accuracy is especially important.

The most straightforward metric is that, in \textsc{Toto}, 17,365 model energy levels out of 301,245 (or 5.76\%) are replaced by experimentally-derived \textsc{Marvel} energy levels for the main isotopologue. The energy levels that are \textsc{Marvel}ised, however, are the most populated energy levels; \textsc{Marvel}ised energy levels contribute 95\% and 86\% to the total partition function at 2000 K and 3000 K, respectively.

In terms of the transitions themselves, 1,373,936 out of the total 58,983,952 transitions for the main isotopologue in \textsc{Toto} are between two \textsc{Marvel}ised energy levels and thus have very accurately determined line positions.  We can examine the importance of these transitions to the overall line list by considering figure \ref{fig:hires}, which shows the cross section from only \textsc{Marvel}ised transitions compared to the total \textsc{Toto} line list. The completeness of the \textsc{Marvel}-only data varies considerably between spectral bands with some almost entirely complete, e.g. around 0.71, 0.52 and 0.49 $\mu$m, and some almost entirely missing, e.g. around 0.57 $\mu$m. The most important bands missing in the \textsc{Marvel}-only data involve $v=2$ and higher vibrational excitations of the B $^3\Pi$ state, which have not yet been experimentally analysed; in light of their importance to high-resolution Doppler spectroscopy, these levels are thus of most importance for future experimental study.

More quantitatively, we can quantify the number of \textsc{Marvel}ised transitions compared to the total number of transitions in the \textsc{Toto} line list at a given temperature above a certain intensity threshold by computing stick spectra using \textsc{ExoCross}  \citep{ExoCross}.  Approximately 83\% of the 4491 strongest transitions (line intensities above 10$^{-17}$ cm/molecule at 2000 K) are \textsc{Marvel}ised, with this number dropping to 64\% of 50,408 transitions if the threshold is chosen as 10$^{-18}$ cm/molecule.

% \red{Laura to add DISCUSSION OF WHAT COMPARISONS WERE IN THE ORIGINAL TOTO PAPER.

% THOMAS CONVERTS PLEZ 2012 LINE LIST TO EXOMOL FORMAT, ANJALI reproduces the figures.

% ADDRESSING COMMENT:

% — The authors first compare the line list of Schwenke 1998 to the new Toto line list. However it is not clear how much more knowledge is gained compared to what is described in the McKemmish et al. paper introducing the new line list. Although the data is shown differently and in an easier to read format in Fig.1, no new knowledge seem presented here and I would recommend moving this comparison to the appendix. Also why use the S98 and not the Plez 2012 list for comparison ?}

\subsection{Effect of line list choice on Transmission Spectra}
\label{sec:transmission}
The first detections of TiO in an exoplanet atmosphere have recently been made using transit spectra \citep{Haynes2015,Sedaghati2017}. In particular, \citet{Sedaghati2017} have detected TiO in the transmission spectrum of WASP-19b (though \citet{Espinoza2019} do not detect TiO at a later epoch). Here, we investigate the effect of the \textsc{Toto} TiO line list, relative to the S98 line list, on transmission spectra by comparing two spectra for a canonical hot Jupiter. Throughout this work, model spectra are calculated for this canonical hot Jupiter using planetary/stellar properties similar to WASP-12b/WASP-12; the planetary radius and log gravity are taken as 1.79 R$_\mathrm{J}$ and 2.989 (cgs), respectively, and the stellar radius, log gravity, [Fe/H] metallicity and effective temperature are taken as 1.63 R$_\odot$, 4.38 (cgs), 0.3 and 6300 K, respectively \citep{Hebb2009,Stassun2017}. We generate cross sections from the Toto and S98 line lists using the methods described in \citet{Gandhi2017}. Using these, we generate two model transmission spectra which are identical other than the TiO cross sections used in each. The model spectra are generated using the transmission model described in \citet{Welbanks2019} \citep[see also][]{Pinhas2018}. We assume an isothermal pressure-temperature ($P$-$T$) profile at 2000 K and a constant-with-depth TiO mixing ratio of $10^{-7}$. For simplicity, no other chemical species are included in these models.

The two spectra and their residuals are shown in figure \ref{fig:transmission}. The grey line in the lower panel shows the residual for smoothed spectra; spectra in the top panel are smoothed with a Gaussian of width 0.16 nm before calculating the residual, to represent smoothing by the PSF of a ground-based instrument. At this resolution, differences of up to 400 ppm can be seen, which are comparable to observational uncertainties with current instruments. When making inferences from optical transmission spectra, choice of TiO line list can therefore have a significant impact on the conclusions drawn.

\subsection{Effect of line list choice on Emission Spectra}
\label{sec:emission}

Here, we consider the effects of the Toto and S98 TiO cross sections on the emission spectrum of a hot Jupiter. We do this by comparing emission spectra from two models which are identical apart from the TiO cross sections used to generate them.

To generate these spectra, we use the \textsc{GENESIS} model \citep{Gandhi2017}, which calculates full, line-by-line radiative transfer under radiative-convective equilibrium. In order to calculate equilibrium chemical abundances as a function of pressure, temperature and elemental abundances we use the software package \textsc{HSC Chemistry} (version 8). This software has been used in several studies in the field of exoplanets, planet formation and the solar nebula \citep[e.g.][]{Pasek2005,Bond2010,Elser2012, Madhusudhan2012,Moriarty2014,Harrison2018}, and calculates abundances by minimising the total Gibbs free energy of the system using the \textsc{GIBBS} solver \citep{White1958}. We include all the chemical species used in \citet{Bond2010} and \citet{Harrison2018}, plus extra ionic and molecular forms of H, O, C, N, Ti and OH (listed in table \ref{tab:chem}), including H$^-$. In this section, all elemental abundances are taken to be solar.

Throughout this work, the cross sections we use for all molecules are calculated as in \citet{Gandhi2017} from the HITEMP, HITRAN and ExoMol line list databases (H$_2$O, CO and CO$_2$: \citealt{Rothman2010}, CH$_4$: \citealt{Yurchenko2013,Yurchenko2014a}, C$_2$H$_2$: \citealt{Rothman2013,Gordon2017}, NH$_3$: \citealt{Yurchenko2011}, HCN: \citealt{Harris2006,Barber2014}, VO: \citealt{McKemmish2016}, Collision-Induced Absorption (CIA): \citealt{Richard2012}). We also include opacity due to Na and K \citep{Burrows2003,Gandhi2017}. The bound-free and free-free cross sections of H$^-$ are calculated using the prescriptions of \citet{Bell1987} and \citet{John1988} (see also \citealt{Arcangeli2018,Parmentier2018,Gandhi2020}).

We also consider the effect of vertical mixing in the \textsc{GENESIS} code. \citet{Spiegel2009} find that TiO is heavy enough to gravitationally settle if not kept aloft by mixing, and cold traps are known to deplete TiO if it is not replenished \citep[e.g.][]{Burrows2007,Spiegel2009}. We use a simple mixing model in which the equilibrium abundance is added to a `quenched' abundance in the range 0.1-$10^{-3}$ bar. This `quenched' value is taken to be the equilibrium abundance at 0.1 bar, where we assume that vertical mixing begins \citep[e.g.][]{Spiegel2009,Moses2013}. The resulting abundance profile is effectively equal to the larger of the equilibrium or quenched values.

Figure \ref{fig:emission_simple} shows the residuals between spectra of two otherwise identical models generated with each of the Toto and S98 cross sections with the $P$-$T$ profile shown in the left panel. The $P$-$T$ profile used is a self-consistent $P$-$T$ profile generated using the \textsc{Toto} line list assuming a dayside equilibrium temperature of 3000 K as well as equilibrium chemistry and vertical mixing as described above. Note that we define dayside equilibrium temperature as
\begin{equation}
    T_{\mathrm{eq}} = \left(\frac{f}{2}\right)^{1/4}\sqrt{\frac{R_s}{a}} \, T_{\mathrm{eff}},
\end{equation}
(henceforth referred to as equilibrium temperature) where $f$ is the fraction of incident irradiation which remains on the dayside, $R_s$ and $T_{\mathrm{eff}}$ are the radius and effective temperature of the star, respectively, and $a$ is the semi-major axis of the planet. Throughout this work we assume a value of $f=1/2$ (i.e. full redistribution of incident radiation around the planet), unless otherwise stated, and $a$ is chosen such that $T_{\mathrm{eq}}$ has the value quoted in each figure. For the incident stellar irradiation we use ATLAS model spectra \citep{Kurucz1979,Castelli2003} and both the stellar and planetary properties we use are listed in section \ref{sec:transmission}.

The spectra in figure \ref{fig:emission_simple} differ most in the optical range by up to a factor of 300 ppm. Intuitively, this is because TiO is primarily active in the optical range, and as such will have a larger impact on optical spectra. In the near-infrared, the two spectra differ by up to $\sim$100 ppm. However, the differences in both the optical and near-infrared are only visible at higher spectral resolutions. When the two spectra are smoothed to a resolution similar to that of HST/WFC3, the residuals between them are of order $\sim$25 ppm which is comparable to the signal-to-noise achievable with HST/WFC3. When interpreting emission spectra, choice of TiO line list is therefore most important at high resolutions.

In what follows, we choose to use the \textsc{Toto} cross sections in our analysis of thermal inversions caused by TiO. In addition to the advantages discussed in section \ref{sec:marvelisation}, \textsc{Toto} also has better completeness than previous line lists and takes advantage of more recent experimental data in fitting the \textsc{Duo} spectroscopic models \citep{McKemmish2019}.

\section{Influence of Chemistry on Thermal Inversions}
\label{sec:eqtio}
%%%%%%%%%%%%%%%%%%%%% -- Figures -- %%%%%%%%%%%%%%%%%%%%%
\begin{figure*}
    \centering
    \includegraphics[width=0.9\textwidth]{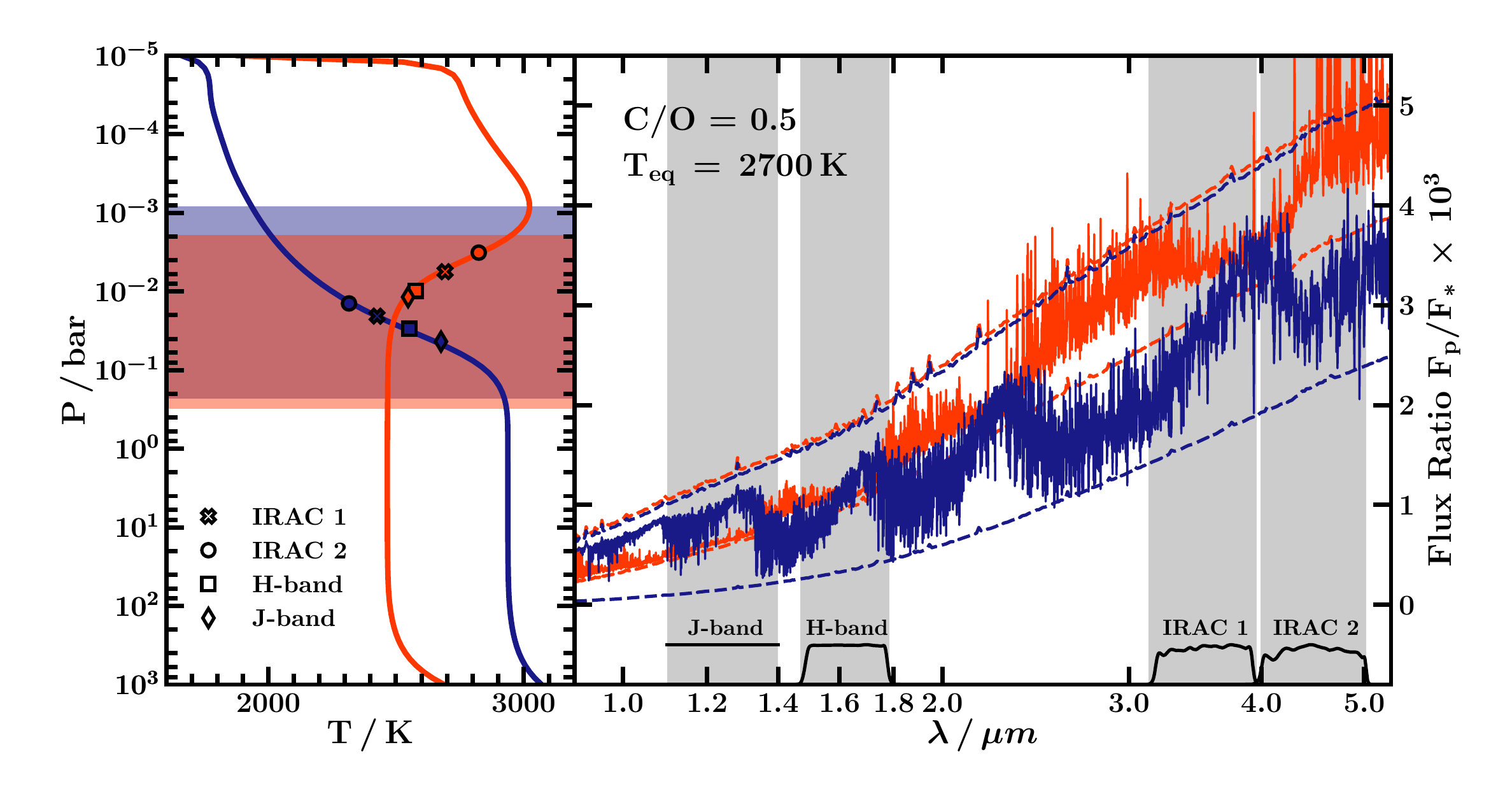}
    \caption{Temperatures and pressures probed by the J-band, H-band, IRAC 1 and IRAC 2 photometric bands. Red and blue spectra in the right panel correspond to the $P$-$T$ profiles of the same colour in the left panel; both are generated using the properties listed in section \ref{sec:crosssec} and the red spectrum includes TiO while the blue one does not. Markers in the left panel show the brightness temperatures (and corresponding pressures) of the IRAC 1, IRAC 2, H-band and J-band photometric points from each of the spectra, respectively. Each of these bands is shown by the grey shaded regions in the right panel and their transmission functions are shown in black. Blue and red shaded regions in the left panel show the pressure ranges of the photospheres for the blue and red $P$-$T$ profiles, respectively. }
    \label{fig:metric}
\end{figure*}

\begin{figure}
    \centering
    \begin{subfigure}[b]{0.49\textwidth}
        \includegraphics[width=\textwidth]{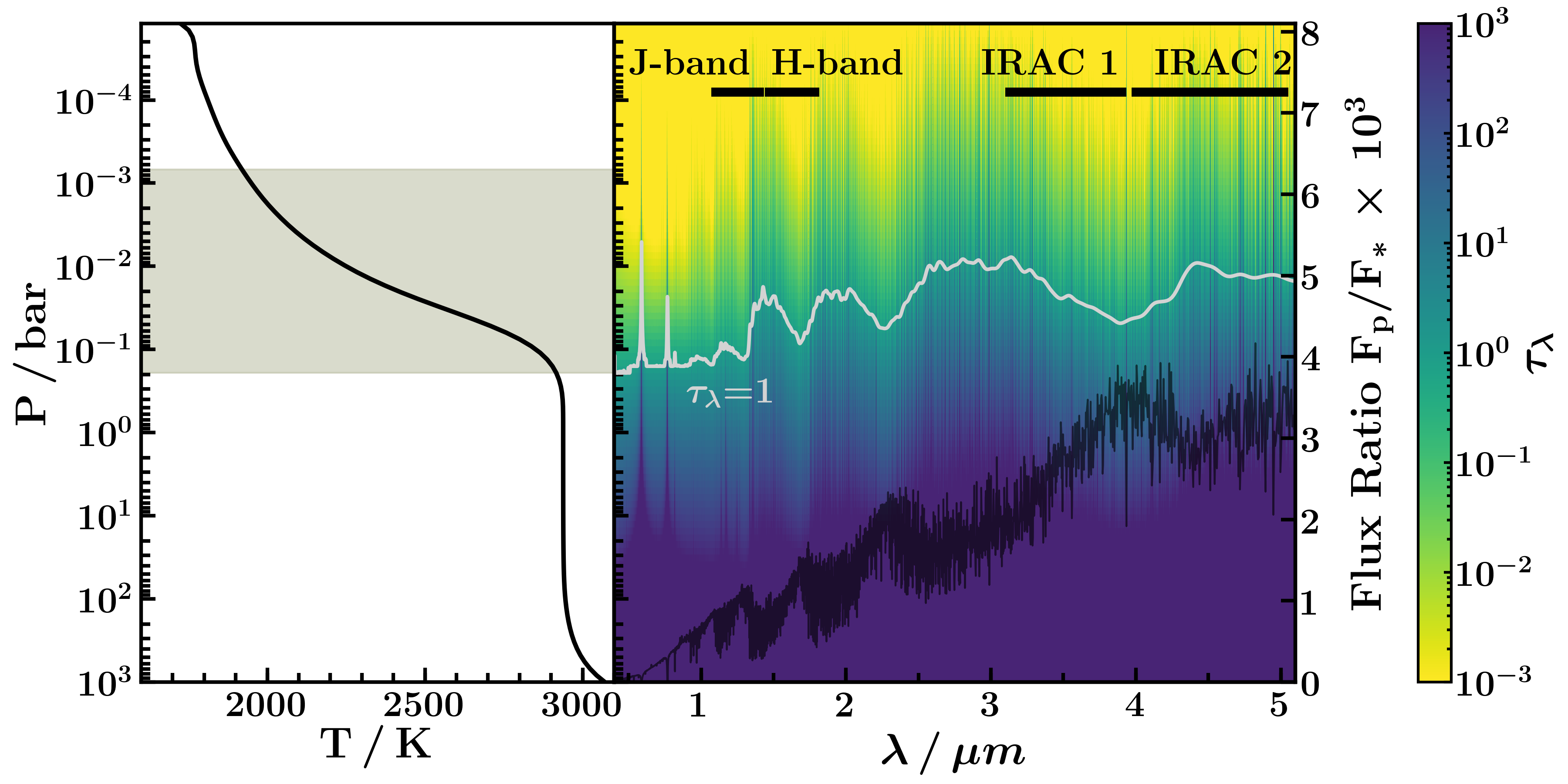}
    \end{subfigure}
    \\
    \begin{subfigure}[b]{0.49\textwidth}
        \includegraphics[width=\textwidth]{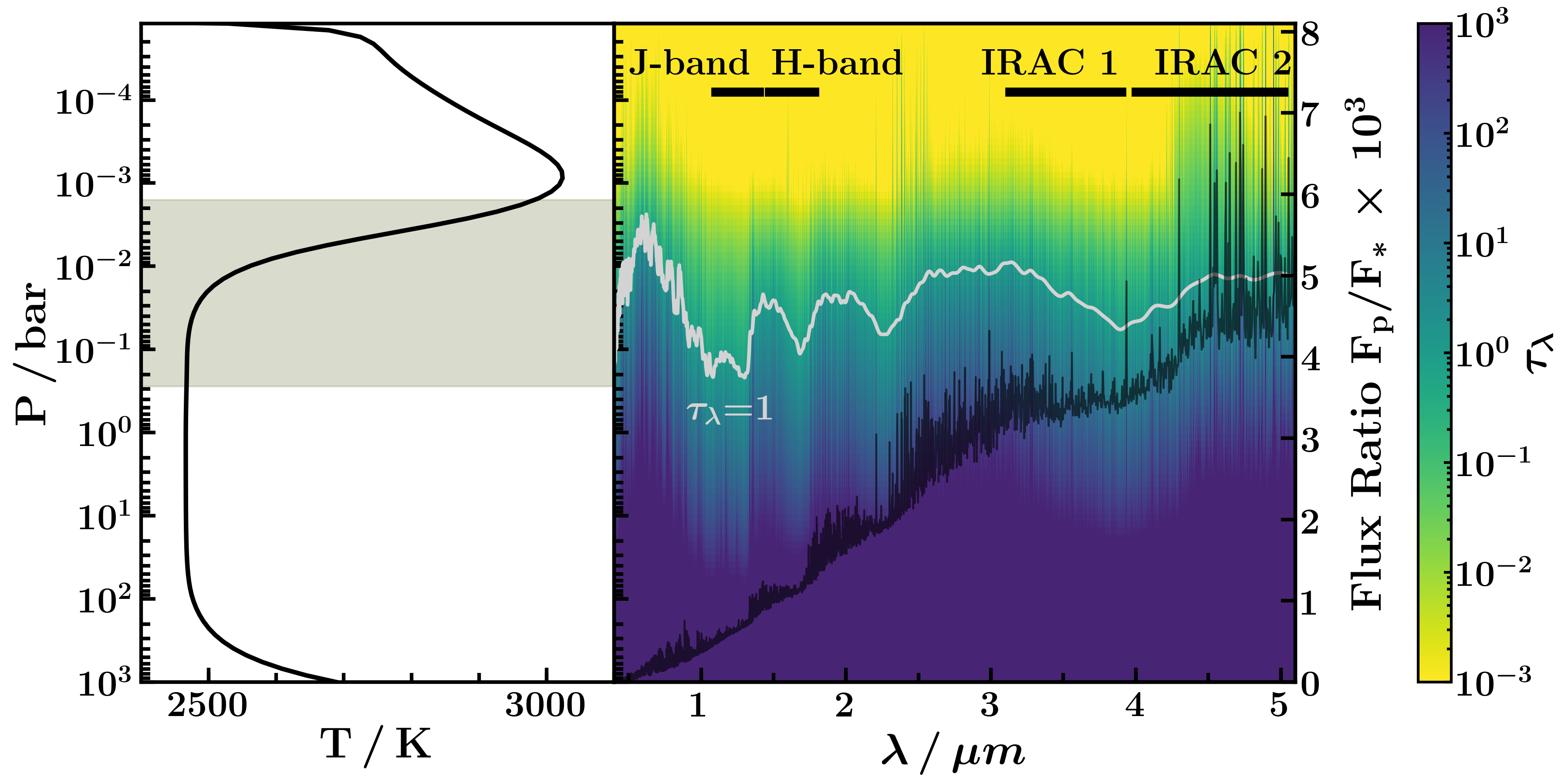}
    \end{subfigure}
    \caption{Optical depth as a function of wavelength and pressure (i.e. altitude) in the atmosphere for the two $P$-$T$ profiles shown in figure \ref{fig:metric}. Left panels show the $P$-$T$ profile with the photosphere (calculated as the extent of the $\tau$=1 surface) shaded in grey. In the right panels, the $\tau$=1 surface is shown as a function of wavelength and pressure by the grey line. Opacity windows, including the J-, H- and IRAC 1 bands occur where the optical depth reaches unity at deeper pressures. The emergent spectrum of the planet is shown in black. Horizontal black lines indicate the J-, H-, IRAC 1 and IRAC 2 bands.}
    \label{fig:tau}
\end{figure}

%%%%%%%%%%%%%%%%%%%%%%%%%%%%%%%%%%%%%%%%%%%%%%%%%%%%%%%%%
The thermal profile in a hot Jupiter atmosphere is dictated by the incident stellar irradiation and the sources of opacity in the atmosphere. In this section, we briefly investigate the relationship between atmospheric chemistry and temperature structure using a semi-analytic approach. A key component in determining the opacities is the abundance of each chemical species. We therefore begin by calculating equilibrium abundances for a range of chemical species as a function of temperature and C/O ratio using the \textsc{HSC Chemistry} software as described in section \ref{sec:emission}. By considering radiative equilibrium, we then use these chemical abundances to explore the range of thermal solutions which are possible under various conditions. We vary the C/O ratio by using solar elemental abundances but varying the C abundance. In later sections, we also vary metallicity by keeping solar elemental ratios but scaling abundances by a constant factor.

Figure \ref{fig:XTCO} shows the calculated equilibrium abundances for H$_2$O, CO and TiO as a function of C/O ratio and temperature. The \textsc{HSC Chemistry} software includes the effects of condensation and thermal dissociation, which can be seen in the top panel of figure \ref{fig:XTCO} as the abundance of TiO decreases rapidly at temperatures below $\sim$1750 K (condensation), and both TiO and H$_2$O are depleted at temperatures above $\sim$3000 K (thermal dissociation). As expected, the abundances of both species drop by several orders of magnitude as the C/O ratio increases above unity, consistent with previous studies \citep{Madhusudhan2011b,Madhusudhan2012,Moses2013}. In contrast, the CO abundance remains high across both low and high C/O ratios, and at cooler temperatures ($\sim$1300 K), CO abundance decreases as CH$_4$ takes over as the dominant carrier of carbon \citep{Madhusudhan2012}.

These abundances affect the temperature in a region of the atmosphere through the ratio of visible to infrared opacity, $\kappa_{\mathrm{vis}}/\kappa_{\mathrm{ir}}$ \citep{Hubeny2003,Guillot2010,Hubeny2017}. $\kappa_{\mathrm{vis}}$ determines the amount of incident radiation which can be absorbed, while $\kappa_{\mathrm{ir}}$ dictates the efficiency with which the gas is able to radiate away energy, both of which affect the steady-state temperature of the gas. These quantities are defined by:
\begin{align}
\label{eq:kappas}
    \begin{split}
        \kappa_{\mathrm{vis}} &= \frac{\int_{0}^{\infty} \kappa_{\nu} J_{\nu} d\nu}{\int_{0}^{\infty} J_{\nu} d\nu}, \\[6pt]
        \kappa_{\mathrm{ir}} &= \frac{\int_{0}^{\infty} \kappa_{\nu} B_{\nu} d\nu}{\int_{0}^{\infty} B_{\nu} d\nu}.
    \end{split}
\end{align}
$B_{\nu}$ is the thermal emission of the gas (a Planck function of temperature $T$) and $\kappa_{\nu}$ is the opacity of the gas as a function of frequency, defined as:
\begin{equation*}
    \kappa_{\nu} = \sum_{i} \sigma_i n_i,
\end{equation*}
where the sum is over all species and $\sigma_i$, $n_i$ are the cross section and number density of each species, respectively.
$J_{\nu}$ is the zeroth moment of spectral intensity of incident irradiation, $J_{\nu} = \frac{1}{2} \int_{-1}^{1}I_{\nu\mu}\mathrm{d}\mu$, where $I_{\nu\mu}$ is the spectral intensity of incident irradiation at an angle $\theta$ ($\mu$=cos$\theta$) to the local normal.

The effect of $\kappa_{\mathrm{vis}}/\kappa_{\mathrm{ir}}$ on the atmospheric temperature gradient can be seen using the analytic $P$-$T$ profile of \citet{Guillot2010}. This profile parameterises atmospheric temperature ($T$), as a function of optical depth ($\tau$), intrinsic planetary temperature ($T_\mathrm{int}$), dayside temperature and $\kappa_{\mathrm{vis}}/\kappa_{\mathrm{ir}}$. As such, $\mathrm{d}T/\mathrm{d}\tau$ can be calculated as a function of $\kappa_{\mathrm{vis}}/\kappa_{\mathrm{ir}}$. Assuming a negligible $T_\mathrm{int}$, $\mathrm{d}T/\mathrm{d}\tau$ is negative (i.e. there is a thermal inversion) when $\kappa_{\mathrm{vis}}/\kappa_{\mathrm{ir}}>1$, for all finite values of $\tau$ and dayside temperature \citep{Guillot2010,Gandhi2019}.

In figure \ref{fig:XTCO}, we show $\kappa_{\mathrm{vis}}/\kappa_{\mathrm{ir}}$ as a function of C/O ratio and temperature. For simplicity, we use a single pressure of 1 mbar to represent the upper atmosphere. To calculate $\kappa_{\mathrm{vis}}/\kappa_{\mathrm{ir}}$, we include opacity contributions from TiO, H$_2$O, CH$_4$, NH$_3$, CO, CO$_2$, HCN, C$_2$H$_2$, Na, K, VO and CIA, using equilibrium chemical abundances from \textsc{HSC Chemistry}. From figure \ref{fig:XTCO} (top panel), we see that $\kappa_{\mathrm{vis}}/\kappa_{\mathrm{ir}}$ only has values above unity for temperatures greater than $\sim$1500. This suggests that thermal inversions caused by TiO can only occur for atmospheric temperatures above 1500~K. This temperature limit corresponds to C/O=0.9, while lower and higher C/O ratios begin to exceed $\kappa_{\mathrm{vis}}/\kappa_{\mathrm{ir}}$ only at higher temperatures. Larger values of $\kappa_{\mathrm{vis}}/\kappa_{\mathrm{ir}}$ coincide with larger values of TiO abundance, which is expected since TiO is a primary visible absorber in this scenario. Outside the range of large $\kappa_{\mathrm{vis}}/\kappa_{\mathrm{ir}}$, condensation and thermal dissociation of TiO both diminish its value at low and high temperatures, respectively. Na and K can also contribute significantly to the optical opacity, and hence thermal inversions, for C/O$\gtrsim$1 \citep[see also][]{Molliere2015}. For example, at C/O$\geq$1, where TiO is strongly depleted, optical opacity due to Na and K results in $\kappa_{\mathrm{vis}}/\kappa_{\mathrm{ir}}$>1 for temperatures greater than $\sim$1750~K.

We can use $\kappa_{\mathrm{vis}}/\kappa_{\mathrm{ir}}$ as shown in figure~\ref{fig:XTCO} to broadly assess the feasibility of thermal inversions for a given $T_\mathrm{eq}$. To do this, we again use the analytic temperature profile from \citet{Guillot2010}, setting $\tau=2/3$ (i.e. corresponding to the photosphere) and assuming that intrinsic heat is negligible compared to the irradiation. For a chosen equilibrium temperature, this results in a relation between $\kappa_{\mathrm{vis}}/\kappa_{\mathrm{ir}}$ and photospheric temperature. This is shown by the dotted black lines in the upper panel of figure \ref{fig:XTCO} for three equilibrium temperatures (1500, 2000 and 2500~K). Solutions for the photospheric temperature occur at the intersection points between these lines and the values of $\kappa_{\mathrm{vis}}/\kappa_{\mathrm{ir}}$ calculated independently from the molecular opacities (coloured lines in lower section of upper panel in figure \ref{fig:XTCO}). Where these intersections happen at $\kappa_{\mathrm{vis}}/\kappa_{\mathrm{ir}}>1$, a thermal inversion is possible.

Figure \ref{fig:XTCO} therefore shows that inversions start to become possible at temperatures in the range $\sim$1500-2000 K, depending on the C/O ratio. For example, for C/O=0.9, thermal inversions are possible at equilibrium temperatures just above 1500~K, while for C/O$\geq$1.0 equilibrium temperatures $\gtrsim$1750~K are required. This is comparable to the equilibrium temperature at which \citet{Fortney2008} begin to see thermal inversions for a Sun-like host star ($\sim$1800 K). To fully understand the effects of equilibrium chemical abundances on the $P$-$T$ profile of an atmosphere, we extend our analysis with full numerical atmospheric models in section \ref{sec:colmaps}.

\section{Metrics for Assessing Thermal Inversions}
\label{sec:h2o_metric}

In this section, we assess metrics for inferring the presence and strength of thermal inversions in exoplanet atmospheres with solar-like compositions. A common way to do this involves comparing two photometric measurements, e.g. using the Spitzer IRAC 1 and IRAC 2 channels \citep[e.g.][]{Burrows2007,Knutson2010,Madhusudhan2010}. Assessing the presence of a thermal inversion with two or more spectral or photometric bands requires at least one band which probes a high-opacity region of the spectrum and an opacity window which probes as little line opacity as possible, i.e. a spectral continuum. The continuum band probes deeper regions of the atmosphere while the high-opacity band probes higher up in the atmosphere. The difference in brightness temperature between these bands provides a measure of the temperature gradient in the photosphere. While a `perfect' opacity window would contain no line opacity at all, this is not realistic. Furthermore, continuum opacity such as CIA opacity is itself wavelength dependent. Therefore, different continuum bands can have varying levels of both line opacity contamination and continuum opacity. A continuum band with less line opacity probes deeper into the atmosphere and therefore has the potential to probe a larger temperature contrast when compared to the `high-opacity' band. Therefore, when looking for metrics to assess thermal inversions it is advantageous to consider the clearest possible continuum bands. For example, \citet{Stevenson2016} use the J-band as a continuum compared to a water absorption band when quantifying absorption features in transmission spectra.

The IRAC 1 and IRAC 2 channels (at 3.6 $\mu$m and 4.5 $\mu$m, respectively) are commonly used to infer the presence of thermal inversions in hot Jupiters as the 3.6 $\mu$m band does not contain very strong molecular features for solar composition atmospheres while the 4.5 $\mu$m band probes a strong CO feature \citep[e.g.][]{Burrows2007,Knutson2009,Madhusudhan2010}. This property has also been used to construct colour-magnitude diagrams for exoplanets and brown dwarfs and to subsequently compare irradiated exoplanets to other types of sub-stellar object \citep{Triaud2014a,Triaud2014b}. For a hot Jupiter with solar composition and a thermal inversion, the brightness temperature contrast between the IRAC 2 and IRAC 1 channels, $T_{4.5\mu\mathrm{m}}-T_{3.6\mu\mathrm{m}}$, is expected to be positive. On the other hand, an atmosphere with solar composition but without a thermal inversion would be expected to have $T_{4.5\mu\mathrm{m}}-T_{3.6\mu\mathrm{m}}<0$. However, in reality, the IRAC 1 band is not a perfect opacity window. In particular, it has contributions from CH$_4$, H$_2$O and HCN \citep{Madhusudhan2012}. This is particularly evident for high C/O ratios close to 1 where CH$_4$ and HCN can have significant opacity contributions. Under such circumstances, an atmosphere without a thermal inversion can still have significantly low flux in the IRAC 1 band due to CH$_4$ and/or HCN absorption compared to the IRAC 2 band with CO. This situation can give rise to a positive $T_{4.5\mu\mathrm{m}}-T_{3.6\mu\mathrm{m}}$ contrast which can mimic the behaviour of a thermal inversion (section \ref{sec:performace_of_metrics}). Other photometric bands may therefore provide a better opacity window compared to the IRAC 1 channel, and planets with different chemistries may require different bands to probe the spectral continuum.

Here, we use theoretical models to asses metrics for temperature gradients in atmospheres with solar-like compositions. We calculate brightness temperatures using the method described by \citet{Garhart2019}; for a spectral bin in the wavelength range $\lambda_{\mathrm{min}}$-$\lambda_{\mathrm{max}}$ and a normalised instrument sensitivity function $\zeta$, the brightness temperature, $T_b$, is chosen such that
\begin{equation*}
    \int_{\lambda_{\mathrm{min}}}^{\lambda_{\mathrm{max}}} \pi \zeta  B_{\lambda}(T_b)  d\lambda =  \frac{F_p}{F_s} \frac{R_s^2}{R_p^2}   \,\, \times \,\, \int_{\lambda_{\mathrm{min}}}^{\lambda_{\mathrm{max}}}  \pi \zeta I_s  d\lambda ,
\end{equation*}
where $\frac{F_p}{F_s}$ is the observed planet-star flux ratio in the photometric band, $\pi I_s$ is the stellar surface flux (we use ATLAS model spectra: \citet{Kurucz1979,Castelli2003}) and $R_p$ is the planetary radius.

We begin by investigating optimal continuum bands for hot Jupiter atmospheres of solar composition. In particular, we compare the H- and J-bands to the IRAC 1 band according to their performance as continuum bands for solar-composition spectra (see figure \ref{fig:metric}). Both the H- and J-bands coincide with local minima in H$_2$O opacity so, for spectra dominated by H$_2$O, they should probe deeper regions of the atmosphere relative to bands which probe stronger features \citep{Stevenson2016}. We demonstrate this property here using two self-consistent atmospheric models for which the $P$-$T$ profiles and spectra are shown in figure \ref{fig:metric}. Both models are generated using the \textsc{GENESIS} self-consistent modelling code \citep{Gandhi2017}. The equilibrium chemical abundances are calculated using \textsc{HSC Chemistry} assuming a C/O ratio of 0.5. The model shown in blue does not include TiO and has a non-inverted profile, while the model shown in red does include TiO and has a thermal inversion. For each model, we compute a simulated photometric measurement for each of the four bands mentioned above, and calculate the brightness temperature of each measurement. These temperatures are plotted over the $P$-$T$ profiles in the left panel of figure \ref{fig:metric} and indicate which part of the photosphere is probed by each band.

We find that the H-band and J-band probe deeper atmospheric pressures compared to the IRAC 1 band for these models. This is clearly demonstrated in figure \ref{fig:tau}, which shows optical depth as a function of wavelength and pressure in the atmosphere (i.e. altitude). The J-, H- and IRAC 1 bands all correspond to windows in opacity, where optical depth reaches unity at deeper pressures. However, in the J- and H- bands, unit optical depth is reached at pressures $\sim$1 order of magnitude greater than that for the IRAC 1 band. Figure \ref{fig:tau} also shows that deeper pressures can be probed at higher spectral resolutions by sampling very narrow windows in opacity and avoiding contamination from higher-opacity regions. We demonstrate this concept in greater detail in appendix \ref{sec:hires_PTgrad} (see also \citealt{deKok2014}). Despite this effect, wider photometric bands are still observationally favourable as they are less sensitive to small wavelength shifts in the detector.

Since, in the cases shown here, the J- and H- bands are better opacity windows than the IRAC 1 band, the brightness temperature contrasts between the IRAC 2 channel and the H-band ($T_{4.5\mu\mathrm{m}}-T_{\mathrm{H}}$) or between the IRAC 2 channel and the J-band ($T_{4.5\mu\mathrm{m}}-T_{\mathrm{J}}$) are greater than the contrast between the two IRAC channels ($T_{4.5\mu\mathrm{m}}-T_{3.6\mu\mathrm{m}}$). For both the inverted and non-inverted model $P$-$T$ profiles, an inversion/non-inversion could be inferred based on the brightness temperatures of the IRAC 2 channel and any of the other three bands, since the IRAC 2 channel probes the shallowest pressures (highest altitudes) in both cases, corresponding to the hottest/coolest brightness temperature for the inverted/non-inverted profile. However, a larger brightness temperature contrast provides more information about the atmospheric temperature gradient and can potentially allow for a more robust inference of an inversion/non-inversion. In this case, the J-band would therefore be the best suited for comparison to the IRAC 2 band, as it probes deeper into the atmosphere. Although the H- and J-bands may not be optimal probes of the continuum for spectra which are not dominated by H$_2$O opacity, this demonstrates that it can be beneficial to consider bands other than the IRAC 1 channel when seeking to determine the presence of a thermal inversion based on photometric measurements alone. In section \ref{sec:colmaps}, we extend this assessment to a wider range of C/O ratios, equilibrium temperatures, metallicities, gravities and stellar types.

\section{Metrics as a function of planetary and host star properties}
\label{sec:colmaps}
%%%%%%%%%%%%%%%%%%%%% -- Figures -- %%%%%%%%%%%%%%%%%%%%%
\begin{figure*}
\centering
	\begin{subfigure}[b]{0.31\textwidth}
	    \includegraphics[width=\textwidth]{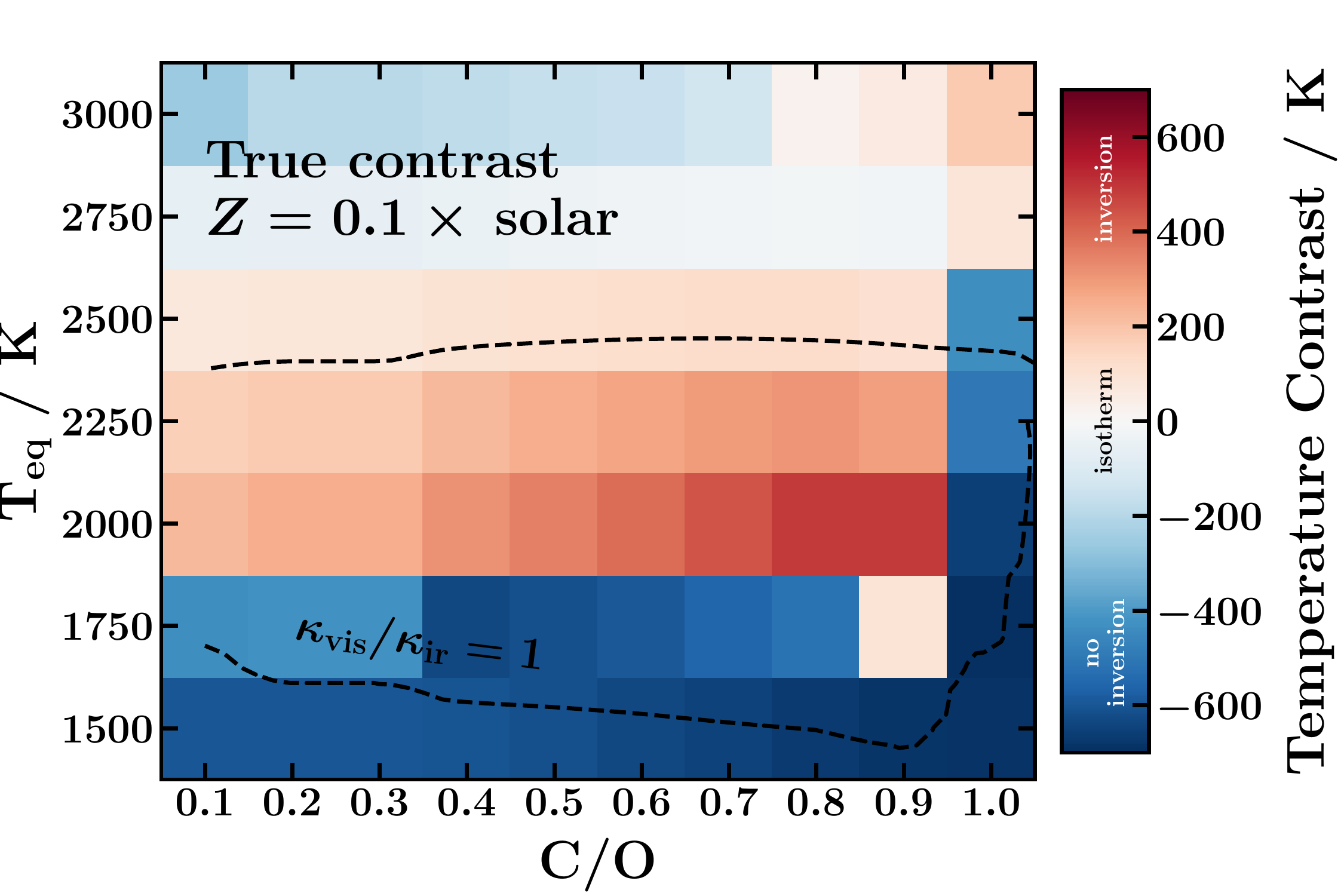}
    \end{subfigure}
    \quad
	\begin{subfigure}[b]{0.31\textwidth}
	    \includegraphics[width=\textwidth]{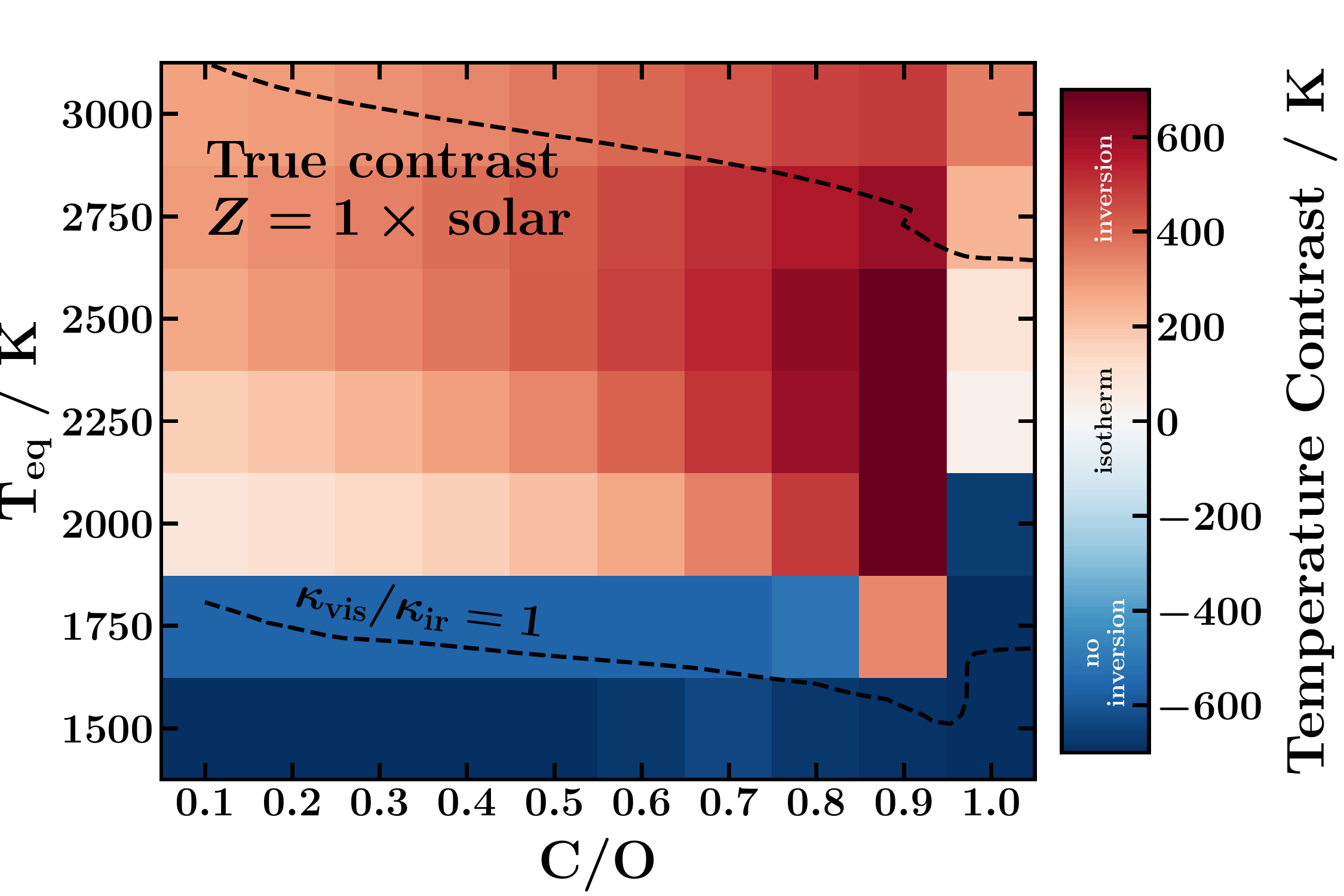}
    \end{subfigure}
    \quad
	\begin{subfigure}[b]{0.31\textwidth}
	    \includegraphics[width=\textwidth]{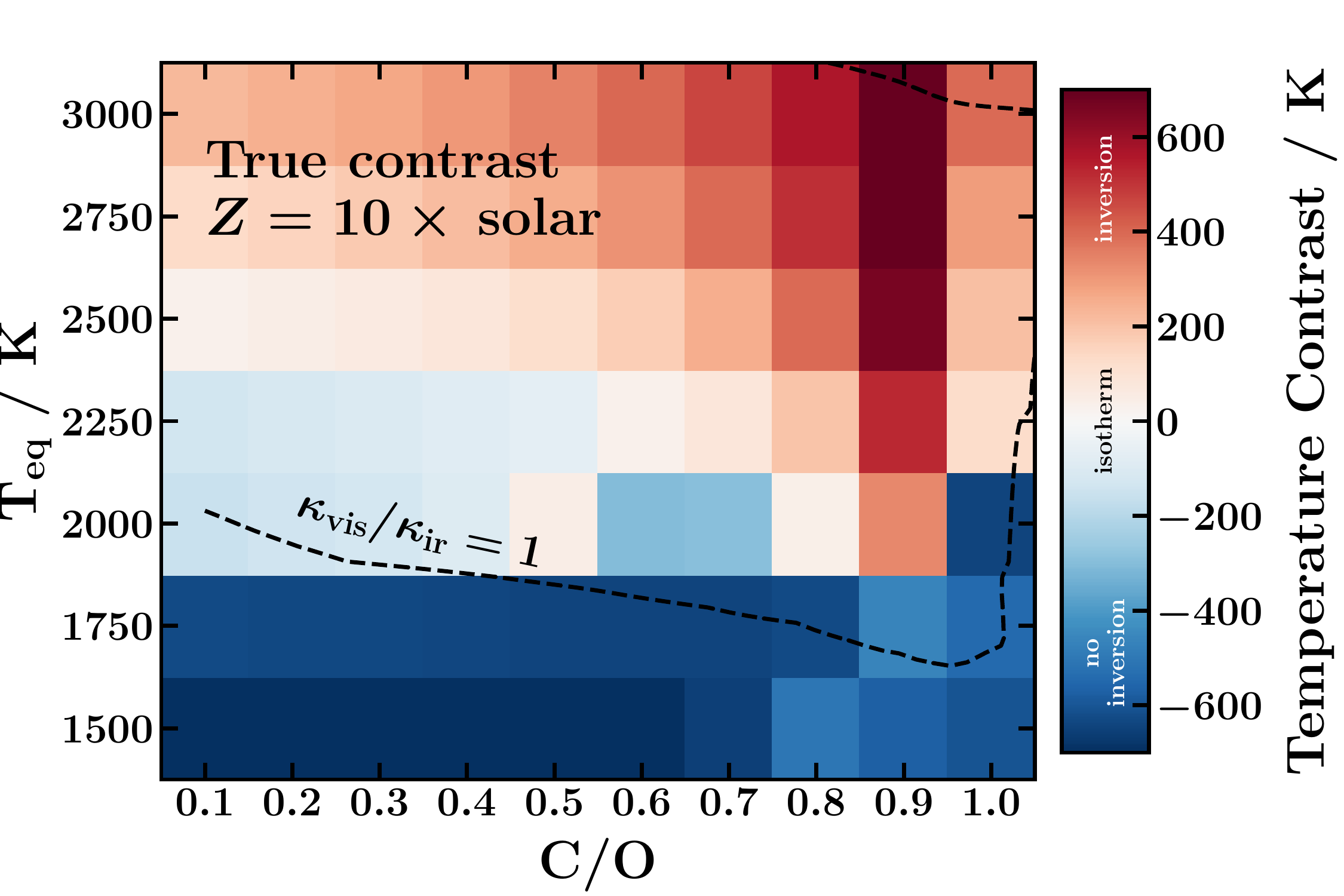}
    \end{subfigure}
	\begin{subfigure}[b]{0.31\textwidth}
	    \includegraphics[width=\textwidth]{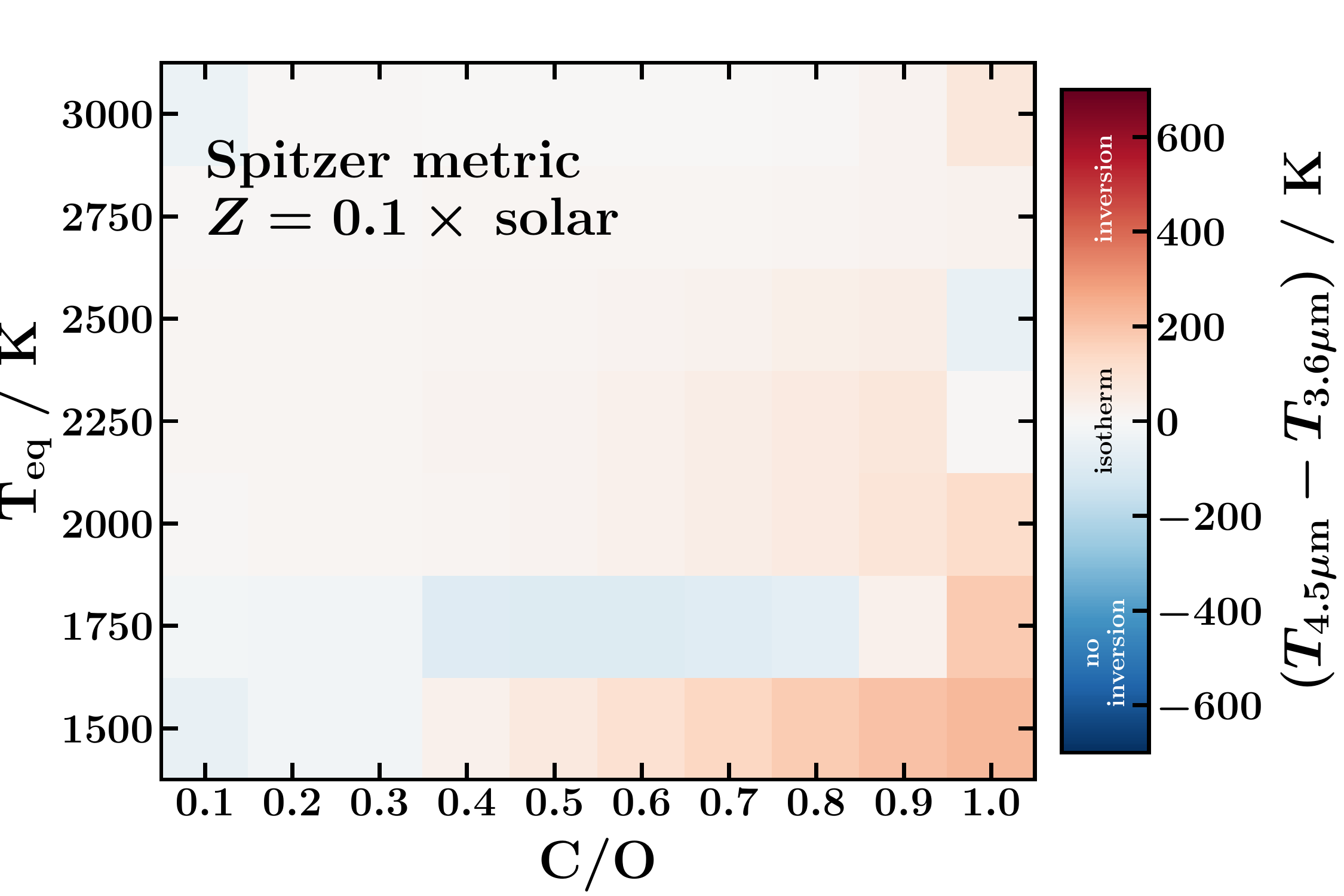}
    \end{subfigure}
    \quad
	\begin{subfigure}[b]{0.31\textwidth}
	    \includegraphics[width=\textwidth]{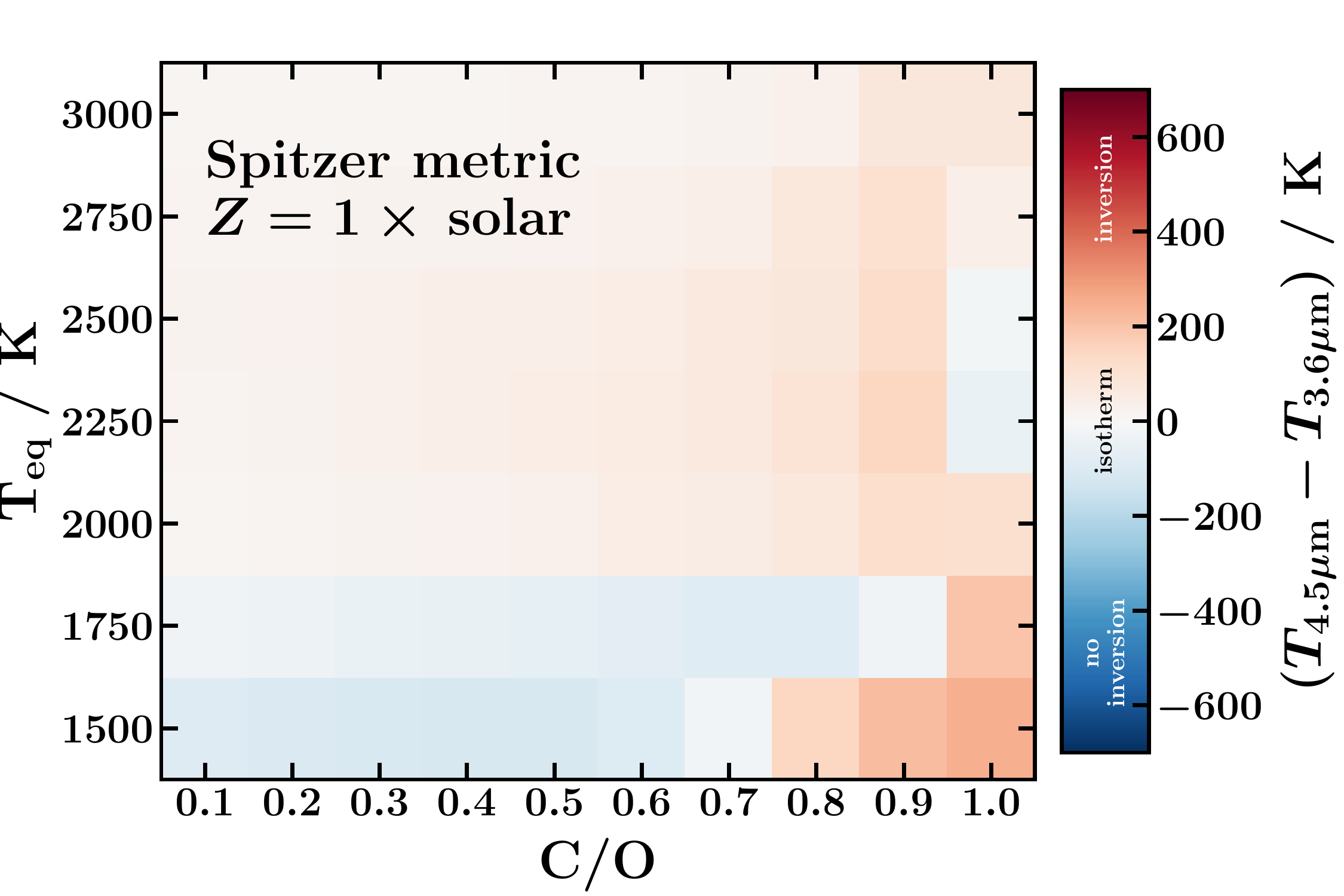}
    \end{subfigure}
    \quad
	\begin{subfigure}[b]{0.31\textwidth}
	    \includegraphics[width=\textwidth]{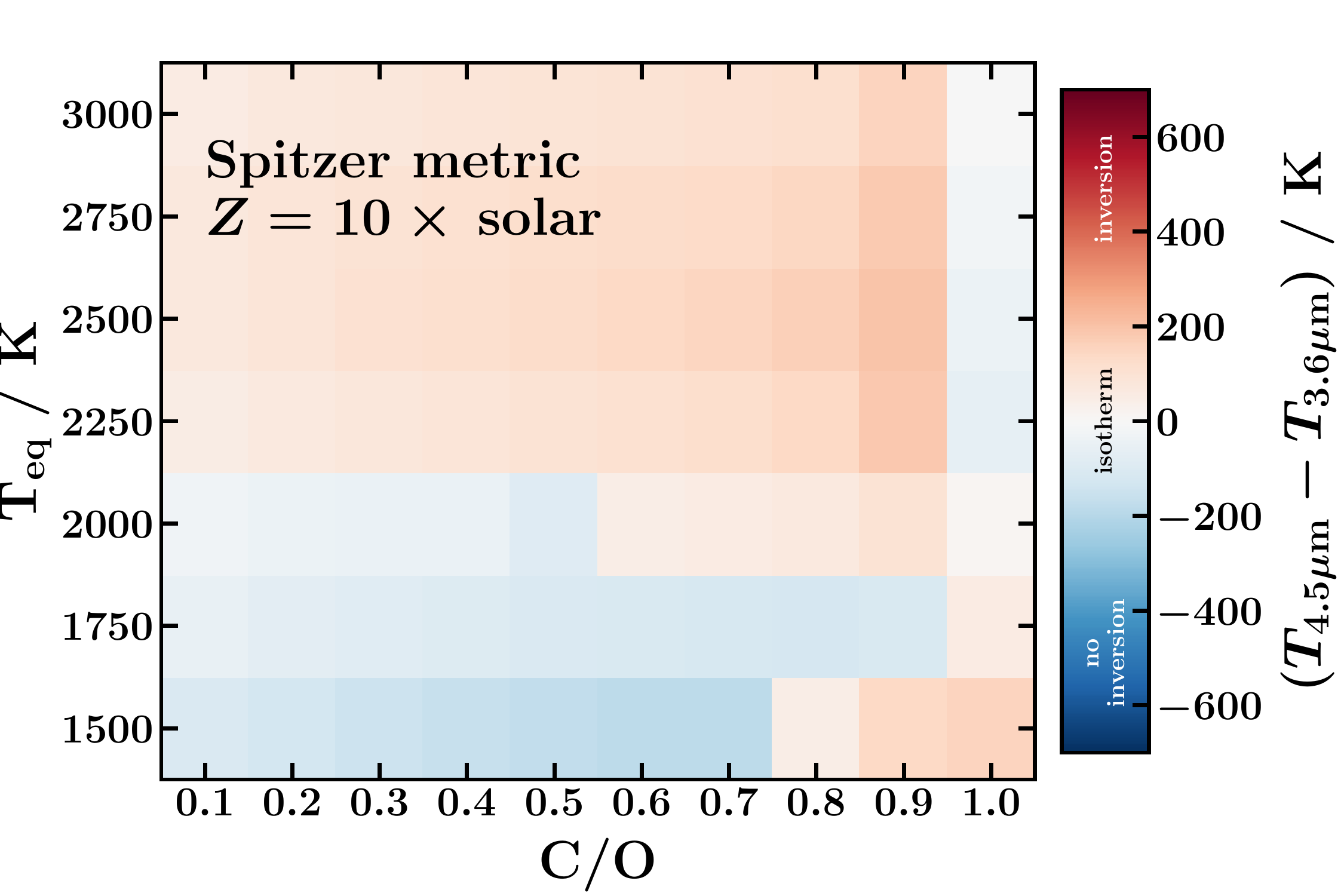}
    \end{subfigure}
	\begin{subfigure}[b]{0.31\textwidth}
	    \includegraphics[width=\textwidth]{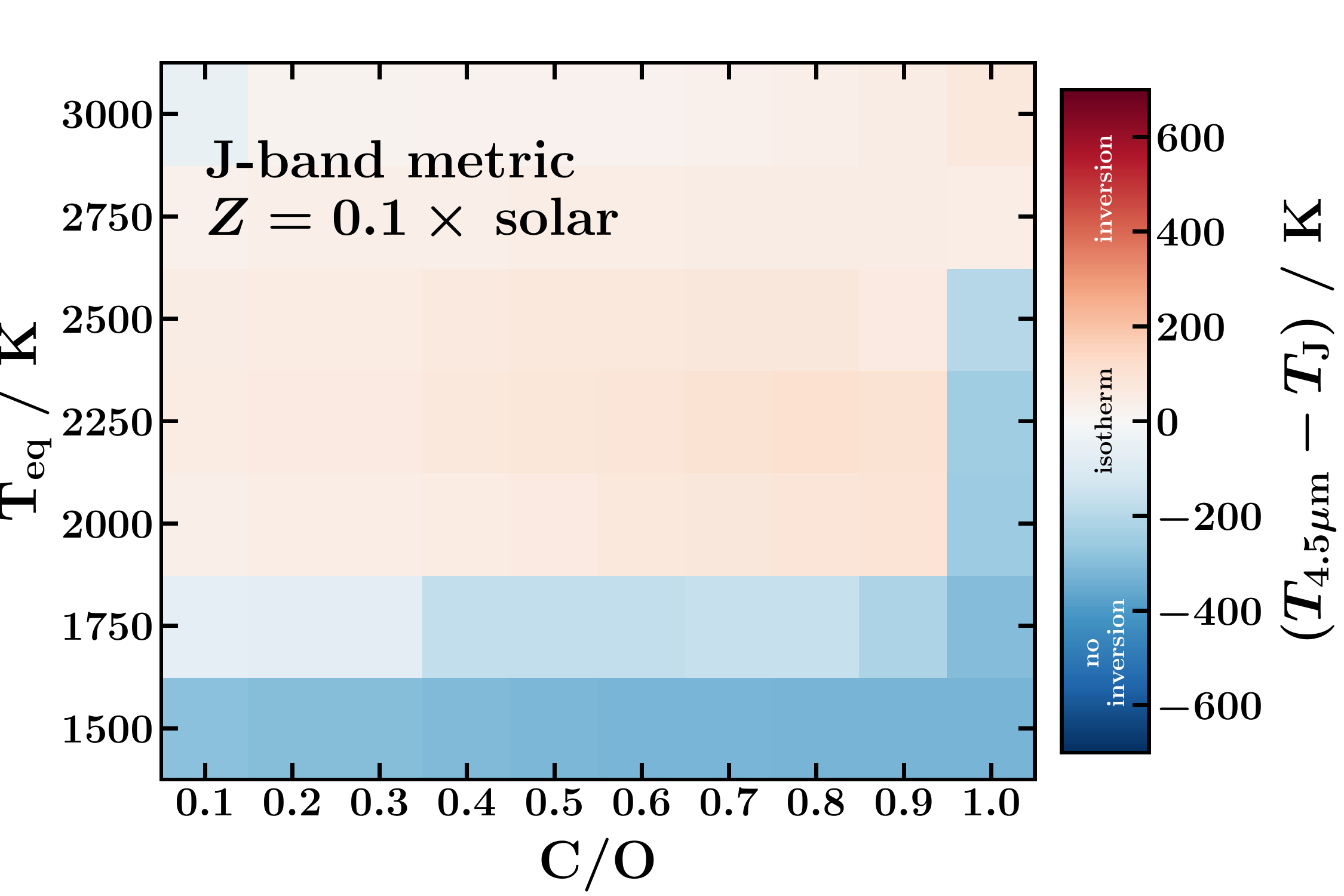}
    \end{subfigure}
    \quad
	\begin{subfigure}[b]{0.31\textwidth}
	    \includegraphics[width=\textwidth]{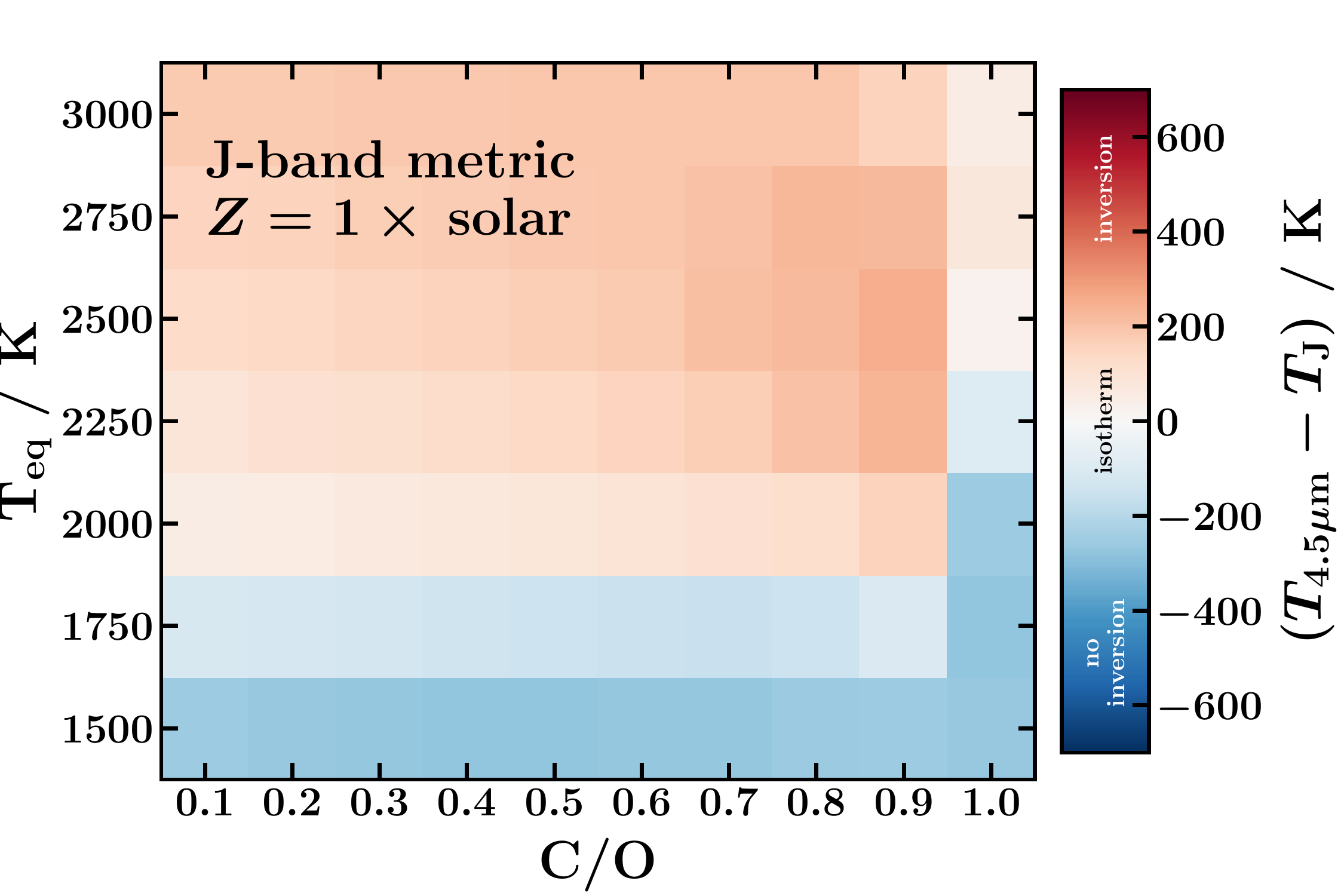}
    \end{subfigure}
    \quad
	\begin{subfigure}[b]{0.31\textwidth}
	    \includegraphics[width=\textwidth]{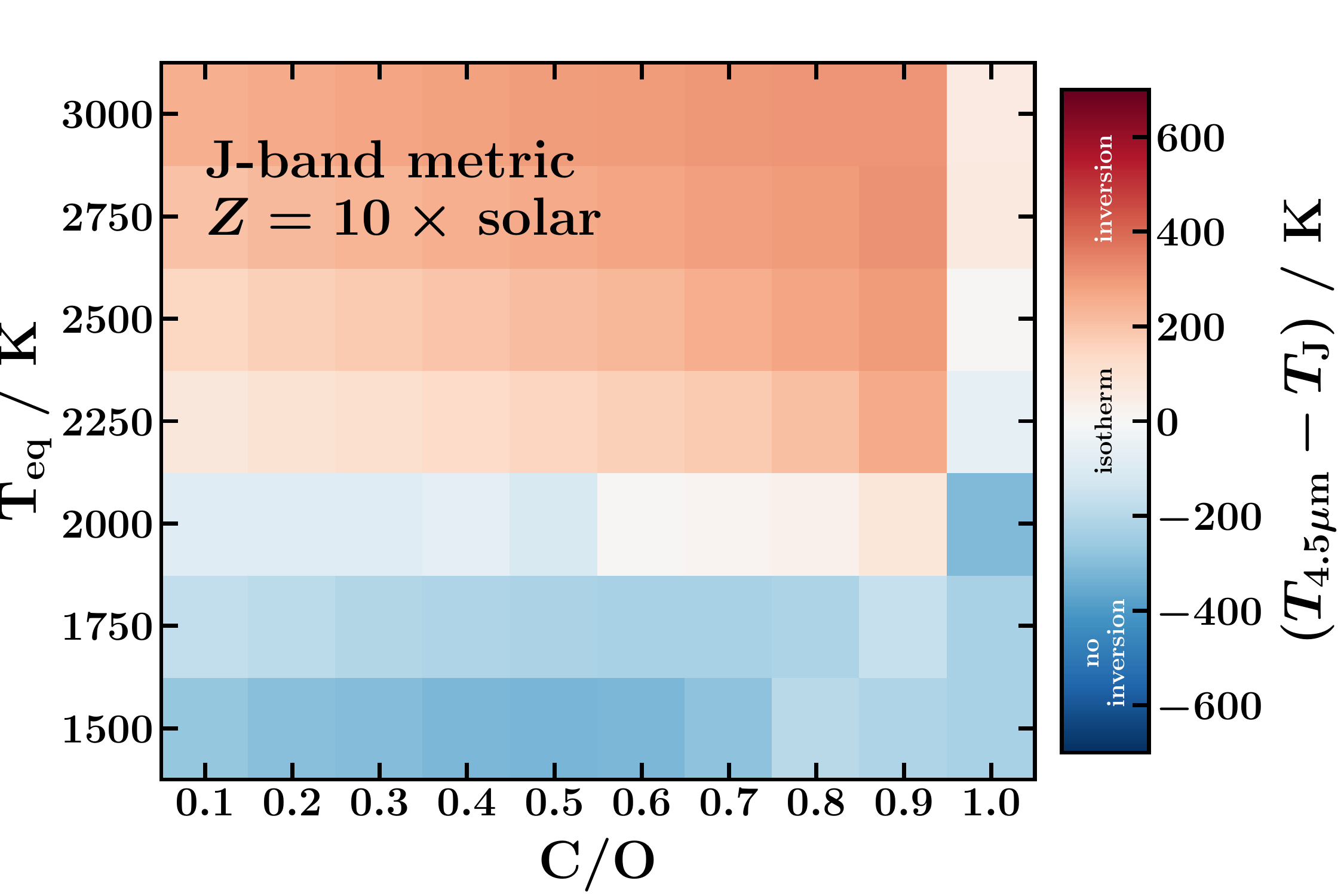}
    \end{subfigure}
\caption{\label{fig:invmaps} Inversion maps showing the strength of the inversion/non-inversion of equilibrium $P$-$T$ profiles as a function of equilibrium temperature and C/O ratio. Equilibrium $P$-$T$ profiles are calculated as described in section \ref{sec:crosssec}. Left, middle and right columns correspond to metallicities of 0.1, 1 and 10$\times$ solar, respectively. Top row: colour scale shows temperature contrast within the pressure range 1-$10^{-3}$ bar (see section \ref{sec:colmaps} for definition of the temperature contrast). Dashed black line shows contour of $\kappa_\mathrm{vis}/\kappa_\mathrm{ir}=1$ at a pressure of 1 mbar (see section \ref{sec:eqtio}). This predicts that models inside the contour should have thermal inversions, though note that this contour does not include effects due to vertical mixing and does not account for the fact that thermal inversions can happen at different pressures for different cases. Middle and bottom rows: performance of the Spitzer and J-band metrics throughout the parameter space. Colour scale shows brightness temperature contrasts between the IRAC~1 and IRAC~2 bands ($T_{4.5\mu\mathrm{m}}-T_{3.6\mu\mathrm{m}}$, middle row) and between the J-band and IRAC~2 band ($T_{4.5\mu\mathrm{m}}-T_{\mathrm{J}}$, bottom row). Positive (negative) values indicate a thermal inversion (non-inversion).}
\end{figure*}

\begin{figure*}
\centering
	\begin{subfigure}[b]{0.31\textwidth}
	    \includegraphics[width=\textwidth]{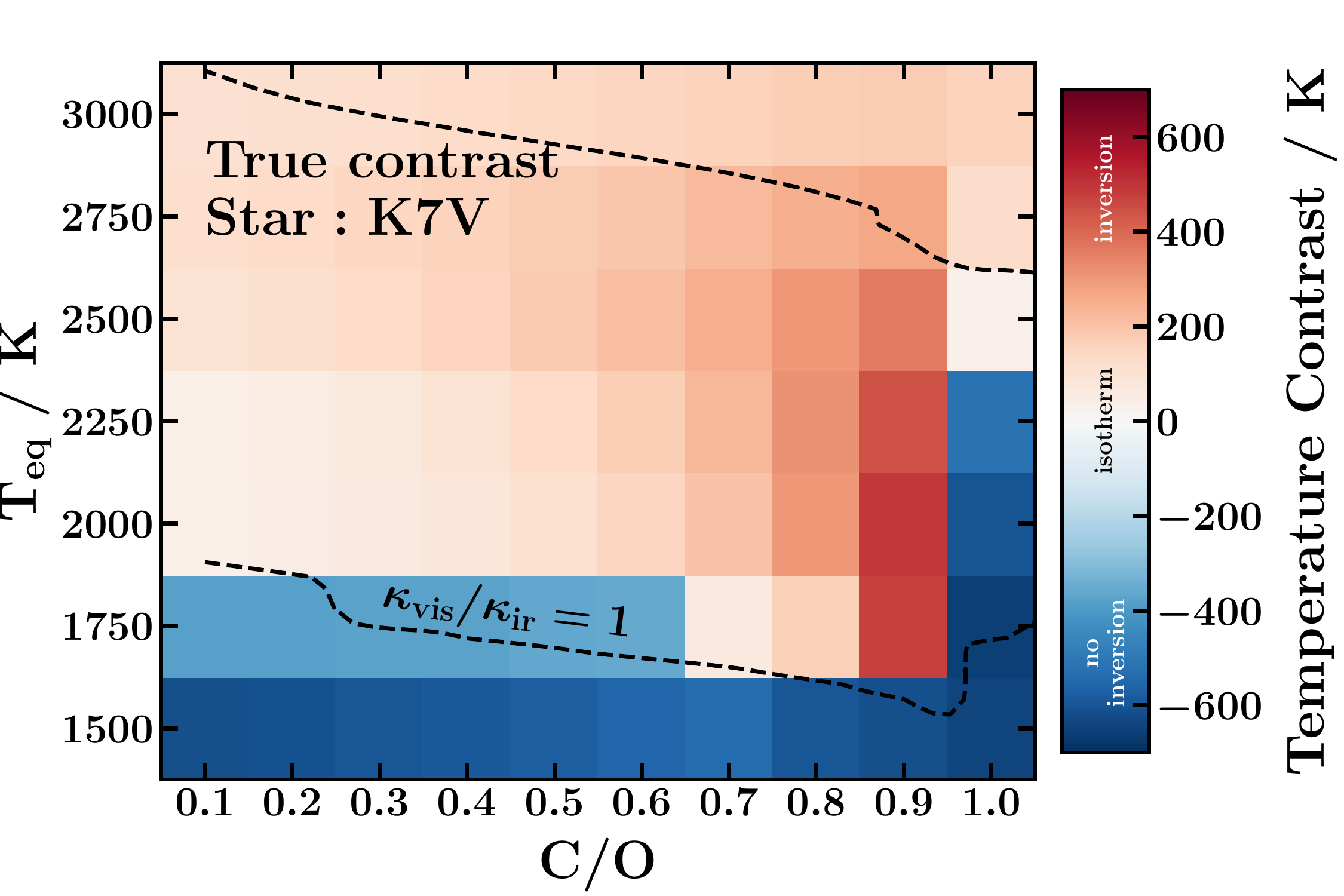}
    \end{subfigure}
    \quad
	\begin{subfigure}[b]{0.31\textwidth}
	    \includegraphics[width=\textwidth]{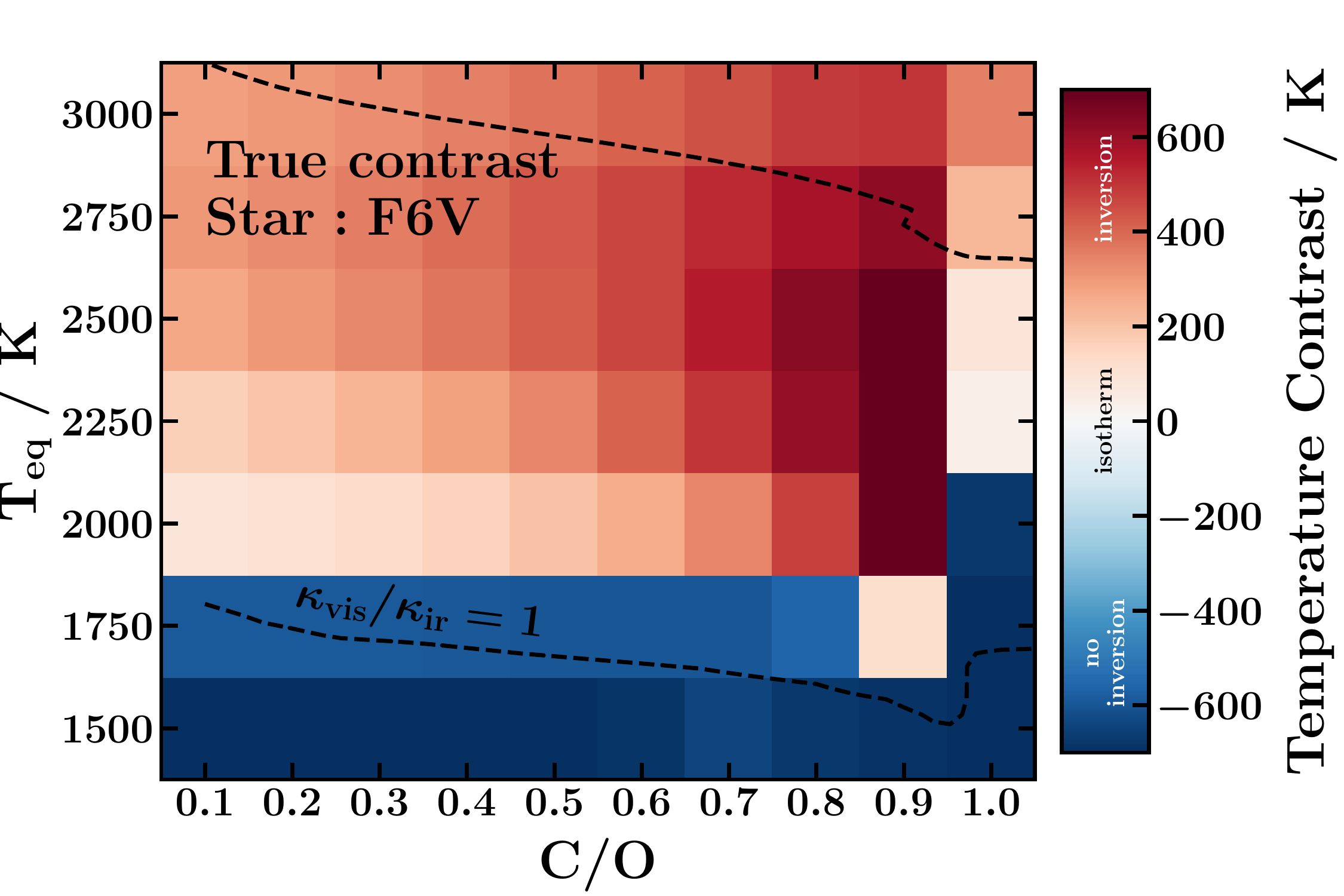}
    \end{subfigure}
    \quad
	\begin{subfigure}[b]{0.31\textwidth}
	    \includegraphics[width=\textwidth]{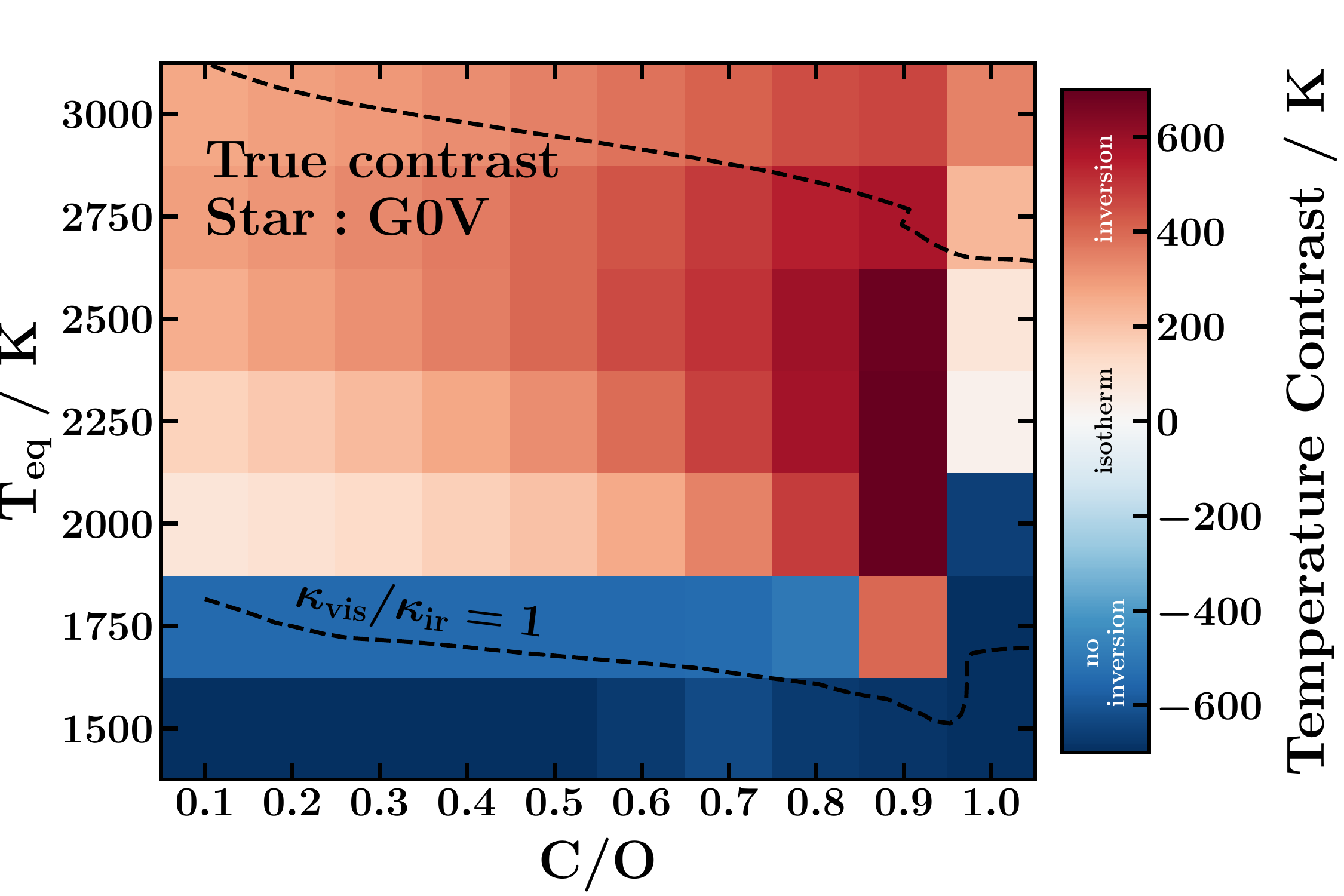}
    \end{subfigure}
	\begin{subfigure}[b]{0.31\textwidth}
	    \includegraphics[width=\textwidth]{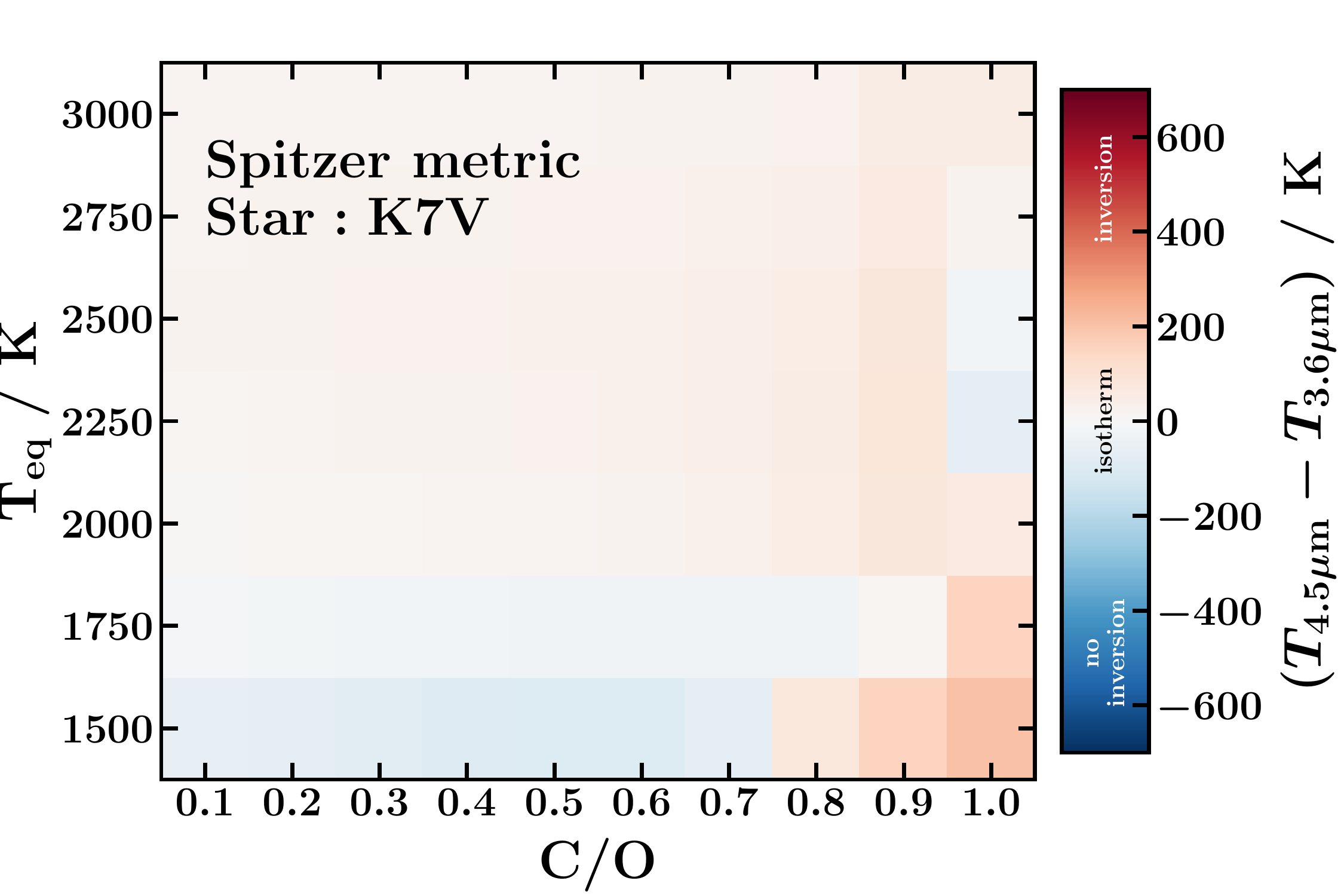}
    \end{subfigure}
    \quad
	\begin{subfigure}[b]{0.31\textwidth}
	    \includegraphics[width=\textwidth]{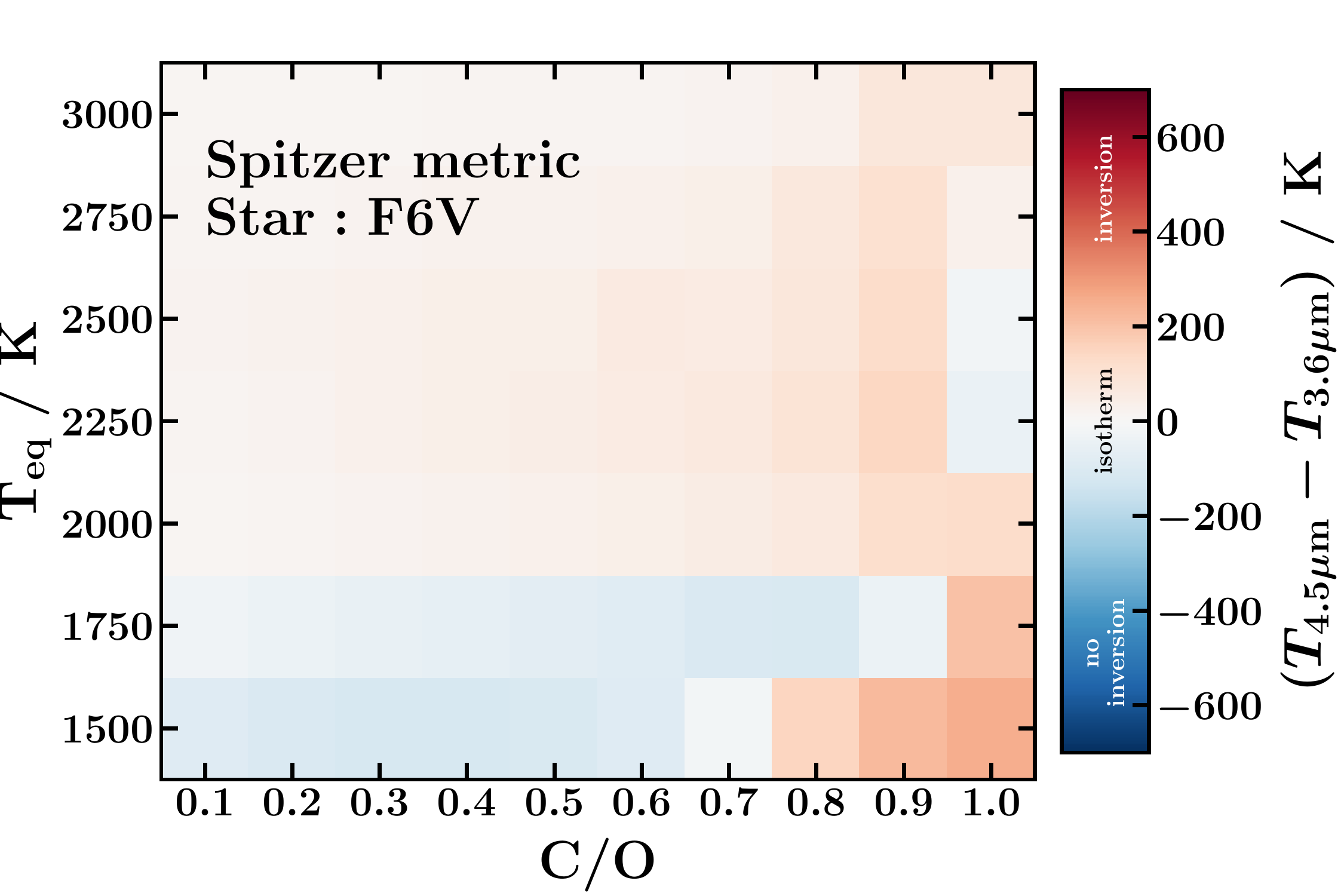}
    \end{subfigure}
    \quad
	\begin{subfigure}[b]{0.31\textwidth}
	    \includegraphics[width=\textwidth]{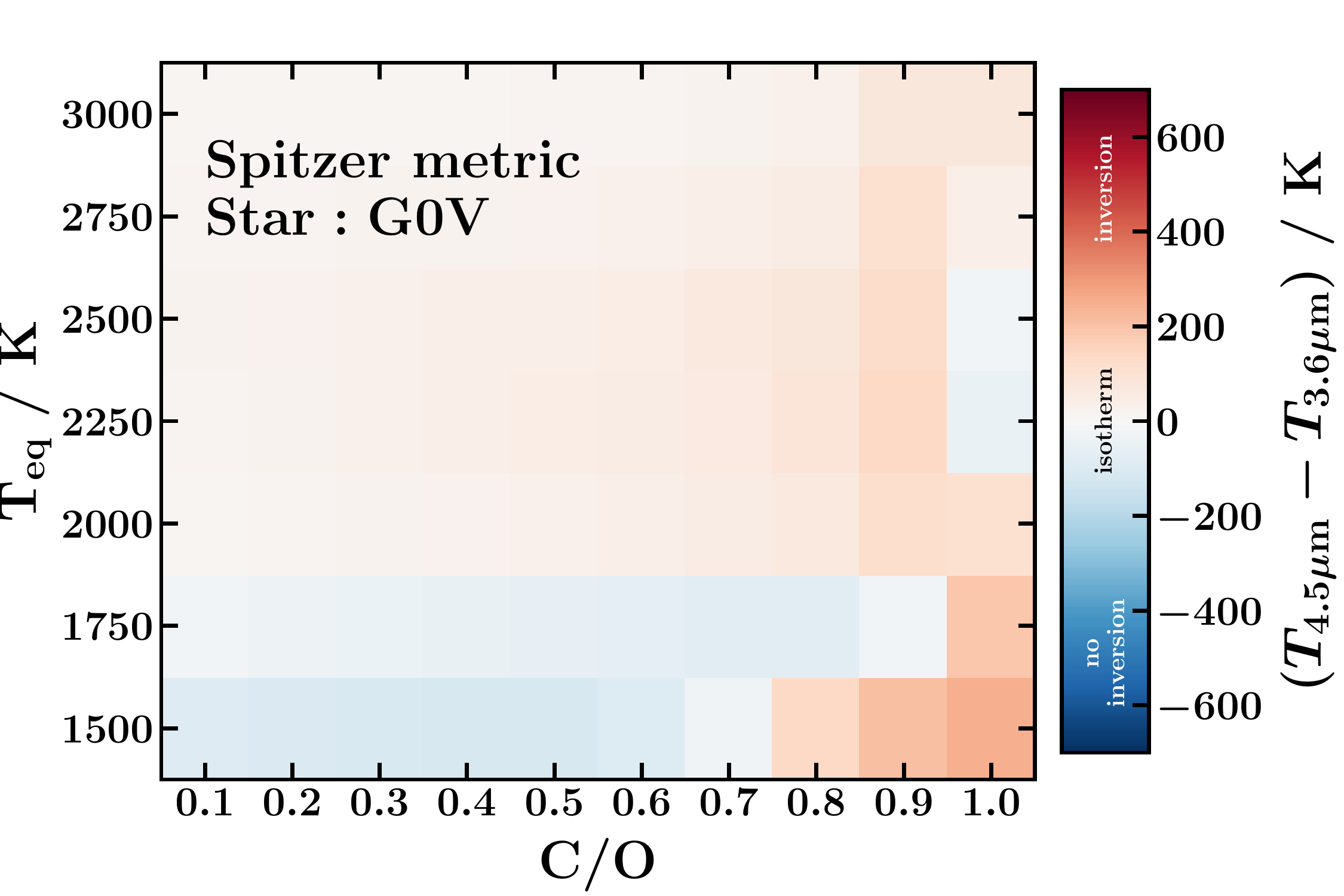}
    \end{subfigure}
	\begin{subfigure}[b]{0.31\textwidth}
	    \includegraphics[width=\textwidth]{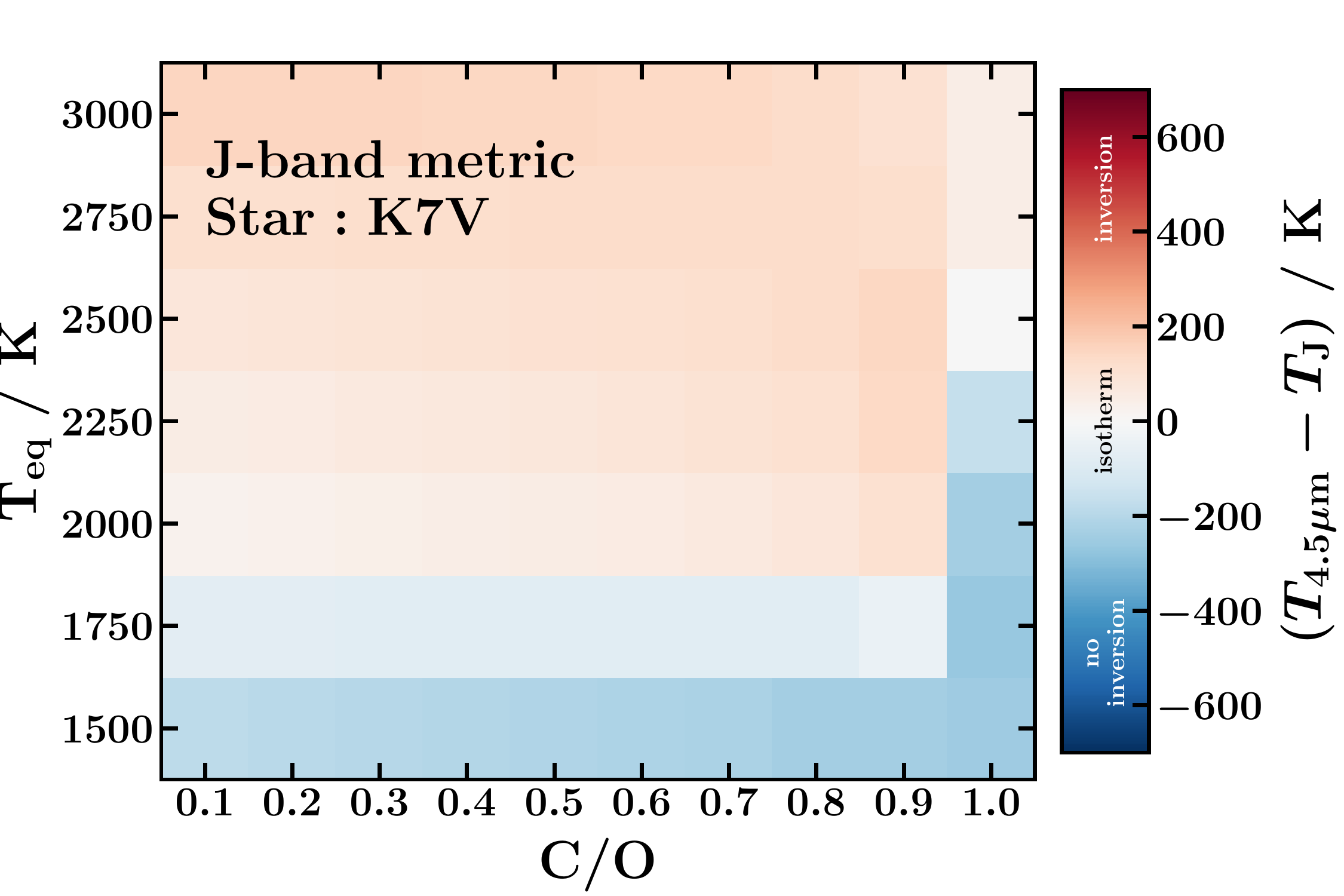}
    \end{subfigure}
    \quad
	\begin{subfigure}[b]{0.31\textwidth}
	    \includegraphics[width=\textwidth]{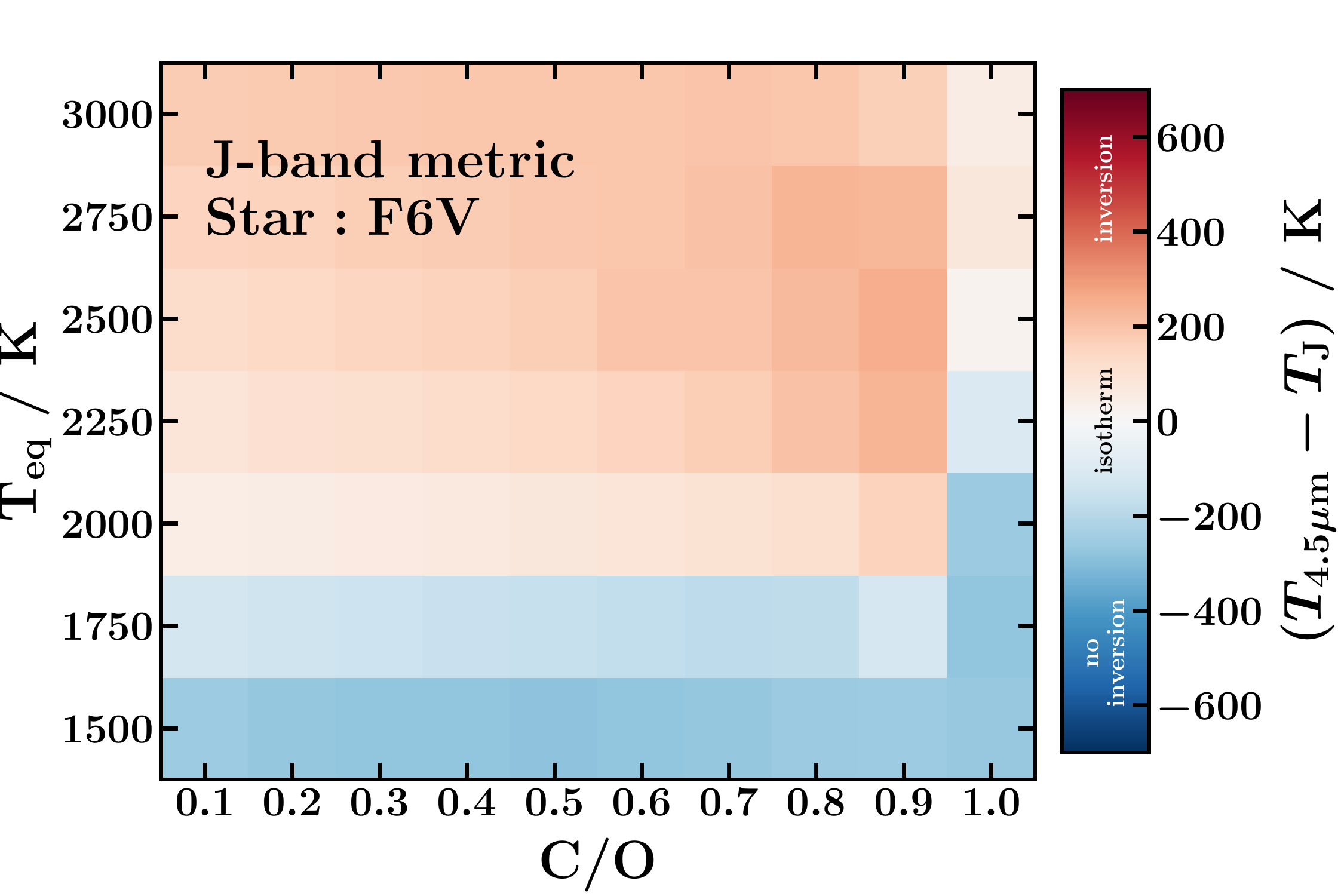}
    \end{subfigure}
    \quad
	\begin{subfigure}[b]{0.31\textwidth}
	    \includegraphics[width=\textwidth]{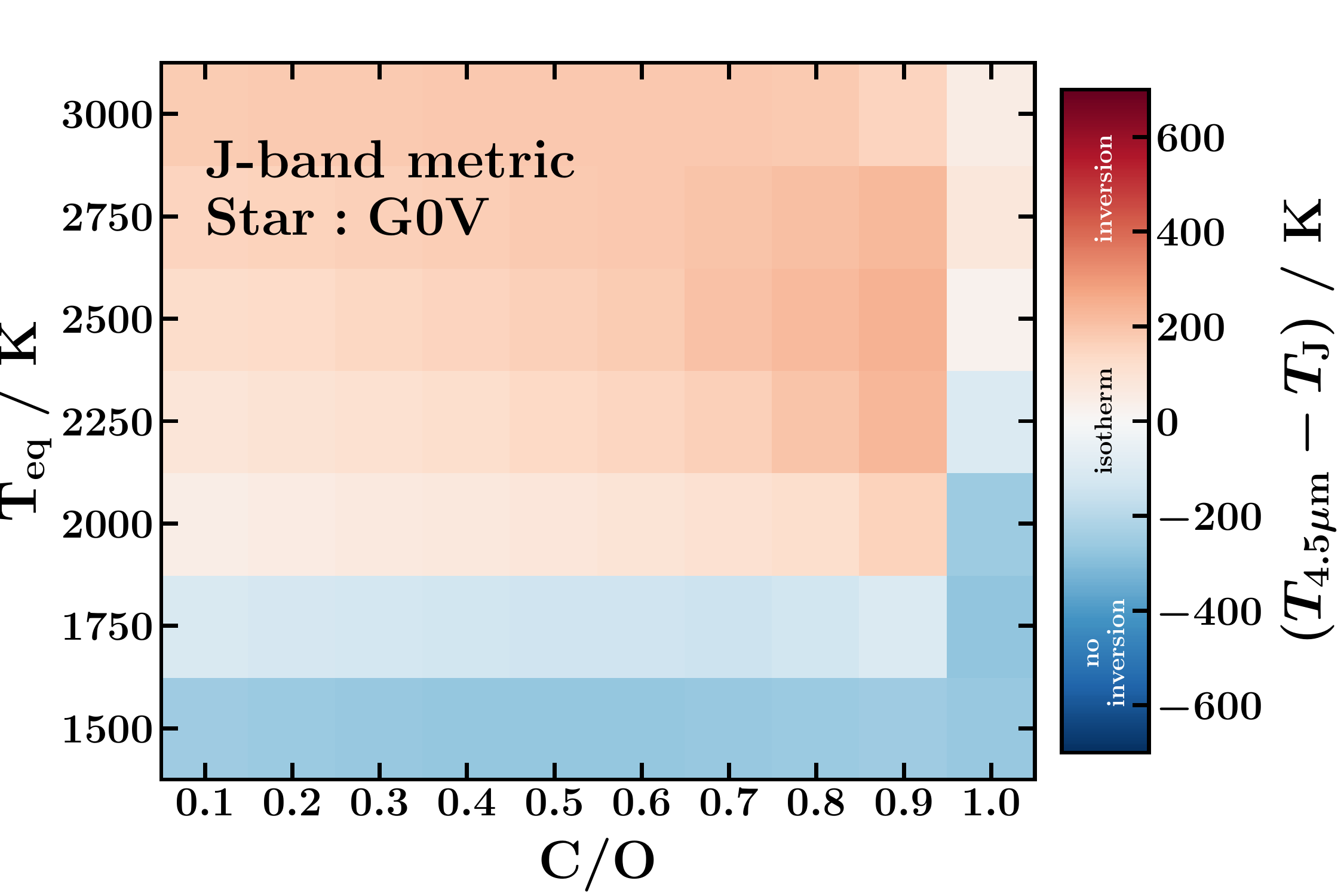}
    \end{subfigure}
\caption{\label{fig:invmaps_star}Inversion strength as a function of equilibrium temperature, C/O ratio and stellar type of the host star. Left, middle and right columns correspond to the stellar types K7V, F6V and G0V, respectively. The stellar properties used for each stellar type are those of WASP-43, WASP-18 and HD~209458, respectively \citep{Hellier2011,Southworth2012,Stassun2017}. Colour scales show the same temperature contrasts as in figure \ref{fig:invmaps}. Contours of $\kappa_\mathrm{vis}/\kappa_\mathrm{ir}=1$ in the top row are calculated in the same way as for figure \ref{fig:invmaps} but using the appropriate host star temperature for each case.}
\end{figure*}

\begin{figure}
\centering
	\begin{subfigure}[b]{0.4\textwidth}
	    \includegraphics[width=\textwidth]{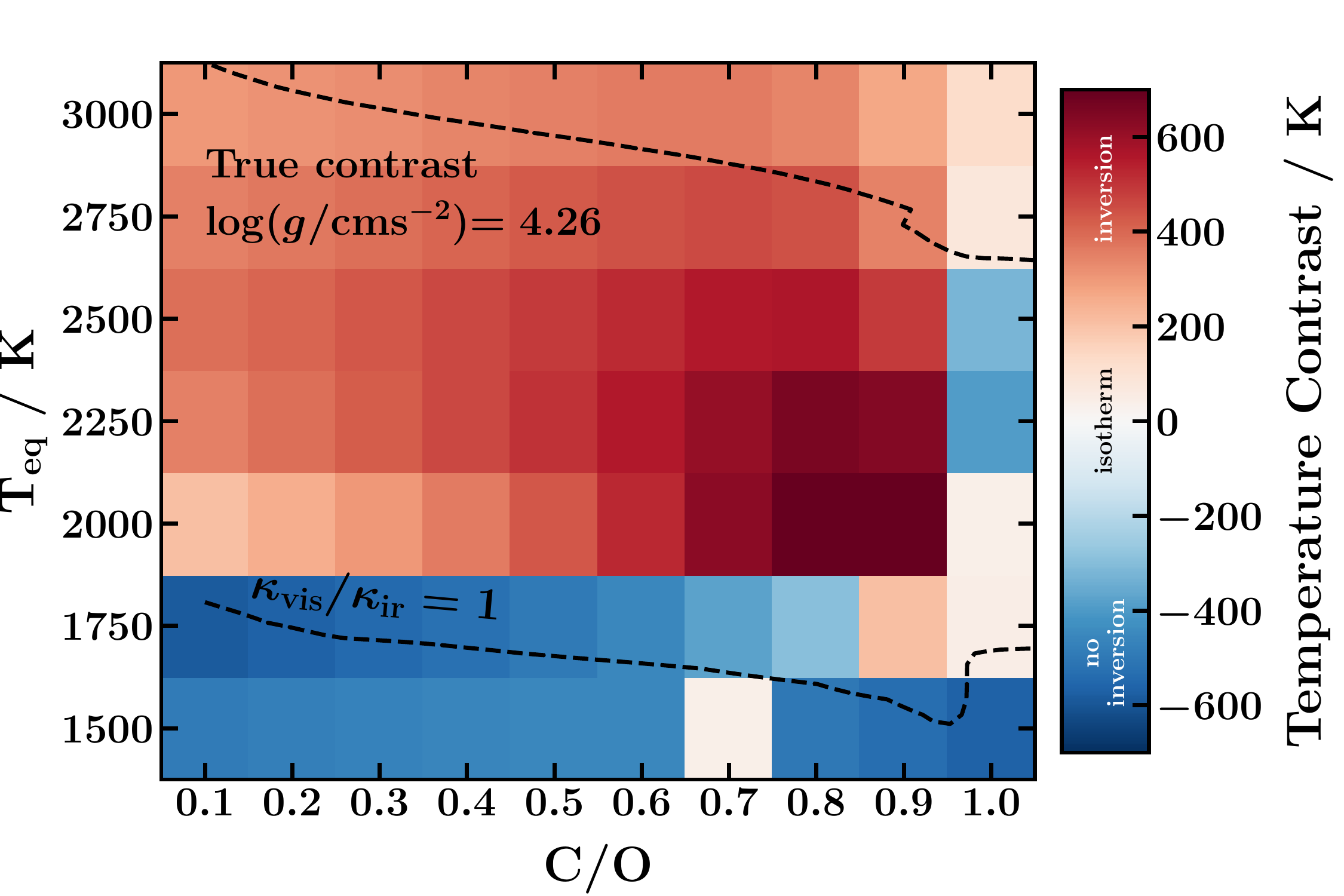}
    \end{subfigure}
	\begin{subfigure}[b]{0.4\textwidth}
	    \includegraphics[width=\textwidth]{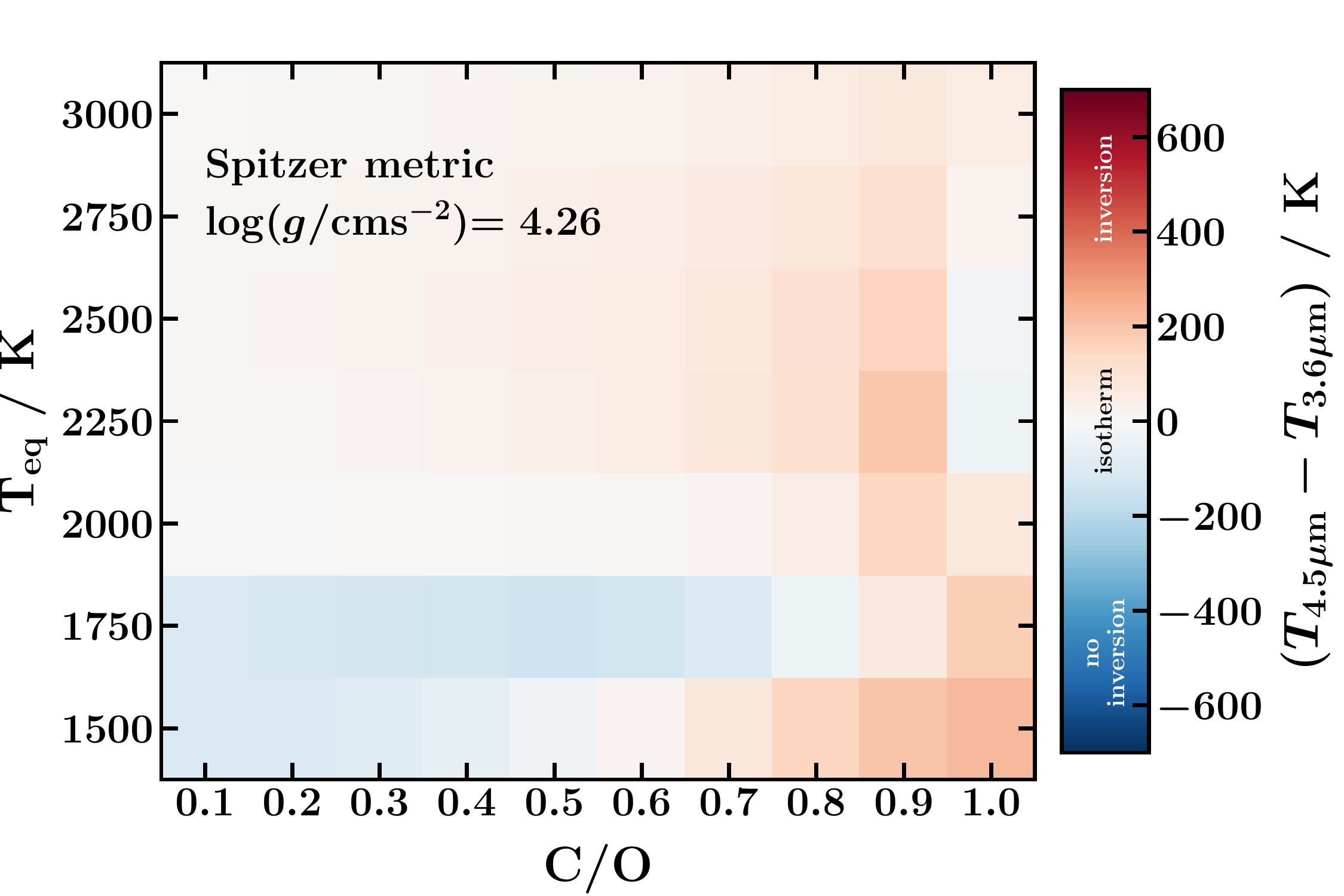}
    \end{subfigure}
	\begin{subfigure}[b]{0.4\textwidth}
	    \includegraphics[width=\textwidth]{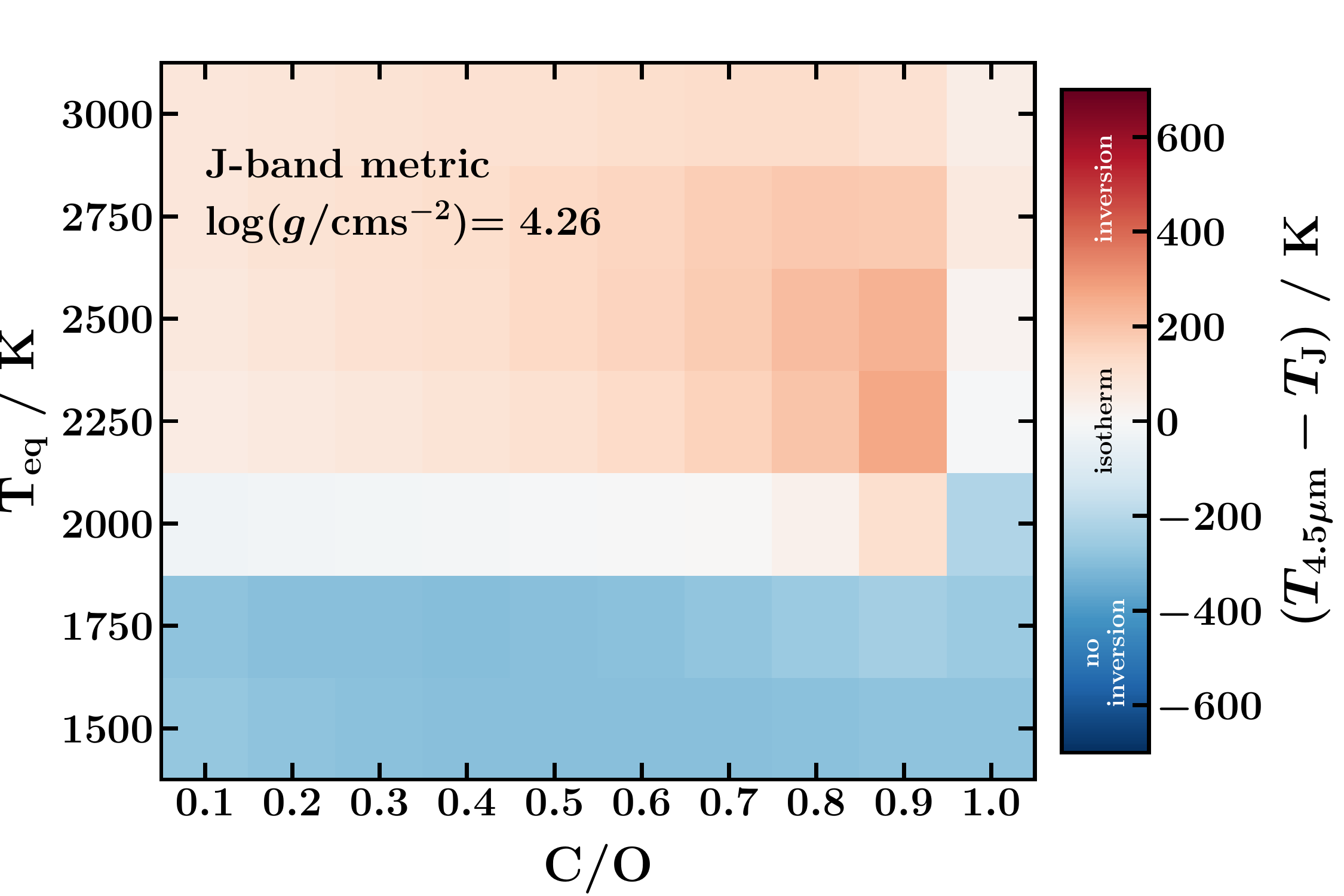}
    \end{subfigure}
\caption{\label{fig:invmaps_modgrav}Same as centre column of figure \ref{fig:invmaps} but for a planet with a higher gravity (log($g$/cm$^{-2}$)=4.26, similar to WASP-18b, compared to log($g$/cm$^{-2}$)=2.99 in figure \ref{fig:invmaps}).}
\end{figure}
%%%%%%%%%%%%%%%%%%%%%%%%%%%%%%%%%%%%%%%%%%%%%%%%%%%%%%%%%

In this section, we use self-consistent models to investigate in detail how chemical composition, incident irradiation and gravity affect the strength of thermal inversions in hot Jupiters as well as the performance of temperature gradient metrics. We use the \textsc{GENESIS} code \citep{Gandhi2017} to generate self-consistent model atmospheres and their spectra for various C/O ratios, equilibrium temperatures, metallicities, stellar types and gravities. The equilibrium chemical abundances are calculated with \textsc{HSC Chemistry}, and we also include the effects of vertical mixing by quenching all chemical abundances between 0.1 and 10$^{-3}$ bar (see section \ref{sec:emission}). In addition to vertical mixing, we also consider the redistribution of incident flux to the nightside of the planet. Following \citet{Garhart2019}, we approximate the redistribution factor, $f$, as a step function where planets with $T_{\mathrm{eq}}\leq 2500$ K transport 50\% of incident flux to the nightside ($f=1/2$), while planets with $T_{\mathrm{eq}} > 2500$ K retain all incident energy on the dayside ($f=1$) \citep[see also][]{Cowan2011}.

Using these equilibrium models, we calculate $P$-$T$ profiles and model spectra for a range of $T_\mathrm{eq}$, C/O ratios, metallicities, stellar types and gravities. Since we are focusing on thermal inversions due to TiO, we explore C/O ratios $\leq 1$ as TiO is significantly depleted at higher C/O ratios. The opacity sources we include in our models are those listed in section \ref{sec:emission}. We note that other sources of optical opacity not considered here (e.g. atomic Fe and Mg, \citealt{Lothringer2019}) can also cause thermal inversions, though we focus on thermal inversions due to TiO in the present work. We then use these models to explore temperature gradient as a function of these parameters, and compare different empirical metrics for assessing this gradient. In order to test the performance of the empirical metrics, we compare them to the true temperature gradient derived from the model $P$-$T$ profiles. This is found by considering the temperature contrast across the photosphere, which we take to be in the range $1-10^{-3}$ bar. For highly irradiated atmospheres, it is well known that the temperature profile becomes isothermal in the high optical depth limit before becoming steeper again at even higher pressures \citep{Guillot2010}. Note that, since we are using an internal temperature of 100~K which is much smaller than the irradiation temperature, the isotherm extends to the bottom of our computational domain and we do not see the steeper gradient. For the dayside of irradiated atmospheres such as those modelled here, the base of the photosphere coincides with the top of this isotherm in the lower atmosphere \citep[e.g.][]{Hubeny2003,Burrows2006b,Guillot2010}. The temperature of the isotherm is therefore a consistent measure of the temperature at the bottom of the photosphere across all the models we consider. For profiles with a thermal inversion, the temperature contrast is defined as the maximum photospheric temperature minus the minimum temperature between the bottom of the photosphere ($\sim$1 bar) and the location of the temperature maximum. For profiles with no inversion, the contrast is defined as the minimum atmospheric temperature minus the temperature at the bottom of the photosphere (at 1 bar, i.e. the temperature of the isotherm). We consider a $P$-$T$ profile to have a thermal inversion if the inversion strength is greater than 30 K. Maps of inversion strength as a function of C/O ratio and equilibrium temperature are shown in figures \ref{fig:invmaps}, \ref{fig:invmaps_star} and \ref{fig:invmaps_modgrav} for varying metallicities, stellar host types and planetary surface gravities.

\subsection{Inversion strength}
\label{sec:inv_strength}
The true temperature contrasts shown in the top rows of figures \ref{fig:invmaps}, \ref{fig:invmaps_star} and \ref{fig:invmaps_modgrav} show that a wide variety of $P$-$T$ profiles are possible in this parameter space, including strong thermal inversions, isotherms and non-inverted profiles. This is in accordance with observations of hot Jupiters, which have revealed spectra with emission features \citep[e.g.][]{Haynes2015,Evans2017,Sheppard2017,Arcangeli2018}, flat spectra \citep[e.g.][]{Swain2013,Cartier2017,Mansfield2018}, and absorption features \citep[e.g.][]{Beatty2017b}. We also plot contours of $\kappa_\mathrm{vis}/\kappa_\mathrm{ir}=1$ (calculated at a nominal pressure of 1 mbar) in the top panels of figure \ref{fig:invmaps} to show the regions of parameter space expected to host thermal inversions (section \ref{sec:eqtio}). Note however that these contours do not include effects due to vertical mixing and are only calculated at a single pressure, whereas thermal inversions can happen at different pressures for different cases. Nevertheless, the $\kappa_\mathrm{vis}/\kappa_\mathrm{ir}=1$ contours roughly outline the regions of the parameter space where thermal inversions occur.

The models in figures \ref{fig:invmaps}, \ref{fig:invmaps_star} and \ref{fig:invmaps_modgrav} suggest that strong inversions are more likely to occur at higher C/O ratios in the C/O$<$1 regime, as has been previously discussed in other works \citep[e.g.][]{Molliere2015,Gandhi2019}. This is consistent with the trend seen in section \ref{sec:eqtio} and in the $\kappa_\mathrm{vis}/\kappa_\mathrm{ir}=1$ contour that thermal inversions are possible for a C/O of 0.9 at a lower equilibrium temperature compared to other C/O ratios. In addition, for a C/O ratio of one, only weak thermal inversions or non-inverted profiles are seen in the models, consistent with the strong depletion of TiO at C/O=1 \citep{Madhusudhan2011b}. As discussed in section \ref{sec:eqtio}, these thermal inversions are caused by optical opacity from Na and K \citep[also see e.g.,][]{Molliere2015} as TiO abundance is strongly depleted for C/O$\geq$1 \citep[e.g.,][]{Madhusudhan2012}.

Another clear trend is that for increasing metallicity, peak inversion strength occurs at higher $T_\mathrm{eq}$ (figure \ref{fig:invmaps}). This is because TiO is depleted by condensation and thermal dissociation at relatively higher temperatures for higher metallicity (figure \ref{fig:met_plot}), so a higher equilibrium temperature is needed to obtain $\kappa_\mathrm{vis}$/$\kappa_\mathrm{ir}$=1 (see appendix \ref{sec:metallicity_plot} and \citet{Molliere2015}). Furthermore, the strongest thermal inversions are stronger for high-metallicity models, as seen in figure \ref{fig:invmaps}. Compared to variation in metallicity, changes in stellar type have a more subtle effect on thermal inversion strength in the parameter space tested here. However, figure \ref{fig:invmaps_star} does show that earlier stellar types can allow somewhat stronger thermal inversions, e.g. inversion strengths for a K7V host star are typically weaker compared to F6V and G0V host stars in this parameter space \citep[see also][]{Lothringer2019}. Furthermore, a later stellar type can allow thermal inversions to occur at lower equilibrium temperatures. In particular, for the K7V host star, thermal inversions start to become possible at 1750~K and C/O=0.7-0.8, which is not the case for the F6V and G0V host stars.

We also test the effect of increasing the planetary surface gravity in figures \ref{fig:invmaps} and \ref{fig:invmaps_modgrav}. In the centre column of figure \ref{fig:invmaps}, the surface gravity is chosen to be similar to WASP-12b (log($g$/cm$^{-2}$)=2.99), while in figure \ref{fig:invmaps_modgrav} we use a larger gravity similar to WASP-18b (log($g$/cm$^{-2}$)=4.26). Between these two extremes, there are only minimal differences in inversion strength in the parameter space investigated here. One difference is that in the high-gravity case, slightly higher inversion strengths are typically found for lower $T_\mathrm{eq}$ and lower C/O ratio. There are also a few cases for which the high- and low-gravity models disagree on the presence of a thermal inversion. However, the $P$-$T$ profiles in these cases are in fact similar between the high- and low-gravity models, and the difference in inversion classification is due only to small differences between the profiles. Since the two gravities investigated here are extremes of the range expected for hot Jupiters, we conclude that gravity has a relatively small impact on inversion strength compared to factors such as metallicity. We also note that in our models, the gravity impacts the thermal profile only through the scale height and the fact that the photosphere is at higher pressure and therefore has greater pressure broadening of the chemical cross sections. However, a varying gravity could impact the gravitational settling of heavier species such as TiO, which we do not consider here \citep{Beatty2017}.

\subsection{Performance of metrics}
\label{sec:performace_of_metrics}
We also use these models to test the performances of the Spitzer and J-band metrics discussed in section \ref{sec:h2o_metric}. As discussed in section \ref{sec:h2o_metric}, each of the photometric bands used by these metrics probes either a spectral feature (i.e. the IRAC 2 band) or a spectral window (e.g. the IRAC 1 and J-bands), allowing different pressures in the atmosphere to be probed. In section \ref{sec:h2o_metric}, the efficacy of these bands in probing spectral features and windows was explored for a solar-like composition and a single equilibrium temperature. Here, we extend this analysis to a range of C/O ratios, equilibrium temperatures, metallicities, stellar types and gravities. Figure \ref{fig:invmaps} shows brightness temperature contrasts for these metrics between equilibrium temperatures of 1500 K to 3000 K, for C/O ratios of 0.1-1.0 and for metallicities of 0.1, 1 and 10$\times$ solar. The middle row of figure \ref{fig:invmaps} shows brightness temperature contrasts between the Spitzer IRAC 1 and IRAC 2 bands, $T_{4.5\mu\mathrm{m}}-T_{3.6\mu\mathrm{m}}$, while the bottom row shows the $T_{4.5\mu\mathrm{m}}-T_{\mathrm{J}}$ contrast. Similarly, these contrasts are shown for different stellar types (K7V, F6V and G0V) and planetary gravities in figures \ref{fig:invmaps_star} and \ref{fig:invmaps_modgrav}, respectively.

Across the parameter space we explore, the IRAC 2 channel probes a CO spectral feature, as CO is abundant at all of the temperatures and C/O ratios modelled (figure \ref{fig:XTCO}). However, the spectral features or windows probed by the IRAC 1 and J-bands can vary depending on chemistry. For example, the IRAC 1 band coincides with CH$_4$ and HCN opacity features. Therefore, for an atmosphere with high CH$_4$ and/or HCN abundance, the IRAC 1 band will no longer probe a spectral window and the performance of the Spitzer metric may not be optimal. This can be seen in figures \ref{fig:invmaps}, \ref{fig:invmaps_star} and \ref{fig:invmaps_modgrav} at low equilibrium temperatures ($T_\mathrm{eq}\approx1500$ K) and high C/O, where $T_{4.5\mu\mathrm{m}}-T_{3.6\mu\mathrm{m}}$ has positive values, despite the lack of thermal inversion in the models in this region of parameter space. Overall, the J-band metric provides a slightly better match to the true contrast than the IRAC 1 band. However, since the J-band is chosen as a continuum relative to H$_2$O absorption, the J-band metric will not necessarily work for atmospheres that are water poor, e.g. for C/O$>$1 or other compositions not considered here. We explore this phenomenon further in section \ref{sec:Tb_contrast}.

While the Spitzer and J-band metrics largely agree as to whether an inversion is present or not, there are two main exceptions to this trend. The first is at low equilibrium temperatures and high C/O, for reasons described above. The second is at $T_\mathrm{eq}\sim$1750 K where, for some C/O ratios, the J-band incorrectly infers a non-inversion. For example, this occurs with a K7V host star (figure \ref{fig:invmaps_star}). Nevertheless, the two metrics agree for most $T_\mathrm{eq}$ and C/O ratios. This is consistent with the fact that our analysis is for oxygen-rich compositions where H$_2$O absorption dominates and the J-band and IRAC 1 band do indeed probe the continuum. Furthermore, where this agreement exists, the $T_{4.5\mu\mathrm{m}}-T_{\mathrm{J}}$ contrast is consistently larger than the $T_{4.5\mu\mathrm{m}}-T_{3.6\mu\mathrm{m}}$ contrast. This effect is visible in figure \ref{fig:metric} for a solar-like composition, and figures \ref{fig:invmaps}, \ref{fig:invmaps_star} and \ref{fig:invmaps_modgrav} confirm that it holds across the range of $T_\mathrm{eq}$, C/O ratios, metallicities, stellar types and gravities explored here. This suggests that, as in figure \ref{fig:metric}, the J-band is consistently probing deeper pressures than the IRAC 1 band. For an oxygen-rich atmosphere with a thermal inversion, one would therefore expect that $T_{\mathrm{J}}<T_{3.6\mu\mathrm{m}}<T_{4.5\mu\mathrm{m}}$. Similarly, $T_{4.5\mu\mathrm{m}}<T_{3.6\mu\mathrm{m}}<T_{\mathrm{J}}$ would be expected for an oxygen-rich atmosphere with no thermal inversion. In section \ref{sec:Tb_contrast}, we compare these brightness temperatures for hot Jupiters with spectral and photometric observations. By looking for outliers to this trend, we can therefore identify which planets have different chemistries or disequilibrium processes compared to those we consider in our models.

\subsection{Comparison of data to models}
\label{sec:Tb_contrast}
\begin{figure}
    \centering
    \includegraphics[width=0.48\textwidth]{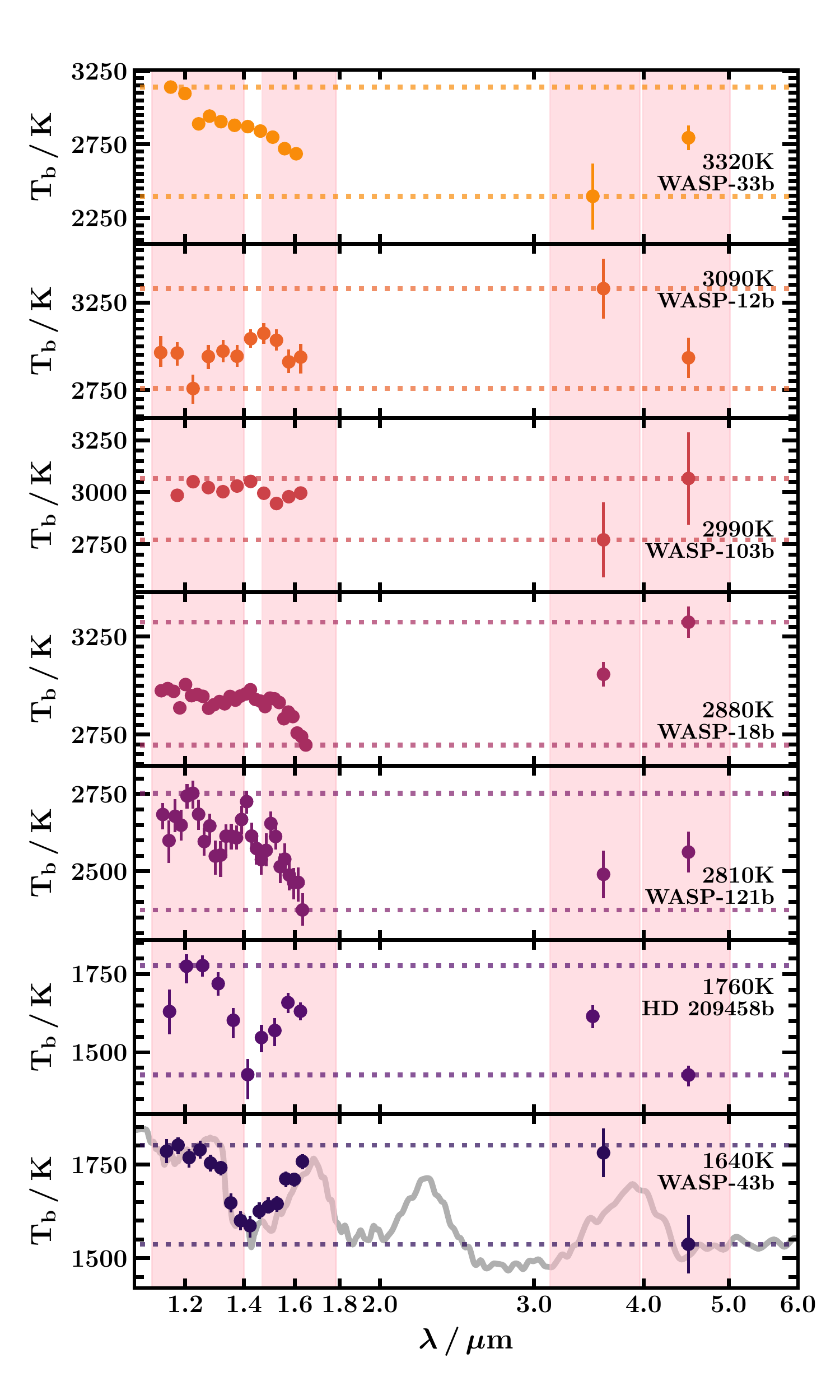}
    \caption{Brightness temperatures probed by HST/WFC3 and Spitzer IRAC 1 and IRAC 2 data for seven planets, ordered by equilibrium temperature. Nominal equilibrium temperatures are calculated assuming dayside only redistribution and are labelled for each planet. The bottom panel shows the spectrum from figure \ref{fig:metric} which corresponds to a non-inverted $P$-$T$ profile, for reference. Spitzer brightness temperatures for WASP-33b and HD~209458b are calculated using data from \citet{Deming2012} and \citet{Diamond-Lowe2014}, respectively. Other Spitzer temperatures are from \citet{Garhart2019}. HST brightness temperatures are calculated using spectra from \citet{Haynes2015} (WASP-33b) \citet{Kreidberg2014} (WASP-43b), \citet{Evans2017} (WASP-121b), \citet{Sheppard2017} (WASP-18b), \citet{Kreidberg2018} (WASP-103b), \citet{Stevenson2014} (WASP-12b) and \citet{Line2016} (HD~209458b). WASP-33b, WASP-18b, \textrm{WASP-103b} and WASP-121b are reported to have thermal inversions in the literature while WASP-12b, WASP-43b \textrm{and HD~209458b} have non-inverted profiles (see references above). Planetary and stellar properties used to calculate the brightness temperatures are from \citet{Collier-Cameron2010,Collins2017,Gillon2014,Hellier2009,Southworth2010,Southworth2012,Delrez2016,Hellier2011,Gillon2012,Stassun2017}.}
    \label{fig:Tblam}
\end{figure}

While we have so far discussed photometric bands for assessing thermal inversions in oxygen-rich spectra, we now consider a more general approach by directly considering brightness temperatures probed by existing data. We do this by plotting the brightness temperatures of HST/WFC3 and Spitzer IRAC 1 and IRAC 2 observations for 7 planets (figure \ref{fig:Tblam}). Here, we choose to use IRAC 1 and IRAC 2 brightness temperatures from \citet{Garhart2019} as they perform a uniform analysis across many Spitzer secondary eclipse observations. However, we note that Spitzer data for a given object can differ depending on the way it is analysed (e.g. \citealt{Garhart2019} compared to \citealt{Kreidberg2018}).

As discussed in section \ref{sec:colmaps}, oxygen-rich atmospheres are expected to have $T_{\mathrm{J}}<T_{3.6\mu\mathrm{m}}<T_{4.5\mu\mathrm{m}}$ or $T_{4.5\mu\mathrm{m}}<T_{3.6\mu\mathrm{m}}<T_{\mathrm{J}}$ for cases with and without a thermal inversion, respectively, and under the assumptions of our models (e.g. chemistry and prescriptions for vertical mixing). Here, we investigate whether the planets in figure \ref{fig:Tblam} also exhibit this trend. To do this, we use HST/WFC3 brightness temperatures in the J-band as a proxy for J-band photometric brightness temperatures and compare these to Spitzer brightness temperatures. We also consider how this trend compares to inferences of thermal (non-)inversions from retrievals in the literature.

WASP-43b and HD~209458b are ideal cases to demonstrate the trend found in our models as they have water detections \citep{Kreidberg2014,Line2016}, suggesting an oxygen-rich composition. Both planets have also been found not to host thermal inversions using retrieval techniques \citep{Blecic2014,Kreidberg2014,Stevenson2014,Diamond-Lowe2014,Line2016}. Therefore, we expect both planets to have $T_{4.5\mu\mathrm{m}}<T_{3.6\mu\mathrm{m}}<T_{\mathrm{J}}$ as described above. Indeed, figure \ref{fig:Tblam} shows that this is the case. The HST/WFC3 brightness temperatures in the J-band are consistent with or hotter than that for IRAC 1 (within error bars), and both are higher than the IRAC 2 brightness temperature. WASP-18b shows the opposite trend, $T_{\mathrm{J}}<T_{3.6\mu\mathrm{m}}<T_{4.5\mu\mathrm{m}}$, which would suggest a thermal inversion and indeed agrees with the inference of a thermal inversion by \citet{Sheppard2017}. WASP-103b also has an inferred thermal inversion \citep{Kreidberg2018}. The brightness temperatures shown in figure \ref{fig:Tblam} are inconclusive as the errorbars on the Spitzer brightness temperatures are quite large, though note that \citet{Kreidberg2018} reduce the Spitzer data differently to \citet{Garhart2019}.

In contrast, WASP-121b, WASP-12b and WASP-33b show evidence for different chemistries/disequilibrium processes to our models. WASP-121b has been inferred to host a thermal inversion \citep{Evans2017}, but has a high J-band brightness temperature relative to IRAC 1 and IRAC 2, which are consistent within the errorbars. Furthermore, for WASP-12b the J-band and IRAC 2 brightness temperatures are significantly lower that that for IRAC 1. In the case of WASP-33b, the opposite is true: the J-band and IRAC 2 brightness temperatures are both significantly higher than that for IRAC 1. Thus, all three planets show a different brightness temperature trend compared to our oxygen-rich models. This could be due to chemical species which we have not considered here (e.g. FeH, TiH), other chemical disequilibrium mechanisms, a C/O ratio greater than 1 or offsets between instruments that are not considered here. Ultimately, while photometric metrics for assessing temperature gradients provide a useful initial assessment, their dependence on chemistry and the inherent complexity of planetary atmospheres is such that atmospheric retrievals are needed to confidently assess the presence of a thermal inversion or lack thereof.

Another use for comparing brightness temperatures for observed spectra is to provide a quick and empirical lower limit on the photospheric temperature gradient in the atmosphere. The larger the contrast between the maximum and minimum measured brightness temperatures, and the more precise they are, the stronger the constraints that can be placed on the photospheric temperature contrast. In particular, HST/WFC3 data can significantly improve the temperature contrasts determined with IRAC 1 and IRAC 2 data alone. For example, in the case of WASP-121b, the brightness temperature contrast based on the IRAC 1 and IRAC 2 measurements alone is 70 K $\pm$ 100 K, which is consistent with an isotherm. However, the range of brightness temperatures probed by the HST/WFC3 data is wider than that for the Spitzer data, and results in a temperature contrast of 380 K $\pm$ 70 K. Similarly, for WASP-33b, WASP-12b and WASP-18b, HST/WFC3 data also makes significant improvements to the brightness temperature contrast. Note that uncertainty in the host star temperature can affect the errors in these brightness temperatures. In the cases shown here, we find that the largest error contributions from the stellar temperature are comparable to those from the flux uncertainties alone (i.e. the errorbars shown in figure \ref{fig:Tblam}) and therefore do not impact our assessment significantly. However, this effect should be treated rigorously for more detailed studies.

One of the reasons that the HST/WFC3 data help to place tighter constraints on some brightness temperature contrasts is that the signal-to-noise of the data is very good and results in tight constraints on brightness temperature. This is despite the fact that, as shown by the blackbody curves in figure \ref{fig:metric}, a greater precision in flux is needed at shorter wavelengths for the same precision in brightness temperature. As a result, fairly precise spectral/photometric measurements are needed at shorter wavelengths in order to place tight constraints on brightness temperature. However, since the precision available with HST/WFC3 allows excellent brightness temperature precision (reaching $\sim$10 K for some cases in figure \ref{fig:Tblam}), this effect does not preclude shorter wavelengths from this type of analysis. When using observations from other instruments, however, it is important to consider how the uncertainties in the fluxes will correspond to uncertainty in brightness temperature, which will depend on the wavelength range of the observations.

While we have used examples with HST/WFC3 and Spitzer data here, this analysis can also be performed with other observations and at different wavelengths. It is important to note that when comparing brightness temperatures from different instruments (including the analysis above), systematic offsets between the instruments can bias the temperature contrast obtained. One way to test for and characterise such offsets is by performing an atmospheric retrieval which includes a freely-varying `offset' parameter, which can then be retrieved from the data. Furthermore, future instruments such as the James Webb Space Telescope's (JWST's) NIRSpec will have much larger spectral ranges, allowing brightness temperatures to be measured across a wide range of wavelengths with the same instrument.

\section{Summary and Discussion}
\label{sec:summary}
In this work, we investigate the importance of TiO in the spectral appearance and temperature structures of hot Jupiter atmospheres. To this end, we first examine the importance of the accuracy of TiO line lists for determining its observable spectral signatures in transmission and emission spectra, both at low and high resolution. We further investigate the effect of TiO on the temperature structure of an atmosphere. To do this, we investigate the strength of thermal inversions as a function of C/O ratio, equilibrium temperature, metallicity, stellar type and gravity, considering TiO as the inversion-causing molecule. In addition, we assess the performance of temperature gradient metrics as a function of these parameters.

We begin by comparing the recent \textsc{Toto} TiO line list \citep{McKemmish2019} to those of \citet{Schwenke1998} and Plez (2012) \citep[sourced from][]{Ryabchikova2015} (hereafter S98 and P12 line lists, respectively), focusing on the consequences for the interpretation of transmission, emission and high-resolution Doppler spectra. Interpretation of high-resolution Doppler spectra using cross-correlation methods relies on accurate line positions and sufficiently strong planetary spectral features. We assess the accuracy of line positions in the \textsc{Toto} line list by considering the fraction of energy levels and transitions which are taken from experimental data. For example, at 2000 K, there are 4491 transitions with line intensities above $10^{-17}$ cm/molecule, of which 83\% are from experimental data and have high accuracy. Furthermore, experimental data for the $v=2$ and higher vibrational excitations of the B $^3\Pi$ state are needed to fill in the strong spectral bands which are currently calculated theoretically.

For both transmission and emission spectra we find that the differences between the \textsc{Toto} and S98 line lists have the greatest impact on model spectra in the optical range. As a result, optical transmission spectra are strongly affected and the spectra generated with each line list can differ significantly depending on the spectral resolution (almost 1000 ppm at R$\sim 10^4$). Since thermal emission spectra of exoplanets are largely observed in the infrared spectral range, the differences between the \textsc{Toto} and S98 line lists have a more subtle effect. For two model spectra generated with the same $P$-$T$ profile but using each of the TiO line lists, the differences are only observable at higher resolutions (e.g. up to $\sim$100 ppm at a resolution of R$\sim 10^4$ in the near-infrared).

Beyond the effects of TiO line lists, we also explore how TiO equilibrium chemistry can shape the thermal profile of an atmosphere. We investigate the equilibrium abundances of chemical species (calculated using the \textsc{HSC Chemistry} software package) as a function of equilibrium temperature, $T_{\mathrm{eq}}$, C/O ratio, metallicity, stellar host type and planetary surface gravity. In turn, this allows an exploration of thermal inversions in this parameter space. Through semi-analytic considerations of radiative equilibrium, we find that thermal inversions can occur at the coolest equilibrium temperature for a C/O ratio of $\sim0.9$, consistent with previous studies \citep[e.g.][]{Molliere2015,Gandhi2019}.

We further calculate self-consistent 1D model atmospheres and their spectra for a range of C/O ratios $T_{\mathrm{eq}}$, metallicities, stellar types and gravities. We include the effects of vertical mixing, as well as the variation of the energy redistribution factor as a function of $T_{\mathrm{eq}}$. These models confirm the trend found by our simple analysis by which thermal inversions happen at cooler $T_{\mathrm{eq}}$ at C/O$\approx$0.9. Furthermore, the transition from non-inverted to inverted $P$-$T$ profiles happens at the expected temperature range of $\sim$1750-2000 K for models with solar-like composition, where TiO is known to condense. We also find that for higher-metallicity models, a higher equilibrium temperature is needed to cause a thermal inversion. Conversely, a later stellar type can result in thermal inversions occurring at cooler $T_{\mathrm{eq}}$. We find that gravity has a smaller effect on thermal inversion strength compared to other parameters.

In order to characterise the strength of the inversions/non-inversions in our models, we assess a range of metrics based on the contrast in brightness temperature between two photometric bands. Although the brightness temperature contrast between Spitzer's IRAC 1 and IRAC 2 channels is commonly used as an indication of inversion strength \citep[e.g.][]{Burrows2007,Knutson2010,Madhusudhan2010}, we find that in the case of H$_2$O-dominated spectra a stronger temperature contrast can be observed by comparing the IRAC 2 channel to the H-band or J-band. Furthermore, we find that for high C/O ratios and low equilibrium temperatures, the IRAC 1 band contains significant opacity from CH$_4$ and HCN and, as such, does not provide a good measure of the spectral continuum. In this regime, an atmosphere without a thermal inversion can have a larger IRAC 2 flux compared to the IRAC 1 flux, mimicking the signature of a thermal inversion for lower C/O ratios. Similarly, the J-band incorrectly infers a non-inversion at some C/O ratios for $T_\mathrm{eq}\sim$1750 K, for example with a later stellar host type. Therefore, neither metric works optimally across the whole parameter space considered. We investigate this for observed spectra by finding the brightness temperature contrasts between the WFC3 and Spitzer IRAC 1 and IRAC 2 bands for various planets. While our oxygen-rich models show the trend that the J-band is a clearer continuum band than the IRAC 1 band, several of these planets do not follow the same pattern. This is likely due to differences in chemistry and/or disequilibrium effects not considered in our models. Overall, we suggest that continuum opacity bands depend on the atmospheric chemistry and that temperature gradient metrics should therefore be used with caution as a single metric does not necessarily apply to all atmospheric chemistries.

Thermal inversions and the role of TiO in producing them remains a key part in understanding exoplanet atmospheres and the processes which shape them. As exoplanetary atmospheric observations continue to improve, it will be necessary to use the highest-quality chemical cross-sections and understand the role of the many parameters which dictate the structure of an atmosphere in order to demystify these exotic worlds.

\section*{Acknowledgements}
A.A.A.P. thanks John Harrison for help with using \textsc{HSC Chemistry}, and acknowledges financial support from the Science and Technology Facilities Council (STFC), UK, towards her doctoral programme. SG acknowledges support from the UK Science and Technology Facilities Council (STFC) research grant ST/S000631/1. We thank the anonymous reviewer for their very helpful comments on our manuscript.

%%%%%%%%%%%%%%%%%%%%%%%%%%%%%%%%%%%%%%%%%%%%%%%%%%



\bibliographystyle{mnras}
\bibliography{refs}

%%%%%%%%%%%%%%%%%%%%%%%%%%%%%%%%%%%%%%%%%%%%%%%%%%

%%%%%%%%%%%%%%%%% APPENDICES %%%%%%%%%%%%%%%%%%%%%

\appendix

\section{Equilibrium Chemistry Calculations}
In the equilibrium chemistry calculations performed using the \textsc{HSC Chemistry} software package, we use the species given in \citet{Bond2010} and \citet{Harrison2018} and include additional relevant species, as listed in table \ref{tab:chem}. Although potassium and vanadium are not present in the equilibrium chemistry calculations, we include them in our atmospheric models by giving them the same vertical profiles as sodium and titanium, respectively, scaled according to the solar K/Na and V/Ti ratios, respectively.

\begin{table}
    \centering
    \begin{tabular}{c | c}
        Element & Species added \\
        \hline
        H & H$^-$, H$^+$, H$_2^-$, H$_2^+$, H$_3^+$, H$_3$ , $(\mathrm{H}_3)_2$, H(H$_3$) \\[3pt]
        O & O$^-$, O$^+$, O$^-2$, O$_2^-$, O$_2^+$, O$_2^-2$, O$_3$ \\[3pt]
        C & C$^-$, C$^+$, C$_2^-$, C$_2^+$, C$_2$, C$_3$, C$_4$, C$_5$, C$_6$, C$_7$, C$_8$, C$_{60}$, C$_{70}$ \\[3pt]
        N & N$^-$, N$^+$, N$_2^-$, N$_2^+$ N$_3^-$, N$_3$ \\[3pt]
        Ti & Ti$^-$, Ti$^+$, Ti$^3+$, Ti$_2$ \\[3pt]
        Na & Na, Na$^-$, Na$^-$, Na$_2$ \\[3pt]
        Other & OH, OH$^-$, OH$^+$, C$_2$H$_2$, C$_2$H$_4$, e$^-$

    \end{tabular}
    \caption{Species added to the \textsc{HSC Chemistry} equilibrium chemistry calculations, on top of those used in \citet{Bond2010} and \citet{Harrison2018}. All species listed here are in the gas phase.}
    \label{tab:chem}
\end{table}

\section{Effect of Temperature Gradient on High-Resolution Spectra}
\label{sec:hires_PTgrad}
%%%%%%%%%%%%%%%%%%%%% -- Figures -- %%%%%%%%%%%%%%%%%%%%%
\begin{figure}
    \centering
    \includegraphics[width=0.5\textwidth]{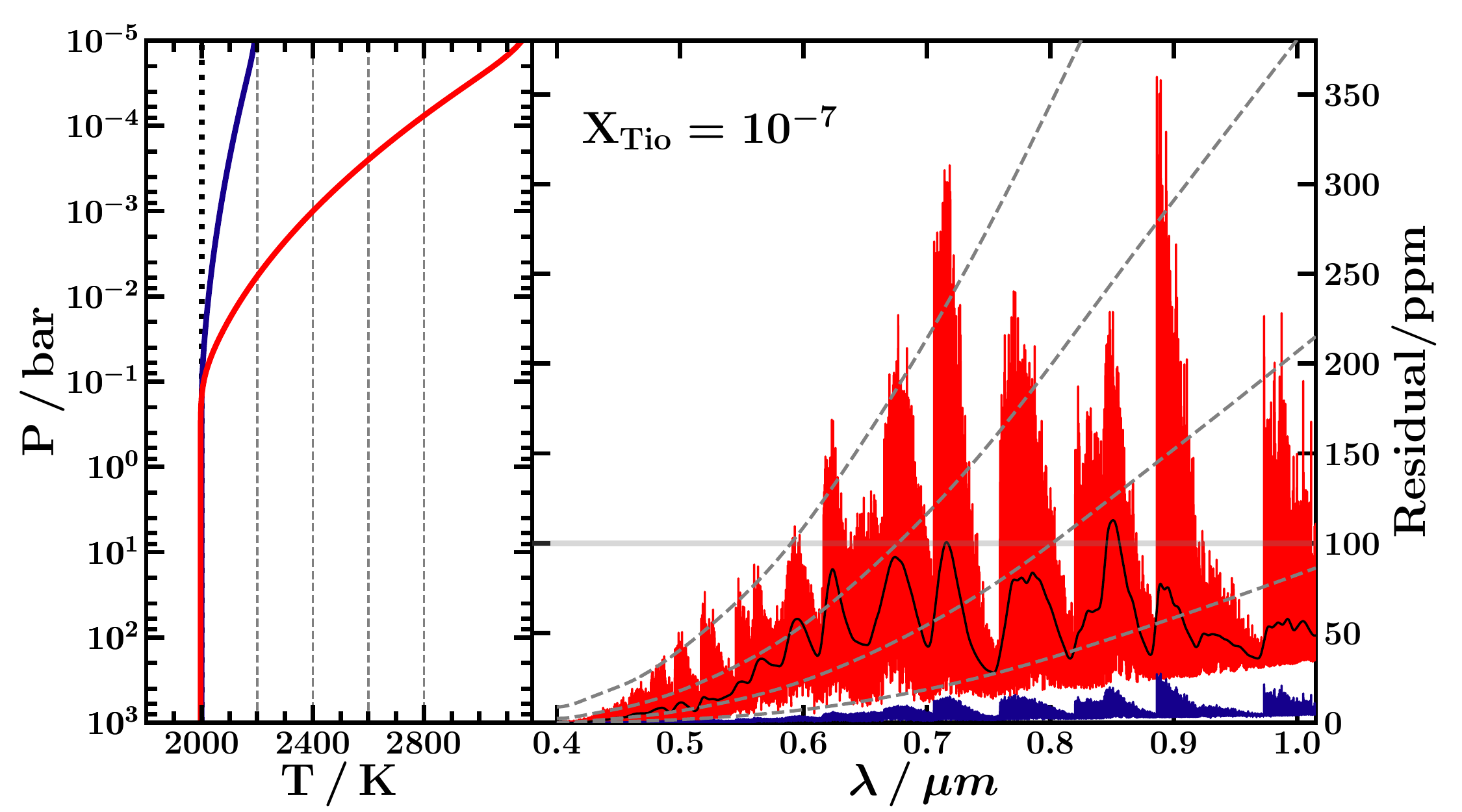}
    \caption{Line strengths of high-resolution (R$\sim 10^5$) spectra compared to a blackbody spectrum for different temperature gradients. Residuals are calculated as the planet-star flux ratio of the atmosphere in question minus the planet-star flux ratio of an atmosphere whose $P$-$T$ profile is a 2000 K isotherm (dashed black line in left panel). Right panel: red and blue residuals correspond to the red and blue $P$-$T$ profiles in the left panel, respectively. Dashed grey lines correspond to isothermal $P$-$T$ profiles at 2200 K, 2400 K, 2600 K and 2800 K, respectively (shown by dashed grey lines in left panel). The solid back line is the residual for the same spectra as the red line, but with both spectra smoothed by a Gaussian of width 2.3 nm (similar to the PSF of HST/WFC3) to show representative line strengths for low-resolution spectra. The 100 ppm level is shown by the solid light grey line. These models only include opacity from TiO and CIA to highlight the TiO spectral lines. A constant-with-depth TiO mixing ratio of $10^{-7}$ is used.}
    \label{fig:PTres_hires}
\end{figure}

\begin{figure*}
    \centering
    \includegraphics[width=\textwidth]{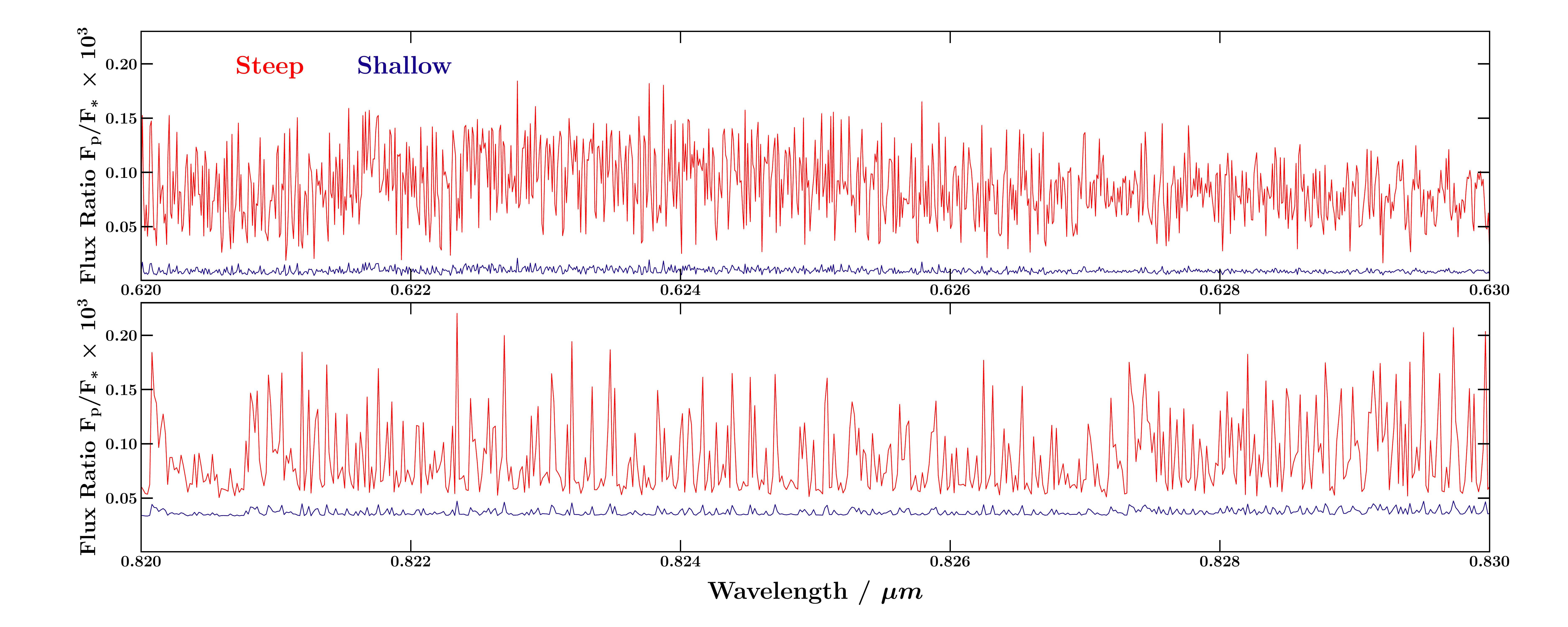}
    \caption{High-resolution (R$\sim 10^5$) model spectra corresponding to a steep (red) and shallow (blue) $P$-$T$ profile (steep and shallow $P$-$T$ profiles are shown in the left panel of figure \ref{fig:PTres_hires} in red and blue, respectively). The spectra are calculated in the same way as in figure \ref{fig:PTres_hires}. Each panel spans 0.01 $\mu$m in the range 0.62-0.83 $\mu$m to give an overview of the differences in line strength between the two spectra in this range.}
    \label{fig:highres_forest}
\end{figure*}
%%%%%%%%%%%%%%%%%%%%%%%%%%%%%%%%%%%%%%%%%%%%%%%%%%%%%%%%%

Beyond line position accuracy, detecting a chemical signature from high-resolution Doppler spectra with cross-correlation methods also requires the planetary spectral features to be sufficiently strong. A significant factor in determining the strength of features in a spectrum is the atmospheric temperature gradient; a steeper temperature slope results in stronger absorption or emission lines (depending on the direction of the gradient).

Here, we consider the effect of the temperature gradient on line strength in high-resolution spectra. Figure \ref{fig:PTres_hires} shows high-resolution spectra for two different temperature gradients relative to an isotherm (red and blue lines correspond to the steep and shallow temperature gradients, respectively). It is clear that the steeper temperature gradient results in significantly stronger features. This suggests that chemical signatures are acutely sensitive to the temperature gradient, especially at high resolution. For example, a 400 K temperature difference at 1 mbar can cause differences in the spectrum of up to 200 ppm, which may be observable. Conversely, the absence of any detectable chemical signatures in a high-resolution Doppler spectrum could be the result of a $P$-$T$ profile which is very close to isothermal. Although this is also true of low-resolution spectra, high-resolution spectra are more sensitive to temperature gradient and a lack of spectral features could potentially place a tighter constraint on this gradient compared to low-resolution spectra. This can be seen from figure \ref{fig:PTres_hires}; the solid black line in the right panel shows the residual spectrum at lower-resolution (to generate this, the spectra used to calculate the red line are smoothed by a Gaussian of width 2.3 nm before calculating the residual, to represent smoothing by the PSF of HST/WFC3). By comparing the high-resolution (red) and low-resolution (black) residuals, we see that the features are smoothed out and weakened at low-resolution and, as such, high-resolution spectra are more sensitive to shallow temperature gradients given a fixed observation sensitivity. This effect can be seen in greater detail in figure \ref{fig:highres_forest}, which shows a zoomed-in view of the high-resolution spectra corresponding to the blue and red $P$-$T$ profiles from figure 6.

Figure \ref{fig:PTres_hires} also demonstrates that high-resolution spectra are sensitive to very low pressures in the atmosphere. The dashed grey lines in the right panel correspond to isotherms at 2200 K, 2400 K, 2600 K and 2800 K, and their crossing points with each residual spectrum indicate the temperatures probed by the spectrum at those wavelengths. For example, the strongest features in the red spectrum cross the spectrum corresponding to 2800 K, showing that this high-resolution spectrum is sensitive down to pressures of $\sim 10^{-4}$ bar or lower (i.e. the pressure at which the temperature of the red $P$-$T$ profile is 2800 K). Conversely, the low-resolution residual spectrum does not probe temperatures hotter than $\sim$ 2600 K, which corresponds to a pressure of $\sim 3\times 10^{-4}$ bar for this $P$-$T$ profile, and does so with a significantly smaller signal.

\section{Effect of metallicity on equilibrium chemistry}
\label{sec:metallicity_plot}
\begin{figure}
    \centering
    \includegraphics[width=0.5\textwidth]{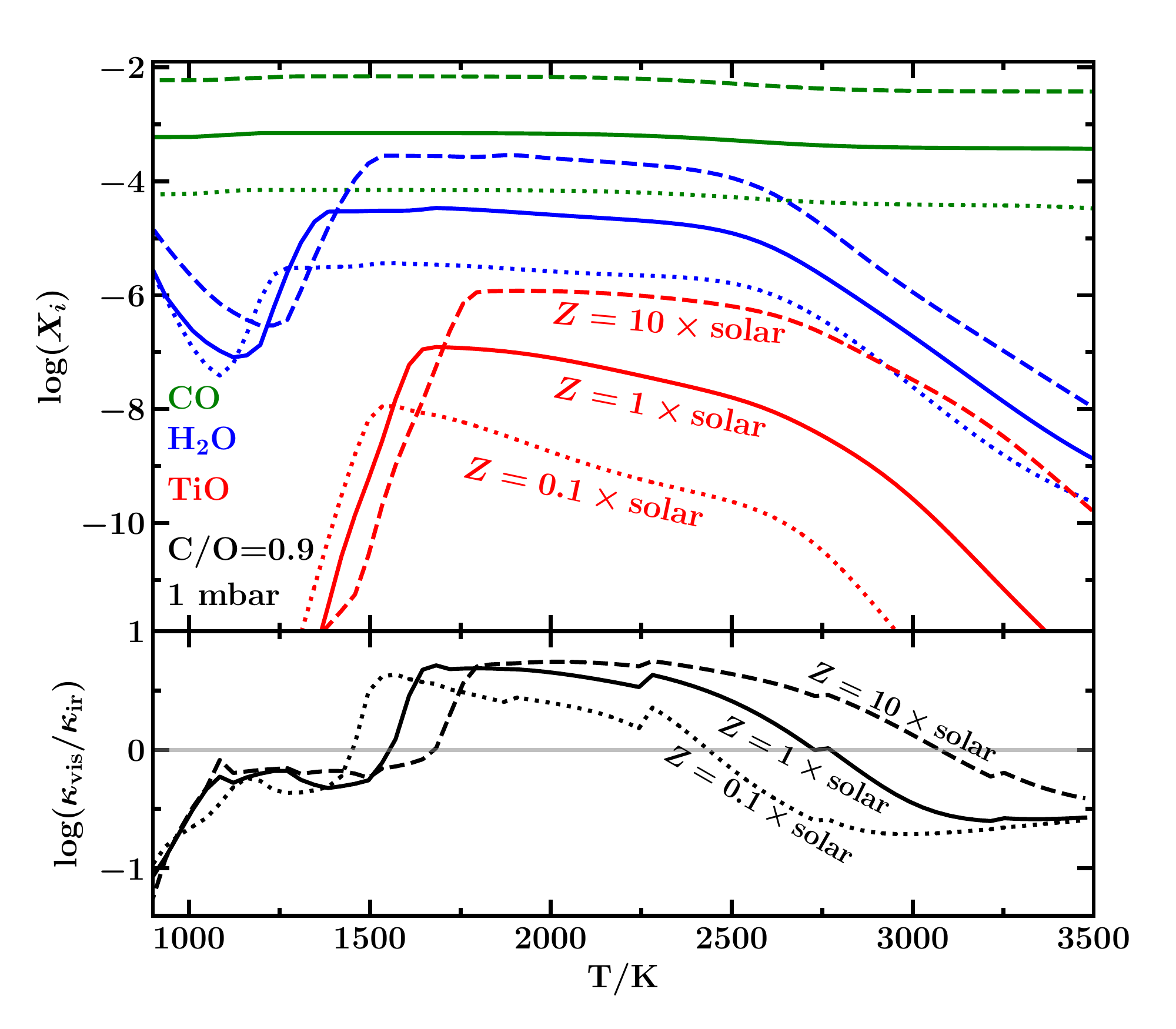}
    \caption{Top panel: abundances of TiO (red), H$_2$O (blue) and CO (green) as a function of temperature for metallicities of 0.1, 1 and 10$\times$ solar (dotted, solid and dashed lines, respectively). Abundances are shown for a pressure of 1 mbar and a C/O ratio of 0.9. Lower panel: $\kappa_\mathrm{vis}$/$\kappa_\mathrm{ir}$ for metallicities of 0.1, 1 and 10$\times$ solar (dotted, solid and dashed lines, respectively), also evaluated at 1 mbar and C/O=0.9.}
    \label{fig:met_plot}
\end{figure}

Figure \ref{fig:met_plot} shows the effect of metallicity on the equilibrium abundances of TiO, H$_2$O and CO as well as $\kappa_\mathrm{vis}$/$\kappa_\mathrm{ir}$. As expected, the abundances increase with metallicity. Furthermore, as metallicity increases, the partial pressure of TiO also increases and the temperatures at which TiO is depleted by condensation or thermal dissociation also increase. As a result, $\kappa_\mathrm{vis}$/$\kappa_\mathrm{ir}$ peaks at higher temperature for higher metallicity.
%%%%%%%%%%%%%%%%%%%%% -- Figures -- %%%%%%%%%%%%%%%%%%%%%

%%%%%%%%%%%%%%%%%%%%%%%%%%%%%%%%%%%%%%%%%%%%%%%%%%

% Don't change these lines
\bsp	% typesetting comment
\label{lastpage}
\end{document}